\documentclass [10pt, a4paper]{article}
\usepackage {etex}

\usepackage[english]{babel}
\usepackage [T1]{fontenc}
\usepackage [utf8] {inputenc}
\usepackage {lmodern}
\usepackage [left=2cm, right=2cm, top=2cm, bottom=2cm]{geometry}
\usepackage {amsfonts}
\usepackage {amsmath}	
\usepackage {amssymb}
\usepackage {amsxtra}
\usepackage {wasysym}
\usepackage {mathtools}
\usepackage {calrsfs}
\usepackage {dsfont}
\usepackage {fancyhdr}
\usepackage {graphicx}
\usepackage [amsthm,thmmarks,amsmath]{ntheorem}
\usepackage {pstricks}
\usepackage {pst-plot}
\usepackage {pst-vue3d}
\usepackage {exscale}
\usepackage {cite}
\usepackage {setspace}
\usepackage {color}
\usepackage [labelfont=bf,labelsep=endash,width=.8\textwidth]{caption}
\usepackage [bottom]{footmisc}
\usepackage [all,cmtip]{xy}
\usepackage{hyperref}
\usepackage[affil-it, blocks]{authblk}
\usepackage {diagbox}
\usepackage {float}
\usepackage [nottoc,numbib]{tocbibind}

\allowdisplaybreaks
\singlespace
\input diagxy

\makeatletter
\@addtoreset{equation}{section}
\renewcommand\theequation{\thesection.\@arabic\c@equation}
\renewcommand\thefigure{\thesection.\@arabic\c@figure}
\renewcommand\thetable{\thesection.\@arabic\c@table}
\makeatother

\DeclareFontFamily{OT1}{pzc}{}
\DeclareFontShape{OT1}{pzc}{m}{it}{<-> s * [1.30] pzcmi7t}{}
\DeclareMathAlphabet{\mathpzc}{OT1}{pzc}{m}{it}

\DeclareMathOperator{\R}{\mathbb R}
\DeclareMathOperator{\C}{\mathbb C}
\DeclareMathOperator{\Cyl}{\textup{Cyl}}

\DeclareMathOperator{\Flux}{\textup{Flux}}
\DeclareMathOperator{\LQG}{\textup{LQG}}

\DeclareMathOperator{\uA}{\textup A}
\DeclareMathOperator{\uB}{\textup B}
\DeclareMathOperator{\ud}{\textup d}
\DeclareMathOperator{\uR}{\textup R}
\DeclareMathOperator{\uL}{\textup L}
\DeclareMathOperator{\uG}{\textup G}
\DeclareMathOperator{\uK}{\textup K}
\DeclareMathOperator{\uP}{\textup P}
\DeclareMathOperator{\fA}{\mathfrak A}

\DeclareMathOperator{\fg}{\mathfrak g}
\DeclareMathOperator{\fk}{\mathfrak k}

\DeclareMathOperator{\fH}{\mathfrak H}
\DeclareMathOperator{\fG}{\mathfrak G}
\DeclareMathOperator{\cG}{\mathcal G}
\DeclareMathOperator{\cP}{\mathcal P}
\DeclareMathOperator{\aut}{\textup{Aut}}
\DeclareMathOperator{\End}{\textup{End}}
\DeclareMathOperator{\hol}{\textup hol}

\DeclareMathOperator*{\slim}{\textup{s-lim}}
\DeclareMathOperator{\vol}{vol}
\DeclareMathOperator{\id}{id}

\DeclareMathOperator{\tr}{tr}

\newcommand{\BIGOP}[1]{\mathop{\mathchoice
{\raise-0.22em\hbox{\huge $#1$}}
{\raise-0.05em\hbox{\Large $#1$}}{\hbox{\large $#1$}}{#1}}}

\renewcommand{\L}{\mathcal L}
\renewcommand{\div}{\mathrm{div}}
\renewcommand{\hom}{\textup{Hom}}

\newtheoremstyle{breakdef}%
  {\item[\rlap{\vbox{\normalfont\bfseries\hbox{\llap{##2}\hskip\labelsep
          ##1:}\hbox{\\[0.1cm]}}}]}%
  {\item[\rlap{\vbox{\normalfont\bfseries\hbox{\llap{##2}\hskip\labelsep
          ##1 (##3):}\hbox{\\[0.1cm]}}}]}
\newtheoremstyle{breaksatz}%
  {\item[\rlap{\vbox{\normalfont\normalsize\bfseries\hbox{\llap{##2}\hskip\labelsep
          ##1:}\hbox{\\[0.1cm]}}}]}%
  {\item[\rlap{\vbox{\normalfont\normalsize\bfseries\hbox{\llap{##2}\hskip\labelsep
          ##1 (##3):}\hbox{\\[0.1cm]}}}]}
\newtheoremstyle{breaklem}%
  {\item[\rlap{\vbox{\normalfont\normalsize\bfseries\hbox{\llap{##2}\hskip\labelsep
          ##1:}\hbox{\\[0.1cm]}}}]}%
  {\item[\rlap{\vbox{\normalfont\normalsize\bfseries\hbox{\llap{##2}\hskip\labelsep
          ##1 (##3):}\hbox{\\[0.1cm]}}}]}
\newtheoremstyle{breakprop}%
  {\item[\rlap{\vbox{\normalfont\normalsize\bfseries\hbox{\llap{##2}\hskip\labelsep
          ##1:}\hbox{\\[0.1cm]}}}]}%
  {\item[\rlap{\vbox{\normalfont\normalsize\bfseries\hbox{\llap{##2}\hskip\labelsep
          ##1 (##3):}\hbox{\\[0.1cm]}}}]}
\newtheoremstyle{breakbem}%
  {\item[\rlap{\vbox{\hbox{\hskip\labelsep\normalfont\bfseries
          ##1 ##2:}\hbox{\\[0.1cm]}}}]}%
  {\item[\rlap{\vbox{\hbox{\hskip\labelsep\normalfont\bfseries
          ##1 ##2 (##3):}\hbox{\\[0.1cm]}}}]}
\newtheoremstyle{breakbsp}%
  {\item[\rlap{\vbox{\hbox{\hskip\labelsep\normalfont\bfseries
          ##1 ##2:}\hbox{\\[0.2cm]}}}]}%
  {\item[\rlap{\vbox{\hbox{\hskip\labelsep\normalfont\bfseries
          ##1 ##2 (##3):}\hbox{\\[0.2cm]}}}]}
\newtheoremstyle{breakkor}%
  {\item[\rlap{\vbox{\hbox{\hskip\labelsep\normalfont\bfseries
          ##1 ##2:}\hbox{\\[0.1cm]}}}]}%
  {\item[\rlap{\vbox{\hbox{\hskip\labelsep\normalfont\bfseries
          ##1 ##2 (##3):}\hbox{\\[0.1cm]}}}]}
\newtheoremstyle{proof}%
  {\item[\rlap{\vbox{\hbox{\hskip\labelsep\normalfont\bfseries
          \underline{##1:}}\hbox{\\[0.1cm]}}}]}%
  {\item[\rlap{\vbox{\hbox{\hskip\labelsep\normalfont\bfseries
          \underline{##1 (##3):}}\hbox{\\[0.1cm]}}}]}
\theoremstyle{breakkor} 
\newtheorem{Definition}{Definition}[section] 
\theoremstyle{breakkor}
\newtheorem{Theorem}[Definition]{Theorem}
\theoremstyle{breakkor}
\newtheorem{Lemma}[Definition]{Lemma}
\theoremstyle{breakkor}
\newtheorem{Proposition}[Definition]{Proposition}
\theoremstyle{breakkor}
\newtheorem{Corollary}[Definition]{Corollary}
\theoremstyle{breakkor}
\theorembodyfont{\normalfont}
\newtheorem{Remark}[Definition]{Remark}

\theoremstyle{proof}
\theoremsymbol{\fbox{}}
\newtheorem{Proof}{Proof}

\begin{document}

\lhead[\thepage]{\rightmark} 
\chead[]{}
\rhead[\leftmark]{\thepage}
\lfoot[]{}
\cfoot[]{}
\rfoot[]{}

\thispagestyle{empty}

\title{\Large\bf Structural aspects of loop quantum gravity and loop quantum cosmology from an algebraic perspective}
\author[]{Alexander Stottmeister\thanks{{\sf 
\hspace{0.1cm}alexander.stottmeister@gravity.fau.de}}\ }
\author[]{Thomas Thiemann\thanks{{\sf 
\hspace{0.1cm}thomas.thiemann@gravity.fau.de}}}
\affil[]{Institut für Quantengravitation, Lehrstuhl für Theoretische Physik III, \\ Friedrich-Alexander-Universtität Erlangen-Nürnberg, \\ Staudtstraße 7/B2, D-91058 Erlangen, Germany}

\maketitle

\begin{abstract}
We comment on structural properties of the algebras $\mathfrak{A}_{\textup{LQG/LQC}}$ underlying loop quantum gravity and loop quantum cosmology, especially the representation theory, relating the appearance of the (dynamically induced) superselection structure ($\theta$-sectors) in loop quantum cosmology to recently proposed representations with non-degenerate background geometries in loop quantum gravity with Abelian structure group. To this end, we review and employ the concept of extending a given (observable) algebra with possibly non-trivial centre to a (charged) field algebra with (global) gauge group. We also interpret the results in terms of the geometry of the structure group $\uG$. Furthermore, we analyze the Koslowski-Sahlmann representations with non-degenerate background in the case of a non-Abelian structure group. We find that these representations can be interpreted from two different, though related, points view: Either, the standard algebras of loop quantum gravity need to be extended by a (possibly) central term, or the elementary flux vector fields need to acquire a shift related to the (classical) background to make these representations well-defined. Both perspectives are linked by the fact that the background shift is not an automorphism of the algebras, but rather an affine transformation. A third perspective is offered by the recent construction of the holonomy-background flux-exponential algebra due to Campiglia and Varadarajan, which modifies the structure group of the standard holonomy-flux algebra by an additional $U(1)^{N}$-factor such that the Koslowski-Sahlmann representations are applicable. Finally, we show how similar algebraic mechanisms that are used to explain the breaking of chiral symmetry and the occurrence of $\theta$-vacua in quantum field theory extend to loop quantum gravity. Thus, opening a path for the discussion of these questions in loop quantum gravity. 
\end{abstract}

\tableofcontents

\section{Introduction}
\label{sec:intro}
Loop quantum gravity is based on a Hamiltonian formulation of general relativity in terms of a constrained Yang-Mills-type theory, i.e. in a field theoretic description the phase space of the classical theory is given by the (densitiezed) cotangent bundle $|\Lambda|^{1}T^{*}\mathcal{A}_{\uP}$ to the space of connections on a given (right) principal $\uG$-bundle $\uP\stackrel{\pi}{\rightarrow}\Sigma$, where $\Sigma$ is the spatial manifold in a 3+1-splitting of a (globally hyperbolic) spacetime $\textup{M}\cong\R\times\Sigma$. In general relativity, we have $\uG=\textup{SU(2)}, \textup{Spin}_{4}$, or central quotients of these groups.\\[0.1cm]
The basic variables, the theory is phrased in, are the Ashtekar-Barbero connection $A\in\mathcal{A}_{\uP}$ and its conjugate momentum $E\in\Gamma\left(T\Sigma\otimes\textup{Ad}^{*}(\uP)\otimes|\Lambda|^{1}(\Sigma)\right)$. Strictly speaking, we further require $E$ to be non-degenerate as a (densitiezed) section of the bundle of linear operators $\textup{L}(\textup{Ad}(\uP),T\Sigma)$. In general relativity, the existence of $E$ is ensured by the triviality of the orthogonal frame bundle $P_{\textup{SO}}(\Sigma)$. This mathematical setup also appears to be valid in the context of the new variables proposed in \cite{BodendorferNewVariablesFor1, BodendorferNewVariablesFor2}. Here, $\textup{Ad}^{*}(\uP)=\uP\times_{\textup{Ad}^{*}}\fg^{*}$ and $|\Lambda|^{1}(\Sigma)$ denotes the bundle of 1-densities on $\Sigma$. Since $\mathcal{A}_{\uP}$ is an affine space modeled on $\Omega^{1}(\textup{Ad}(\uP))=\Gamma(T^{*}\Sigma\otimes\textup{Ad}(\uP))$, $\textup{Ad}(\uP)=\uP\times_{\textup{Ad}}\fg$ , the following Poisson structure
\begin{align}
\label{eq:poisson}
\{E^{a}_{i}(x),A^{j}_{b}(y)\}=\delta^{a}_{b}\delta^{j}_{i}\delta(x,y)
\end{align}
is meaningful in local coordinates $\phi:U\subset\Sigma\rightarrow V\subset\R^{3}$ subordinate to a local trivialization $\psi:\uP_{|U}\rightarrow U\times\uG$, i.e.
\begin{align}
\label{eq:loccoord}
((\phi\circ\psi)^{-1})^{*}A_{|P_{|U}}=A^{j}_{b}dx^{b}\otimes\tau_{j},\ (\phi\circ\psi)_{*}E_{|P_{|U}}=E^{a}_{i}\frac{\partial}{\partial x^{a}}\otimes\tau^{*i}.
\end{align}
Here, $\{\tau_{j}\}_{j}$ is a basis of $\fg$ and $\{\tau^{*i}\}_{i}$ its dual in $\fg^{*}$.\\[0.1cm]
The variables $(A,E)$ are directly related to the Arnowitt-Deser-Misner variables $(q,P)$. Namely, $E^{a}_{i}$ is a densitiezed dreibein for the spatial metric $q_{ab}E^{a}_{i}E^{b}_{j}=\det(q)\delta_{ij}$, and $A^{i}_{a}=\Gamma^{i}_{a}+K^{i}_{a}$ is built out of the Levi-Civita connection $\Gamma$ of the spatial metric $q$ and the extrinsic curvature $K$ determined by the momentum $P$.\\[0.1cm]
What makes the variables $(A,E)$ special, is that they allow to carry out a canonical quantization of general relativity, i.e. loop quantum gravity (cf. \cite{RovelliQuantumGravity, ThiemannModernCanonicalQuantum} for general accounts on the topic). Especially, it is possible to construct mathematically well-defined operators for all constraints acting in a suitable Hilbert space within this approach, most prominently the Wheeler-DeWitt constraint (cf. \cite{ThiemannQuantumSpinDynamics1, ThiemannQuantumSpinDynamics2, ThiemannQuantumSpinDynamics3, ThiemannQuantumSpinDynamics4, ThiemannQuantumSpinDynamics5, ThiemannQuantumSpinDynamics6, ThiemannQuantumSpinDynamics7}). \\[0.1cm]
The process of canonical quantization of constrained system in the sense of Dirac can roughly be divided into four steps: First, a point-separating Poisson algebra of function(al)s on the classical phase space is identified. Second, an abstract quantum *-algebra based on the Poisson algebra is defined. Third, a representation of the quantum *-algebra is chosen. Fourth, operators corresponding to the constraint are constructed in the chosen representation, and invariant (sub)spaces w.r.t. to these are selected as physical Hilbert spaces.\\[0.1cm]
In this article, we focus on the third step of this program. That is, we will analyze structures of loop quantum gravity related to the representation theory of a choice of quantum *-algebra. We will mainly work in the setting of the F/LOST theorem \cite{LewandowskiUniquenessOfDiffeomorphism, FleischhackRepresentationsOfThe}, which is an analog of the von Neumann uniqueness theorem for diffeomorphism invariant theories. Therefore, the classical Poisson algebra will be given by the Ashtekar-Corichi-Zapata algebra \cite{AshtekarQuantumTheoryOf1, AshtekarQuantumTheoryOf2, AshtekarQuantumTheoryOf3}, which is based on the Ashtekar-Isham configuration space \cite{AshtekarRepresentationsOfThe} of generalized connections and its associated differential calculus \cite{AshtekarDifferentialGeometryOn} (see also \cite{AbbatiOnDifferentialStructure}). As quantum *-algebra, we will use the holonomy-flux algebra in the semi-analytic category, which was defined in \cite{LewandowskiUniquenessOfDiffeomorphism}, or a certain Weyl form of this algebra \cite{SahlmannOnTheSuperselection}. Although, the F/LOST theorem states the uniqueness of a diffeomorphism invariant, pure state on the holonomy-flux algebra or a (concrete) Weyl form of it, which leads to a unitary implementation of the diffeomorphisms in the associated Gel’fand-Naimark-Segal (GNS) representation, the Ashtekar-Isham-Lewandowski representation, it was pointed out by several authors \cite{VaradarajanTowardsNewBackground, DziendzikowskiNewDiffeomorphismInvariant, SahlmannOnLoopQuantum, KoslowskiLoopQuantumGravity} that some of the underlying assumptions of the theorem have a rather technical flavor and could be weakened from a physical perspective while others are not strictly necessary to from a mathematical point of view to achieve a unitary implementation of the diffeomorphisms. Another issue, which was raised in \cite{KoslowskiDynamicalQuantumGeometry} and followed upon in \cite{SahlmannOnLoopQuantum, KoslowskiLoopQuantumGravity}, is the peculiar nature of the GNS vacuum of the Ashtekar-Isham-Lewandowski representation describing the extremely degenerate situation of an empty geometry. While this appears to be a valid ground state for the deep quantum regime of a quantum theory of gravity, where geometry is built from excitations of the gravitational field, such a state is not well suited for semi-classical considerations, where a classical background geometry needs to be approximated. Therefore, candidates for representations with ground states capturing information on a fixed background geometry were proposed: The Koslowski-Sahlmann representations. Quite recently \cite{VaradarajanTheGeneratorOf, CampigliaTheKoslowskiSahlmann1, CampigliaTheKoslowskiSahlmann2, CampigliaAQuantumKinematics, SenguptaQuantumGeometryWith, SenguptaAsymptoticFlatnessAnd}, these candidates were analyzed with a focus on their applicability to asymptotically flat boundary conditions for the gravitational field, which require a non-degenerate geometry at spatial infinity.\\[0.1cm]
Although, we will discuss certain aspects of the Koslowski-Sahlmann representations, and point out the need to extend the standard holonomy-flux algebra to make these representations well-defined, e.g. by admitting additional ``central terms'' in the commutation relations of the fluxes or by the use of the holonomy-background flux-exponential algebra, as recently pointed out by Campiglia and Varadarajan (cf. especially \cite{CampigliaTheKoslowskiSahlmann2}), the main focus of the article lies on structural aspects of the quantum *-algebras, which are related to non-trivial geometrical and topological features of the structure group of the underlying Yang-Mills-type theory. More precisely, we observe that the use of a compact structure group $\uG$ leads to a non-trivial center in the Weyl form of the holonomy-flux algebra, which clearly affects the representation theory, because central elements need to be given by multiples of the identity in irreducible representations. Similar features are known in quantum mechanical models \cite{StrocchiAnIntroductionTo, LoeffelholzMathematicalStructureOf, StrocchiNonPerturbativeFoundations, MorchioQuantumMechanicsOn}. Moreover, we point out distinctive features between the cases where $\uG$ is Abelian or non-Abelian, and find that the representation theory is severely more constrained in the latter case. The Koslowski-Sahlmann representation can be interpreted in this setting, as well. We also identify a purely topological feature, which leaves its imprint in the representation theory. Namely, the existence of a sequence of coverings
\begin{align}
\label{eq:coveringsequence}
\tilde{\uG}\rightarrow...\rightarrow\uG\rightarrow...\rightarrow\tilde{\uG}/Z(\tilde{\uG}),
\end{align}
where $\tilde{\uG}$ is the simply connected cover of $\uG$ and $Z(\tilde{\uG})$ its center, accompanied by a sequence of non-trivial bundle coverings
\begin{align}
\label{eq:bundlecoveringsequence}
\uP_{\tilde{\uG}}\rightarrow...\rightarrow\uP\rightarrow...\rightarrow\uP_{\tilde{\uG}/Z(\tilde{\uG})}
\end{align}
allows for the construction of a sequence of extensions of *-algebras
\begin{align}
\label{eq:extensionsequence}
\mathfrak{A}_{\tilde{\uG}}\rightarrow...\rightarrow\mathfrak{A}\rightarrow...\rightarrow\mathfrak{A}_{\tilde{\uG}/Z(\tilde{\uG})}.
\end{align}
Such a sequence of extensions gives rise to another type of candidates for new representations of the quantum *-algebra, which are in some sense complementary to the Koslowski-Sahlmann representations.\\[0.1cm]
These structures resemble in many aspects a rigorous, fully quantum theoretical discussion of chiral symmetry breaking and the related $\theta$-vacua in quantum field theory \cite{JackiwTopologicalInvestigationsOf, MorchioChiralSymmetryBreaking}. That is, the existence of large gauge transformations, $\pi_{3}(G)\neq\{1\}$, is reflected in a non-trivial center of the (observable) algebra, and the anomalous chiral symmetry does not leave the center point-wise invariant, thus leading to a spontaneous breakdown of the chiral symmetry and the appearance of the $\theta$-sectors. Interestingly, the main arguments of \cite{MorchioChiralSymmetryBreaking} can be transferred to the framework of loop quantum gravity, if the existence of an anomalous chiral symmetry is assumed. This provides a first step towards a discussion of anomalies in loop quantum gravity, which is a important issue in the analysis of the semi-classical limit of the theory, especially in the presence of additional matter degrees of freedom. More precisely, since anomalies lead to non-trivial prediction concerning the matter content of quantum field theory, it is necessary to establish a relation to such results in this limit. Thus, our observation will allow to draw conclusions in loop quantum gravity similar those of quantum field theory, if the presence of a chiral anomaly is achieved, either in full quantum theory or the semi-classical limit only. An arena for detailed investigations of these issue could be given by the so-called deparametrized models (see \cite{GieselScalarMaterialReference} for an overview).\\[0.25cm]
The article is organized as follows:\\[0.1cm]
In section \ref{sec:pre}, we provide a review of the mathematical background required to give precise definitions of the algebraic structures employed in loop quantum gravity. Specifically, we use subsection \ref{sec:bundles} to recall some facts from the theory of (principal) fibre bundles, which are the basis for the (classical) phase space formulation of loop quantum gravity. In subsection \ref{sec:prelqg} and \ref{sec:prelqc}, we introduce the (quantum) algebras and states, which form the standard setting of loop quantum gravity. Readers, which are familiar with these topics and/or are mainly interested in the results, can skip this section and use it as a reference.\\[0.1cm]
In section \ref{sec:centers}, we show that those algebras possess non-trivial centers, which are closely related to geometric and topological properties of the structure group, and affect their representation theory, e.g. by the appearance of the Koslowski-Sahlmann representations. Moreover, if the structure group is not simply connected, $\pi_{1}(\uG)\neq\{1\}$, we provide a mechanism to construct extended (field) algebras, which admit automorphic actions by the centers, and contain the original algebras in their fix-point algebras w.r.t. these actions (cp. \cite{MorchioQuantumMechanicsOn, StrocchiAnIntroductionTo}). This, in turn, allows us to understand parts of the representation theory from a constructive point of view.\\[0.1cm]
In section \ref{sec:KSreps}, we analyze the Koslowski-Sahlmann representations in more detail, and point out the necessity to extend the algebras if $\uG$ is non-Abelian, e.g. by admitting central terms in the basic commutation relations. We comment on the interpretation of the Koslowski-Sahlmann representations in terms of the holonomy-background flux-exponential algebra in section \ref{sec:conclusions}.
The case, when $\uG$ is Abelian, can be understood in terms of the results of section \ref{sec:centers}.\\[0.1cm]
In section \ref{sec:theta}, we explain, how the algebraic explanation of chiral symmetry breaking and the occurrence of the $\theta$-vacua in quantum field theory (cf. \cite{JackiwTopologicalInvestigationsOf, MorchioChiralSymmetryBreaking}) can be imported into the framework of loop quantum gravity. Again, the non-trivial structure in the representation theory, i.e. the $\theta$-sectors, manifests itself as a consequence of a non-trivial center of the (quantum) algebra, which is closely related to topological properties of the structure group.\\[0.25cm]
Throughout the whole article, we choose units such that $G=\hbar=c=1$. Furthermore, we fix the Barbero-Immirzi parameter $\beta=1$, although everything applies to the case $\beta\in\R_{\neq0}$, as well.
\section{Preliminaries}
\label{sec:pre}
In this section, we review the definition of the (quantum) algebras  $\mathfrak{P}_{\textup{LQG}}$ and $\mathfrak{A}_{\textup{LQG/LQC}}$ (cf. \cite{ThiemannModernCanonicalQuantum, FleischhackRepresentationsOfThe, AshtekarLoopQuantumCosmology}) based on the (classical) variables $(A,E)$, and provide the necessary formalism for the analysis of the following sections.
\subsection{Some fibre bundle theoretic digressions}
\label{sec:bundles}
Before we explain the construction of the algebras $\mathfrak{P}_{\textup{LQG}}$ \& $\mathfrak{A}_{\textup{LQG/LQC}}$ , we need some formalism from the theory of principal fibre bundles. \\[0.1cm]
As above, let $\uP\stackrel{\pi}{\rightarrow}\Sigma$ be a principal $\uG$-bundle. Since $A\in\mathcal{A}_{\uP}$, it defines a parallel transport (or holonomy)
\begin{align}
\label{eq:paralleltrans}
h^{A}_{e}:\uP_{|e(0)}=\pi^{-1}(e(0))\longrightarrow\uP_{|e(1)}=\pi^{-1}(e(1))
\end{align}
for every (broken, $C^{\infty}$) path $e:[0,1]\rightarrow\Sigma$.
\begin{Definition}[cf. \cite{KobayashiFoundationsOfDifferential1, BishopGeometryOfManifolds}]
\label{def:holonomy}
Given a path $e:[0,1]\rightarrow\Sigma$, for every $p\in\uP_{e(0)}$ we consider the unique, horizontal (w.r.t. $A$) lift $\tilde{e}:[0,1]\rightarrow\uP$ defined by
\begin{align}
\label{eq:horizontalpath}
& 1.\ \ \forall t\in[0,1]: A_{|\tilde{e}(t)}(\tilde{e}'(t))=0 \\
& 2.\ \ \pi\circ\tilde{e}=e \\
& 3.\ \ \tilde{e}(0)=p.
\end{align}
The parallel transport (or holonomy) of $A$ along $e$ is the map
\begin{align}
\label{eq:holdef}
\xymatrix@R=0.1cm{
h^{A}_{e}:\uP_{|e(0)}\ar[r]& \uP_{|e(1)}\hspace{1.3cm}\\
\hspace{0.25cm}p \ar@{|->}[r] & h^{A}_{e}(p)=\tilde{e}(1).
}
\end{align}
\end{Definition}
Clearly, the parallel transport is right equivariant, because the connection $A$ is $\textup{Ad}$-equivariant, i.e.
\begin{align}
\label{eq:holequi}
\forall g\in\uG: h^{A}_{e}\circ R_{g}=R_{g}\circ h^{A}_{e},
\end{align}
and satisfies
\begin{align}
\label{eq:holprop}
h^{A}_{e_{2}\circ e_{1}}=h^{A}_{e_{2}}\circ h^{A}_{e_{1}},\ h^{A}_{e^{-1}}=(h^{A}_{e})^{-1},
\end{align}
where $e_{2}\circ e_{1}$ is the composition of the paths $e_{1},\ e_{2}$  ($e_{1}(1)=e_{2}(0)$), and $e^{-1}$ is the reversion of the path $e$.\\[0.1cm]
To set up a correspondence between parallel transports, $h^{A}_{e}:\uP_{|e(0)}\rightarrow\uP_{|e(1)}$, and elements $g\in\uG$, we fix a set of reference points $p_{x}\in\uP_{|x},\ x\in\Sigma$, and use the relation
\begin{align}
\label{eq:grouphol}
h^{A}_{e}(p_{e(0)})=R_{g(e,A,\{p_{x}\}_{x\in\Sigma})}(p_{e(1)})
\end{align}
to define the element $g(e,A,\{p_{x}\}_{x\in\Sigma})\in\uG$, which is well-defined by the free and fibre transitive action of $\uG$ on $\uP$ (cf. \cite{LewandowskiUniquenessOfDiffeomorphism, FleischhackRepresentationsOfThe}).
\begin{Definition}
\label{def:grouphol}
Given a path $e:[0,1]\rightarrow\Sigma$ and set of reference points $\{p_{x}\}_{x\in\Sigma}\subset\uP$ \eqref{eq:grouphol} defines the map
\begin{align}
\label{eq:groupholmap}
\xymatrix@R=0.1cm{
g(e,\ .\ ,\{p_{x}\}_{x\in\Sigma}):\mathcal{A}_{P}\ar[r]& \uG\hspace{2.25cm}\\
\hspace{2.5cm}A \ar@{|->}[r] & g(e,A,\{p_{x}\}_{x\in\Sigma}).
}
\end{align}
\end{Definition}
This map inherits the properties \eqref{eq:holprop} in the following sense:
\begin{align}
\label{eq:groupholprop}
g(e_{2}\circ e_{1},A,\{p_{x}\}_{x\in\Sigma}) & =g(e_{2},A,\{p_{x}\}_{x\in\Sigma})g(e_{1},A,\{p_{x}\}_{x\in\Sigma}),\\ \nonumber
g(e^{-1},A,\{p_{x}\}_{x\in\Sigma}) & =g(e,A,\{p_{x}\}_{x\in\Sigma})^{-1}.
\end{align}
Due to the equivariance of the parallel transport, a change of reference points $\{p_{x}\}_{x\in\Sigma}\mapsto\{p'_{x}=p_{x}g_{x}\}_{x\in\Sigma}$, where the set $\{g_{x}\}_{x\in\Sigma}\subset\uG$ is, again, well-defined by the free and fibre transitive action of $\uG$ on $\uP$, effects the map \eqref{eq:groupholmap} in a equivariant way:
\begin{align}
\label{eq:holrefchange}
g(e,A,\{p'_{x}\}_{x\in\Sigma})=g^{-1}_{e(1)}g(e,A,\{p_{x}\}_{x\in\Sigma})g_{e(0)}.
\end{align}
For the purposes of loop quantum gravity, it is important that the map \eqref{eq:groupholmap} separates points in $\mathcal{A}_{\uP}$, if we allow the path $e:[0,1]\rightarrow\Sigma$ to vary among a suitable class of paths $\cP_{\Sigma}$ (cf. \cite{GilesReconstructionOfGauge}). Furthermore, \eqref{eq:groupholmap} allows to identify the space of generalized connections $\overline{\mathcal{A}}$ with the groupoid homomorphisms $\hom\left(\cP_{\Sigma},\uG\right)$,
\begin{align}
\label{eq:genholequivalence}
\overline{\mathcal{A}}\cong\hom(\cP_{\Sigma},G)
\end{align}
where $\uG$ is considered as the action groupoid over a single object $\{*\}$. \\[0.1cm]
Recall, that elements of $\overline{\mathcal{A}}$ are defined as sets of parallel transports w.r.t. the class of paths $\cP_{\Sigma}$.
\begin{Definition}
\label{def:genhol}
A generalized connection $A\in\overline{\mathcal{A}}$ is given by maps
\begin{align}
\label{eq:genhol}
h^{A}_{e}:\uP_{|e(0)}\longrightarrow\uP_{|e(1)}
\end{align}
for every $e\in\cP_{\Sigma}$ with the properties
\begin{align}
\label{eq:genholprop}
h^{A}_{e_{2}\circ e_{1}}=h^{A}_{e_{2}}\circ h^{A}_{e_{1}},\ h^{A}_{e^{-1}}=(h^{A}_{e})^{-1}.
\end{align}
\end{Definition}
The space of connections $\mathcal{A}_{\uP}$ is naturally identified with a subset of $\overline{\mathcal{A}}$ via the holonomies.\\[0.1cm]
Next, let us consider the conjugate momentum $E\in\Gamma\left(T\Sigma\otimes\textup{Ad}^{*}(\uP)\otimes|\Lambda|^{1}(\Sigma)\right)$, which similar to a connection $A\in\mathcal{A}_{\uP}$ that is given as an $\textup{Ad}$-equivariant 1-form on $\uP$ with values in $\fg$, i.e. an element of $\Lambda^{1}(\uP,\fg)_{\textup{Ad}}$, has an interpretation as a geometric entity on $\uP$ rather than on $\Sigma$. To this end, we need the following proposition (cf. \cite{KobayashiFoundationsOfDifferential1, BleeckerGaugeTheoryAnd}).
\begin{Proposition}
\label{prop:horizontalforms}
A section $\omega\in\Gamma(\Lambda^{k}\Sigma\otimes(\uP\times_{\rho}V))=:\Omega^{k}(\uP\times_{\rho}V)$, where $\uP\times_{\rho}V$ is the bundle associated with $\uP$ via a (linear) representation $\rho:\uG\rightarrow\aut(V)$, corresponds in a one-to-one fashion to an element $\tilde{\omega}\in\overline{\Lambda}^{k}(\uP,V)_{\rho}$, the horizontal, $\rho$-equivariant $k$-forms on $\uP$ with values in $V$, or shortly $\rho$-tensorial $k$-forms on $\uP$.
\begin{Proof}
Given an element $\omega\in\Gamma(\Lambda^{k}\Sigma\otimes(\uP\times_{\rho}V))$, we define $\tilde{\omega}$ in the following way:
\begin{align}
\label{eq:horizontaldef1}
\tilde{\omega}_{|p}(\tilde{X}_{1},...,\tilde{X}_{k})& =p^{-1}\omega_{\pi(p)}(d\pi_{|p}(\tilde{X}_{1}),...,d\pi_{|p}(\tilde{X}_{k})),\ \ \ p\in\uP,\ \tilde{X}_{1},...,\tilde{X}_{k}\in T_{p}\uP,
\end{align}
where $p^{-1}$ is the inverse of $p:V\rightarrow(\uP\times_{\rho}V)_{|\pi(p)},\ p(v)=[(p,v)]_{\rho}$. Clearly, $\tilde{\omega}$ is well-defined and horizontal, as $d\pi_{|p}:T_{p}\uP\rightarrow T_{\pi(p)}\Sigma$ vanishes on vertical vectors, i.e. elements of $T_{p}(\uP)_{|\pi(p)}$. Furthermore, it is $\rho$-equivariant:
\begin{align}
\label{eq:horizontalequivariance}
((R_{g})^{*}\tilde{\omega})_{|p}(\tilde{X}_{1},...,\tilde{X}_{k}) & =\tilde{\omega}_{|pg}(dR_{g|p}(\tilde{X}_{1}),...,dR_{g|p}(\tilde{X}_{k})) \\ \nonumber
 & =(pg)^{-1}\omega_{\pi(pg)}(d\pi_{|pg}(dR_{g|p}(\tilde{X}_{1})),...,d\pi_{|pg}(dR_{g|p}(\tilde{X}_{k}))) \\ \nonumber
 & =\rho(g^{-1})\cdot p^{-1}\omega_{\pi(p)}(d\pi_{|p}(\tilde{X}_{1}),...,d\pi_{|p}(\tilde{X}_{k})) \\ \nonumber
 & =\rho(g^{-1})\cdot\tilde{\omega}_{|p}(\tilde{X}_{1},...,\tilde{X}_{k}),
\end{align}
since $d\pi_{|pg}\circ dR_{g|p}=d\pi_{|p}$ and $(pg)(v)=p(\rho(g)v)$. \\[0.1cm]
Conversely, if $\tilde{\omega}\in\overline{\Lambda}^{k}(\uP,V)_{\rho}$ we construct $\omega$ by:
\begin{align}
\label{eq:horizontaldef2}
\omega_{|x}(X_{1},...,X_{k}) & =p\tilde{\omega}_{|p}(\tilde{X}_{1},...,\tilde{X}_{k}),
\end{align}
for any $p\in\uP_{|x}$ and any $\tilde{X}_{1},...,\tilde{X}_{k}\in T_{p}\uP$, s.t. $d\pi_{|p}(\tilde{X}_{i})=X_{i},\ i=1,...,k$, which is well-defined, because $\tilde{\omega}$ is horizontal and $\rho$-equivariant.
\end{Proof}
\end{Proposition}
Similarly, we may set up a correspondence between sections $X\in\Gamma(T\Sigma\otimes(\uP\times_{\rho}V))$ and horizontal, $\rho$-equivariant vector fields on $\uP$ with values in $V$, $\tilde{X}\in\overline{\mathfrak{X}}(P,V)_{\rho}$, or $\rho$-tensorial vector fields on $\uP$ for short. In contrast to $\overline{\Lambda}^{k}(\uP,V)_{\rho}$, $\overline{\mathfrak{X}}(P,V)_{\rho}$ requires a connection $A\in\mathcal{A}_{P}$ to be defined, as only horizontal $k$-forms and vertical vector fields on $\uP$ are defined naturally. On the other hand, we expect this to be the case, as we expect the Ashtekar-Barbero connection $A$ and its conjugate momentum $E$ to provide coordinates for $T^{*}\mathcal{A}_{\uP}$, and we have
\begin{align}
\label{eq:cotangenthorizontal}
T_{A}\mathcal{A}_{\uP}=\overline{\Lambda}^{1}(P,\fg)_{\textup{Ad}}=(\overline{\mathfrak{X}}(P,\fg^{*})_{\textup{Ad}^{*}})^{*}.
\end{align}
\begin{Proposition}
\label{prop:horizontalvf}
If we fix a connection $A\in\mathcal{A}_{\uP}$, there is a one-to-one correspondence between sections $X\in\Gamma(T\Sigma\otimes(\uP\times_{\rho}V))$ and elements $\tilde{X}$ of $\overline{\mathfrak{X}}(P,V)_{\rho}$.
\begin{Proof}
Given $X\in\Gamma(T\Sigma\otimes(\uP\times_{\rho}V))$, let $\overline{X}\in\Gamma(T\uP\otimes(\uP\times_{\rho}V))$ be its unique horizontal lift w.r.t. to $A$ (cf. \cite{KobayashiFoundationsOfDifferential1}), which is right invariant,
\begin{align}
\label{eq:rightinvariancehorizontallift}
((R_{g})_{*}\overline{X})_{|p}=\overline{X}_{|p},\ p\in\uP, g\in\uG,
\end{align}
by the $\textup{Ad}$-equivariance of $A$. We define
\begin{align}
\label{eq:horizontalvf1}
\tilde{X}_{|p}=p^{-1}\overline{X}_{|p},\ p\in\uP.
\end{align}
We only have to check $\rho$-equivariance.
\begin{align}
\label{eq:horizontalequivariancevf}
((R_{g})_{*}\tilde{X})_{|p} & = dR_{g|pg^{-1}}(\tilde{X}_{|pg^{-1}}) \\ \nonumber
 & = dR_{g|pg^{-1}}((pg^{-1})\overline{X}_{|pg^{-1}}) \\ \nonumber
 & = \rho(g)\cdot p^{-1}\overline{X}_{|p} \\ \nonumber
 & = \rho(g)\cdot\tilde{X}_{|p}. 
\end{align}
Conversely, let $\tilde{X}\in\overline{\mathfrak{X}}(P,V)_{\rho}$, and set
\begin{align}
\label{eq:horizontalvf2}
X_{|x}=d\pi_{|p}(p\tilde{X}_{|p})
\end{align}
for an arbitrary $p\in\uP_{|x}$. This is well-defined, because
\begin{align}
\label{eq:horizontalvf3}
d\pi_{|p'}(p'\tilde{X}_{|p'}) & = d\pi_{|pg'}((pg')\tilde{X}_{|pg'}) \\ \nonumber
 & = (d\pi_{|pg'}\circ dR_{g'|p})(p(\rho(g')\cdot\rho(g'^{-1})\cdot\tilde{X}_{|p})) \\ \nonumber
 & = d\pi_{|p}(p\tilde{X}_{|p}),
\end{align}
for any pair $p,p'\in\uP_{|x}$.
\end{Proof}
Clearly, \eqref{eq:horizontalvf2} does not depend on the choice of connection $A$, which will be important in the follow-up.
\end{Proposition}
In analogy with the pairing between connections $A\in\mathcal{A}_{\uP}$ and paths $e:[0,1]\rightarrow\Sigma$ yielding group elements $g(e,A,\{p_{x}\}_{x\in\Sigma})\in\uG$,
\begin{align}
\label{eq:holpairing}
\xymatrix@R=0.1cm{
\mathcal{A}_{\uP}\times\cP_{\Sigma} \ar[r] & \uG\hspace{2.25cm} \\
(A,e) \ar@{|->}[r] & g(e,A,\{p_{x}\}_{x\in\Sigma}),
}
\end{align}
there is a pairing
\begin{align}
\label{eq:fluxpairing}
\xymatrix@R=0.1cm{
\Gamma(T\Sigma\otimes\textup{Ad}^{*}(\uP)\otimes|\Lambda|^{1}(\Sigma))\times\Gamma(\textup{Ad}(\uP))\times\mathcal{S}_{\Sigma} \ar[r] & \C\hspace{1.5cm} \\
(E,n,S) \ar@{|->}[r] & \int_{S}*(E(n)),
}
\end{align}
where $\mathcal{S}_{\Sigma}$ is a suitable class of hypersurfaces in $\Sigma$, $E(n)\in\Gamma(T\Sigma\otimes|\Lambda|^{1}(\Sigma))$ denotes the fibrewise pairing between $E$ and $n$, and $*(E(n))$ is the pseudo-2-form associated with $E(n)$:
\begin{align}
\label{eq:pseudo2form}
*(E(n))=\varepsilon_{abc}E(n)^{a}dx^{b}\wedge dx^{c}
\end{align}
in local coordinates $\phi:U\subset\Sigma\rightarrow V\subset\R^{3}$. Here, $\varepsilon_{abc}=\delta^{1}_{[a}\delta^{2}_{b}\delta^{3}_{c]}$ denotes the invariant pseudo tensor density of weight $-1$. Noteworthy, the duality
\begin{align}
\label{eq:vfduality}
\xymatrix@R=0.1cm{
\Gamma(T\Sigma\otimes\textup{Ad}^{*}(\uP)\otimes|\Lambda|^{1}(\Sigma))\times\Gamma(\textup{Ad}(\uP)) \ar[r] & \Gamma(T\Sigma\otimes|\Lambda|^{1}(\Sigma)) \\
(E,n) \ar@{|->}[r] & E(n),\hspace{1.7cm}
}
\end{align}
is compatible with the corresponding pairing
\begin{align}
\label{eq:horizontalvfduality}
\xymatrix@R=0.1cm{
\overline{\Gamma}(T\uP\otimes|\Lambda|^{1}(\uP),V)_{\textup{Ad}^{*}}\times\overline{\Lambda}^{0}(P,\fg)_{\textup{Ad}} \ar[r] & \overline{\Gamma}(T\uP\otimes|\Lambda|^{1}(\uP))^{\uG}  \\
(\tilde{E},\tilde{n}) \ar@{|->}[r] & \tilde{E}(\tilde{n}),\hspace{1.9cm}
}
\end{align}
in the sense, that
\begin{align}
\label{eq:compatibleduality}
d\pi_{|p}(\tilde{E}(\tilde{n})_{|p})=E(n)_{|\pi(p)},\ p\in\uP_{|x}.
\end{align}

\subsubsection{Gauge transformations}
\label{sec:gaugetrafo}

In this subsection, we will analyze the behaviour of the variables
\begin{align}
\label{eq:bundleashtekarvar}
(A,\tilde{E})\in|\Lambda|^{1}T^{*}\mathcal{A}_{\uP} & =\bigsqcup_{A\in\mathcal{A}_{\uP}}\overline{\Gamma}(T\uP\otimes|\Lambda|^{1}(\uP),\fg^{*})_{\textup{Ad}^{*}}
\end{align}
under the action of the gauge transformations $\cG_{\uP}$ of $\uP$\footnote{It is possible to consider the action of general bundle automorphism $\aut(\uP)$ on $|\Lambda|^{1}T^{*}\mathcal{A}_{\uP}$ (see e.g. \cite{LewandowskiUniquenessOfDiffeomorphism, BahrAutomorphismsInLoop}). Furthermore, in local trivialization $\psi:\uP_{|U}\rightarrow U\times\uG$ of $\uP$ we have $\aut(\uP_{|U})\cong\textup{Diff}(U)\ltimes\cG_{\uP_{|U}}$ (cf. \cite{ThiemannModernCanonicalQuantum}).}, i.e. right equivariant diffeomorphisms of $\uP$,
\begin{align}
\label{eq:gaugeauto}
\xymatrix@=1.5cm{
\ar@{}[dr] |{\circlearrowright} \uP \ar[r]^{\lambda} \ar[d]_{R_{g}} & \uP \ar[d]^{R_{g}} \\
\uP \ar[r]^{\lambda} & \uP
}
\end{align}
that reduce to the identity $\pi\circ\lambda=\pi$ on $\Sigma$. First, let us derive some properties of gauge transformations.
\begin{Lemma}[cf. \cite{BleeckerGaugeTheoryAnd}]
\label{lem:gaugeauto}
There is an isomorphism between the group of gauge transformations $\cG_{\uP}$ and the group of $\alpha$-equivariant maps from $\uG$ to $\uP$, $C(\uP,\uG)_{\alpha}$, where
\begin{align}
\label{eq:groupconjugation}
\xymatrix@R=0.1cm{
\alpha_{g}:\uG \ar[r] & \uG \hspace{2.35cm} \\
\hspace{0.75cm}g' \ar@{|->}[r] & \alpha_{g}(g')=gg'g^{-1}.
}
\end{align}
In analogy with proposition \ref{prop:horizontalforms}, we also have the isomorphism $C(\uP,\uG)_{\alpha}\cong\Gamma(\uP\times_{\alpha}\uG)$.
\begin{Proof}
Let $\lambda\in\cG_{\uP}$ and define $f_{\lambda}\in C(\uP,\uG)_{\alpha}$ by
\begin{align}
\label{eq:gaugeautoiso1}
\lambda(p)=pf_{\lambda}(p),\ p\in\uP.
\end{align}
$f_{\lambda}$ is well-defined by right equivariance of $\lambda$ and the free and fibre transitive action of $\uG$ on $\uP$. \\[0.1cm]
Conversely, for $f\in C(\uP,\uG)_{\alpha}$ we obtain $\lambda_{f}\in\cG_{\uP}$ by
\begin{align}
\label{eq:gaugeautoiso2}
\lambda_{f}(p)=pf(p),\ p\in\uP.
\end{align}
Similarly, we get $s_{f}\in\Gamma(\uP\times_{\alpha}\uG)$ for the second isomorphism,
\begin{align}
\label{eq:gaugeautoiso3}
s_{f}(x)=pf(p)
\end{align}
for any $p\in\uP_{|x}$. Here $p:\uG\rightarrow\uP\times_{\alpha}\uG$ is the map we get from the associated bundle construction. \eqref{eq:gaugeautoiso3} is independent of the choice of $p$ by $\alpha$-equivariance,
\begin{align}
\label{eq:gaugeautoiso4}
p'f(p') & = (pg')f(pg') = p\alpha_{g'}\cdot\alpha_{g'^{-1}}\cdot f(p) \\ \nonumber
 & = pf(p)
\end{align}
for any pair $p,p'\in\uP_{|x}$. The inverse of the second isomorphism is
\begin{align}
\label{eq:gaugeautoiso5}
f_{s}(p)=p^{-1}s(\pi(p)),\ p\in\uP
\end{align}
for $s\in\Gamma(\uP\times_{\alpha}\uG)$.
\end{Proof}
\begin{Remark}
\label{rem:gaugeauto}
In general, the right action $R_{g}:\uP\rightarrow\uP$ is not a gauge transformation, as this would require 
\begin{align}
\label{eq:rightgaugeauto}
R_{g}\circ R_{g'} = R_{g'g} = R_{g'}\circ R_{g},\ \forall g,g'\in\uG,
\end{align}
which holds if and only if $\uG$ is Abelian.
\end{Remark}
\end{Lemma}
The $\textup{Ad}$-tensorial $0$-forms on $\uP$, $\overline{\Lambda}^{0}(\uP,\fg)_{\textup{Ad}}$, can be regarded as the Lie algebra of $\cG_{\uP}$.
\begin{Theorem}[cf. \cite{BleeckerGaugeTheoryAnd}]
\label{th:gaugealg}
$\overline{\Lambda}^{0}(\uP,\fg)_{\textup{Ad}}$ has a natural Lie algebra structure inherited from $\fg$,
\begin{align}
\label{eq:gaugealg1}
[\tilde{n},\tilde{n}']_{|p}=[\tilde{n}_{|p},\tilde{n}'_{|p}],\ \tilde{n},\tilde{n}'\in\overline{\Lambda}^{0}(\uP,\fg)_{\textup{Ad}},p\in\uP.
\end{align}
\begin{Proof}
Given $\tilde{n},\tilde{n}'\in\overline{\Lambda}^{0}(\uP,\fg)_{\textup{Ad}}$, we need to verify that $[\tilde{n},\tilde{n}']\in\overline{\Lambda}^{0}(\uP,\fg)_{\textup{Ad}}$.
\begin{align}
\label{eq:gaugealg2}
(R_{g})^{*}[\tilde{n},\tilde{n}']_{|p} & = [\tilde{n},\tilde{n}']_{|pg} = [\tilde{n}_{pg},\tilde{n}'_{|pg}] = [\textup{Ad}_{g^{-1}}\cdot\tilde{n}_{|p},\textup{Ad}_{g^{-1}}\cdot\tilde{n}'_{|p}] \\ \nonumber
 & = \textup{Ad}_{g^{-1}}\cdot[\tilde{n},\tilde{n}']_{|p},\ p\in\uP.
\end{align}
\end{Proof}
\end{Theorem}
\begin{Definition}[cf. \cite{BleeckerGaugeTheoryAnd}]
\label{def:gaugealg}
The gauge algebra $\fG_{\uP}$ of $\uP$ is the space $\overline{\Lambda}^{0}(\uP,\fg)_{\textup{Ad}}$ of $\textup{Ad}$-tensorial $0$-forms on $\uP$ with the Lie algebra structure given in theorem \ref{th:gaugealg}.
\end{Definition}
Furthermore, there is an exponential map
\begin{align}
\label{eq:gaugealgexp}
\exp_{\cG_{\uP}}:\fG_{\uP}\longrightarrow\cG_{\uP}.
\end{align} 
\begin{Theorem}[cf. \cite{BleeckerGaugeTheoryAnd}]
\label{th:gaugealgexp}
There is a map $\exp:\overline{\Lambda}^{0}(\uP,\fg)_{\textup{Ad}}\rightarrow C(\uP,\uG)_{\alpha}$ defined by $\exp(\tilde{n})(p)=\exp_{\uG}(\tilde{n}_{|p})$, $\tilde{n}\in\overline{\Lambda}^{0}(\uP,\fg)_{\textup{Ad}}$, $p\in\uP$ with the properties
\begin{align}
\label{eq:gaugealgexpprop}
1.\ &\ \frac{\ud}{\ud t}_{|t=0}\exp(t\tilde{n})=\tilde{n} \\
2.\ &\ \frac{\ud^{2}}{\ud t\ud s}_{|t,s=0}\alpha_{\exp(t\tilde{n})}(\exp(s\tilde{n}'))=[\tilde{n},\tilde{n}'].
\end{align}
$\exp:\overline{\Lambda}^{0}(\uP,\fg)_{\textup{Ad}}\rightarrow C(\uP,\uG)_{\alpha}$ induces $\exp_{\cG_{\uP}}:\fG_{\uP}\longrightarrow\cG_{\uP}$ by
\begin{align}
\label{eq:gaugealgexpdef}
\exp_{\cG_{\uP}}(\tilde{n})(p)=p\exp(\tilde{n})(p).
\end{align}
\begin{Proof}
Clearly, $\alpha$-equivariance of $\exp(\tilde{n})$ follows from the properties of $\exp_{\uG}:\fg\rightarrow\uG$.
\begin{align}
\label{eq:gaugealgexpequi}
(R_{g})^{*}\exp(\tilde{n})(p) & = \exp(\tilde{n})(pg) = \exp_{\uG}(\tilde{n}_{|pg}) = \exp_{\uG}(\textup{Ad}_{g^{-1}}\cdot\tilde{n}_{|p}) = \alpha_{g^{-1}}(\exp_{\uG}(\tilde{n}_{|p})) \\ \nonumber
 & = \alpha_{g^{-1}}\cdot\exp(\tilde{n})(p).
\end{align}
The properties $1.\ \&\ 2.$ are proved along the same lines, and are omitted at this point. $\exp_{\cG_{\uP}}:\fG_{\uP}\longrightarrow\cG_{\uP}$ is well-defined by appealing to the isomorphism of lemma \ref{lem:gaugeauto}.
\end{Proof}
\end{Theorem}
\begin{Remark}
\label{rem:gaugealg}
The notation $\tilde{n}\in\overline{\Lambda}^{0}(\uP,\fg)_{\textup{Ad}}$ is intentional, when compared with \eqref{eq:vfduality} \& \eqref{eq:horizontalvfduality}, as it will be important to consider $\tilde{n}$ as generator of a gauge transformation in the regularization of the Poisson structure \eqref{eq:poisson}.
\end{Remark}
Next, we define the (left) action of $\cG_{\uP}$ on $|\Lambda|^{1}T^{*}\mathcal{A}_{\uP}$.
\begin{Definition}
\label{def:gaugeaction}
The gauge transformations $\cG_{\uP}$ act on $|\Lambda|^{1}T^{*}\mathcal{A}_{\uP}$ to the left by pullback and pushforward, i.e.
\begin{align}
\label{eq:gaugeaction}
\xymatrix@R=0.1cm{
L_{\lambda}:|\Lambda|^{1}T^{*}\mathcal{A}_{\uP} \ar[r] & |\Lambda|^{1}T^{*}\mathcal{A}_{\uP}\hspace{3.5cm} \\
\hspace{0.6cm}(A,\tilde{E}) \ar@{|->}[r] & L_{\lambda}(A,\tilde{E})=((\lambda^{-1})^{*}A,\lambda_{*}\tilde{E}).
}
\end{align}
\end{Definition}
This action is well-defined by the duality between pullback and pushforward
\begin{align}
\label{eq:gaugeactionhorizontal}
((\lambda^{-1})^{*}A)(\lambda^{\ }_{*}\tilde{E})=A(\lambda^{-1}_{*}(\lambda^{\ }_{*}\tilde{E}))=A(\tilde{E})=0,\ (A,\tilde{E})\in|\Lambda|^{1}T^{*}\mathcal{A}_{\uP}.
\end{align}
By the definition \ref{def:gaugeaction}, we only need the differential $d\lambda:T\uP\rightarrow T\uP$ to obtain an explicit expression for $((\lambda^{-1})^{*}A,\lambda_{*}\tilde{E})$.
\begin{Lemma}[cf. \cite{BleeckerGaugeTheoryAnd}]
\label{lem:gaugeder}
The differential $d\lambda:T\uP\rightarrow T\uP$ of $\lambda\in\cG_{\uP}$ is given by
\begin{align}
\label{eq:gaugeder}
d\lambda_{|p}(\tilde{X}_{|p})=dR_{f_{\lambda}(p)|p}(\tilde{X}_{|p})+(dL_{f_{\lambda}(p)^{-1}|f_{\lambda}(p)}\circ df_{\lambda|p}(\tilde{X}_{|p}))^{*}_{\lambda(p)},\ p\in\uP,\tilde{X}_{|p}\in T_{p}\uP,
\end{align}
where $^{*}:\fg\rightarrow\mathfrak{X}(P)$ gives the fundamental vector fields:
\begin{align}
\label{eq:fundamentalvf}
T^{*}_{|p}=\frac{\ud}{\ud t}_{|t=0}p\exp_{\uG}(tT),\ p\in\uP,T\in\fg,
\end{align}
which have the properties
\begin{align}
\label{eq:fundamentalvfprop}
1.\ &\ (R_{g})_{*}T^{*}_{|p}=(\textup{Ad}_{g^{-1}}\cdot T)^{*}_{|p} \\
2.\ &\ \lambda_{*}T^{*}_{|p}=T^{*}_{|p}.
\end{align}
\begin{Proof}
Let $\lambda\in\cG_{\uP}$ and $\gamma:[0,1]\rightarrow\uP,\ \gamma(0)=p,\ \gamma'(0)=\tilde{X}_{|p}$, then
\begin{align}
\label{eq:gaugederproof}
d\lambda_{|p}(\tilde{X}_{|p}) & = \frac{\ud}{\ud t}_{|t=0}\lambda(\gamma(t)) = \frac{\ud}{\ud t}_{|t=0}\gamma(t)f_{\lambda}(\gamma(t)) \\ \nonumber
 & = dR_{f_{\lambda}(p)|p}(\tilde{X}_{|p}) + \frac{\ud}{\ud t}_{|t=0}pf_{\lambda}(\gamma(t)) = dR_{f_{\lambda}(p)|p}(\tilde{X}_{|p}) + \frac{\ud}{\ud t}_{|t=0}\lambda(p)L_{f_{\lambda}(p)^{-1}}(f_{\lambda}(\gamma(t))) \\ \nonumber
 & = dR_{f_{\lambda}(p)|p}(\tilde{X}_{|p}) + d\lambda(p)_{|e}\circ dL_{f_{\lambda}(p)^{-1}|f_{\lambda}(p)}\circ df_{\lambda|p}(\tilde{X}_{|p}) \\ \nonumber
 & = dR_{f_{\lambda}(p)|p}(\tilde{X}_{|p}) + (dL_{f_{\lambda}(p)^{-1}|f_{\lambda}(p)}\circ df_{\lambda|p}(\tilde{X}_{|p}))^{*}_{|\lambda(p)}.
\end{align}
The properties of the fundamental vector fields are evident from their definition.
\end{Proof}
\end{Lemma}
\begin{Corollary}
\label{cor:gaugeaction}
The action of the gauge transformations $\cG_{\uP}$ on $|\Lambda|^{1}T^{*}\mathcal{A}_{\uP}$ is explicitly given as
\begin{align}
\label{eq:gaugeactionexplicit}
(\lambda^{-1})^{*}A_{|p} & = A_{|\lambda^{-1}(p)}\circ d\lambda^{-1}_{|p} \\ \nonumber
 & = \textup{Ad}_{f_{\lambda}(p)}\cdot A_{|p} + dL_{f_{\lambda}(p)|f_{\lambda}(p)^{-1}}\circ df_{\lambda|p} \\[0.25cm]
\lambda_{*}\tilde{E}_{|p} & = d\lambda_{|\lambda^{-1}(p)}(\tilde{E}_{\lambda^{-1}(p)}) \\ \nonumber
 & = \textup{Ad}^{*}_{f_{\lambda}(p)}\cdot(\tilde{E}_{|p}+(dR_{f_{\lambda}(p)^{-1}|f_{\lambda}(p)}\circ df_{\lambda|p}(\tilde{E}_{|p}))^{*}_{|p}).
\end{align}
\begin{Proof}
Recall that $A(T^{*})=T,\ T\in\fg$ for $A\in\mathcal{A}_{\uP}$.
\end{Proof}
\end{Corollary}
The map $\exp_{\cG_{\uP}}:\fG_{\uP}\longrightarrow\cG_{\uP}$ allows us to derive the (infinitesimal) action of $\fG_{\uP}$ on $|\Lambda|^{1}T^{*}\mathcal{A}_{\uP}$.
\begin{Lemma}
\label{lem:infgaugeaction}
The explicit form of the action of $\fG_{\uP}$ on $|\Lambda|^{1}T^{*}\mathcal{A}_{\uP}$ is
\begin{align}
\label{eq:infgaugeaction}
\frac{\ud}{\ud t}_{|t=0}((\lambda_{t\tilde{n}})^{-1})^{*}A & = -(d\tilde{n}+[A,\tilde{n}])=-d_{A}\tilde{n} \\ \nonumber
\frac{\ud}{\ud t}_{|t=0}(\lambda_{t\tilde{n}})_{*}\tilde{E} & = (d\tilde{n}(\tilde{E}))^{*}+\textup{ad}^{*}_{\tilde{n}}\cdot\tilde{E},
\end{align}
where $\lambda_{t\tilde{n}}=\exp_{\cG_{\uP}}(t\tilde{n})\in\cG_{\uP},\ \tilde{n}\in\fG_{\uP}$, and $\textup{ad}^{*}:\fg\rightarrow\fg^{*}$ is the co-adjoint representation of $\fg$.
\begin{Proof}
Note that for $\gamma:[0,1]\rightarrow\uP,\ \gamma(0)=p,\ \gamma'(0)=\tilde{X}_{|p}$ we have
\begin{align}
\label{eq:infgaugeactionproof}
\frac{\ud}{\ud t}_{|t=0}dR_{\exp_{\uG}(-t\tilde{n}_{|p})|\exp_{\uG}(t\tilde{n}_{|p})}\circ d\exp_{\uG}(t\tilde{n}_{|(\ .\ )})_{|p}(\tilde{X}_{|p}) & = \frac{\ud^{2}}{\ud t\ud s}_{|t,s=0}R_{\exp_{\uG}(-t\tilde{n}_{|p})}(\exp_{\uG}(t\tilde{n}_{|\gamma(s)})) \\ \nonumber
 & = \frac{\ud^{2}}{\ud s\ud t}_{|t,s=0}R_{\exp_{\uG}(-t\tilde{n}_{|p})}(\exp_{\uG}(t\tilde{n}_{|\gamma(s)})) \\ \nonumber
 & = d\tilde{n}_{|p}(\tilde{X}_{|p}).
\end{align}
Then apply corollary \ref{cor:gaugeaction}.
\end{Proof}
\end{Lemma}
For completeness, we also state the transformation behavior of $\rho$-tensorial $k$-forms on $\uP$, since $T_{A}\mathcal{A}_{\uP}\cong\overline{\Lambda}^{1}(\uP,\fg)_{\textup{Ad}}$.
\begin{Lemma}[cf. \cite{BleeckerGaugeTheoryAnd}]
\label{lem:gaugeactiontensorial}
The gauge transformations $\cG_{\uP}$ and the gauge algebra $\fG_{\uP}$ act on $\overline{\Lambda}^{k}(\uP,V)_{\rho}$ (to the left) in the following way:
\begin{align}
\label{eq:gaugeactiontensorial}
(\lambda^{-1})^{*}\tilde{\omega} & = \rho(f_{\lambda})\cdot\tilde{\omega}, & \frac{\ud}{\ud t}_{|t=0}((\lambda_{t\tilde{n}})^{-1})^{*}\tilde{\omega} & = d\rho(\tilde{n})\cdot\tilde{\omega}.
\end{align}
Here $\omega\in\overline{\Lambda}^{k}(\uP,V)_{\rho}$, $\lambda\in\cG_{\uP}$, $\tilde{n}\in\fG_{\uP}$, $\lambda_{t\tilde{n}}=\exp_{\cG}(t\tilde{n})$, and $d\rho:\fg\rightarrow\End(V)$ is the differential of \mbox{$\rho:\uG\rightarrow\aut(V)$.}
\begin{Proof}
Use lemma \ref{lem:gaugeder} and $\rho$-equivariance of $\tilde{\omega}$.
\end{Proof}
\end{Lemma}
Now, that we understand how the gauge transformations $\cG_{\uP}$ act on pairs $(A,\tilde{E})\in|\Lambda|^{1}T^{*}\mathcal{A}_{\uP}$, we are able to derive their action on parallel transports $h^{A}_{e},\ e\in\cP_{\Sigma},$ and projected, $1$-density, vector fields $E\in\Gamma(T\Sigma\otimes\textup{Ad}^{*}(\uP)\otimes|\Lambda|^{1}(\Sigma))$.
\begin{Proposition}[cf. \cite{LewandowskiUniquenessOfDiffeomorphism}]
\label{prop:gaugeactionhol}
A Gauge transformation $\lambda\in\cG_{\uP}$ affects the parallel transports $h^{A}_{e},\ e\in\cP_{\Sigma},$ of a connection $A\in\mathcal{A}_{\uP}$ via conjugation, i.e.
\begin{align}
\label{eq:gaugeactionhol}
h^{(\lambda^{-1})^{*}A}_{e} & =\lambda\circ h^{A}_{e}\circ\lambda^{-1}.
\end{align}
The associated group elements $g(e,A,\{p_{x}\}_{x\in\Sigma})\in\uG$ (see definition \ref{def:grouphol}) behave in an equivariant way, as well:
\begin{align}
\label{eq:gaugeactiongrouphol}
g(e,(\lambda^{-1})^{*}A,\{p_{x}\}_{x\in\Sigma}) & =f_{\lambda}(p_{e(1)})g(e,A,\{p_{x}\}_{x\in\Sigma})f_{\lambda}(p_{e(0)})^{-1},
\end{align}
which is compatible with changes of reference points $\{p_{x}\}_{x\in\Sigma}\rightarrow\{p'_{x}\}_{x\in\Sigma}$.
The corresponding infinitesimal actions of $\tilde{n}\in\fG_{\uP}$ are:
\begin{align}
\label{eq:infgaugeactionhol1}
\frac{\ud}{\ud t}_{|t=0}h^{((\lambda_{t\tilde{n}})^{-1})^{*}A}_{e}(p) & = (\tilde{n}_{|h^{A}_{e}(p)})^{*}_{|h^{A}_{e}(p)}-(\tilde{n}_{|p})^{*}_{|h^{A}_{e}(p)} \\[0.25cm]
\label{eq:infgaugeactionhol2}
\frac{\ud}{\ud t}_{|t=0}g(e,((\lambda_{t\tilde{n}})^{-1})^{*}A,\{p_{x}\}_{x\in\Sigma}) & = \tilde{n}^{i}_{|p_{e(1)}}\uR_{i|g(e,A,\{p_{x}\}_{x\in\Sigma})}-\tilde{n}^{j}_{|p_{e(0)}}\uL_{j|g(e,A,\{p_{x}\}_{x\in\Sigma})},
\end{align}
where $\{\uR_{i}\}, \{\uL_{j}\}\subset\mathfrak{X}(\uG)$ are the right and left invariant vector fields on $\uG$ associated with the generators $\{\tau_{i}\}$.
\begin{Proof}
First, observe that $\lambda\in\cG_{\uP}$ acts on horizontal lifts in the appropriate way, i.e. if $\tilde{e}:[0,1]\rightarrow\uP$ is a horizontal lift of $e:[0,1]\rightarrow\uP$ w.r.t $A$, then $\lambda\circ\tilde{e}:[0,1]\rightarrow\uP$ is a horizontal lift w.r.t $(\lambda^{-1})^{*}A$ by \eqref{eq:gaugeder}.
Second, we have:
\begin{align}
\label{eq:gaugeactionholproof}
h^{A}_{e}(\tilde{e}(0)) & =\tilde{e}(1) = \lambda^{-1}(\lambda(\tilde{e}(1))) \\ \nonumber
 & = \lambda^{-1}(h^{(\lambda^{-1})^{*}A}_{e}(\lambda(\tilde{e}(0)))).
\end{align}
\eqref{eq:gaugeactiongrouphol} and compatibility follow from the right equivariance of $\lambda\in\cG_{\uP}$ resp. $\alpha$-euqivariance of $f_{\lambda}\in C(\uP,\uG)_{\alpha}$. To prove \eqref{eq:infgaugeactionhol1} \& \eqref{eq:infgaugeactionhol2} we merely stick to the definition of fundamental, left invariant and right invariant vector fields.
\end{Proof}
\end{Proposition}
\begin{Remark}
\label{rem:gaugeactionhol}
The action \eqref{eq:gaugeactiongrouphol} is opposite to the one employed in parts of the literature (cf. \cite{FleischhackRepresentationsOfThe, ThiemannModernCanonicalQuantum}), where instead we find
\begin{align}
\label{eq:gaugeactiongroupholop}
g(e,(\lambda^{-1})^{*}A,\{p_{x}\}_{x\in\Sigma}) & =f_{\lambda}(p_{e(0)})g(e,A,\{p_{x}\}_{x\in\Sigma})f_{\lambda}(p_{e(1)})^{-1}.
\end{align}
This could be achieved if we worked with left principal bundles, or if we changed the defining identity \eqref{eq:grouphol} to
\begin{align}
h^{A}_{e}(p_{e(0)}) & = R_{g(e,A,\{p_{x}\}_{x\in\Sigma})^{-1}}(p_{e(1)}).
\end{align}
The former would, in the case of trivial bundles, $\uP\cong\Sigma\times\uG$, lead to a right action of the gauge transformations $\cG_{\uP}\cong C(\Sigma,\uG)$, which is not the typical choice in the majority of the literature. On the other hand, the latter would make the homomorphism \eqref{eq:groupholmap} an anti-homomorphism, i.e. reverse the order in the first line of \eqref{eq:groupholprop}.
\end{Remark}
The actions of the gauge transformations $\cG_{\uP}$ and the gauge algebra $\fG_{\uP}$ on $\Gamma(T\Sigma\otimes\textup{Ad}^{*}(\uP)\otimes|\Lambda|^{1}(\Sigma))$ remain to be discussed.
\begin{Proposition}
\label{prop:gaugeactiontriad}
The compatible actions of $\cG_{\uP}$ and $\fG_{\uP}$ on $\Gamma(T\Sigma\otimes\textup{Ad}^{*}(\uP)\otimes|\Lambda|^{1}(\Sigma))$ are
\begin{align}
\label{eq:gaugeactiontriad}
\lambda\triangleright E_{|x} & = p\ \textup{Ad}^{*}_{f_{\lambda}(p)}\cdot p^{-1}E_{|x}, & \tilde{n}\triangleright E_{|x} & = p\ \textup{ad}^{*}_{\tilde{n}_{|p}}\cdot p^{-1}E_{|x}.
\end{align}
Here $x\in\Sigma$, $p\in\uP_{|x}$ and $\lambda\in\cG_{\uP}$, $\tilde{n}\in\fG_{\uP}$. As before, we regard $p:\uG\rightarrow\textup{Ad}^{*}(\uP)$ as a map.
\begin{Proof}
The actions are well-defined, i.e. independent of the choice of $p\in\uP_{|x}$, because of $\alpha$-equivariance of $f_{\lambda}$ resp. $\textup{Ad}$-equivariance of $\tilde{n}$. To prove compatibility, we only need to combine proposition \ref{prop:horizontalvf}, corollary \ref{cor:gaugeaction}, lemma \ref{lem:infgaugeaction} and the fact that $d\pi:T\uP\rightarrow T\Sigma$ vanishes on vertical vectors.
\begin{align}
\label{eq:gaugeactiontriadproof}
d\pi_{|p}(p(\lambda_{*}\tilde{E})_{|p}) & = d\pi_{|p}(p\ \textup{Ad}^{*}_{f_{\lambda}(p)}\cdot p^{-1}\overline{E}_{|p}) = p\ \textup{Ad}^{*}_{f_{\lambda}(p)}\cdot p^{-1}\ d\pi_{|p}(\overline{E}_{|p}) \\ \nonumber
 & = p\ \textup{Ad}^{*}_{f_{\lambda}(p)}\cdot p^{-1}E_{|x}, \\
\frac{\ud}{\ud t}_{|t=0}d\pi_{|p}(p((\lambda_{t\tilde{n}})_{*}\tilde{E})_{|p}) & = d\pi_{|p}(p\ \textup{ad}^{*}_{\tilde{n}_{|p}}\cdot p^{-1}\overline{E}_{|p}) = p\ \textup{ad}^{*}_{\tilde{n}_{|p}}\cdot p^{-1}\ d\pi_{|p}(\overline{E}_{|p}) \\ \nonumber
 & = p\ \textup{ad}^{*}_{\tilde{n}_{|p}}\cdot p^{-1}E_{|x},
\end{align}
where $\tilde{E}\in\overline{\Gamma}(T\uP\otimes|\Lambda|^{1}(\uP),\fg^{*})_{\textup{Ad}^{*}}$ corresponds to $E$ via proposition \ref{prop:horizontalvf}.
\end{Proof}
\end{Proposition}
In view of lemma \ref{lem:gaugeactiontensorial}, identical formulas hold for $k$-forms in associated bundles.
\begin{Proposition}
\label{prop:gaugeactionforms}
There are compatible actions of $\cG_{\uP}$ and $\fG_{\uP}$ on $\Omega^{k}(\uP\times_{\rho}V)$:
\begin{align}
\label{eq:gaugeactionforms}
\lambda^{-1}\triangleright\omega_{|x} & = p\ \rho(f_{\lambda}(p))\cdot p^{-1}\omega_{|x}, & -\tilde{n}\triangleright\omega_{|x} & = p\ d\rho(\tilde{n})\cdot p^{-1}\omega_{|x},
\end{align}
where $x\in\Sigma$, $p\in\uP_{|x}$ and $\lambda\in\cG_{\uP}$, $\tilde{n}\in\fG_{\uP}$. As before, we regard $p:V\rightarrow\uP\times_{\rho}V$ as a map.
\begin{Proof}
Just apply lemma \ref{lem:gaugeactiontensorial} and proposition \ref{prop:horizontalforms}.
\end{Proof}
\end{Proposition}
These induced actions on spaces of section in associated bundles have propertie that is essential in the following subsection \ref{sec:prelqg}.
\begin{Corollary}
\label{cor:gaugeactionduality}
The actions given in propositions \ref{prop:gaugeactiontriad} \& \ref{prop:gaugeactionforms} are transpose w.r.t. to the duality pairing \eqref{eq:horizontalvfduality}, i.e.
\begin{align}
\label{eq:gaugeactionduality}
(\lambda\triangleright E)(n) & = E(\lambda\triangleright n), &(\tilde{n}'\triangleright E)(n) & = E(\tilde{n}'\triangleright n).
\end{align}
\end{Corollary}
Let us make a closing remark for this subsection regarding the formalism in trivial bundles $\uP\cong\Sigma\times\uG$.
\begin{Remark}
\label{rem:trivialbundles}
If the bundle $\uP$ is isomorphic to the trivial bundle $\Sigma\times\uG$, the gauge transformations $\cG_{\uP}$ are isomorphic to the $\uG$-valued functions on $\Sigma$, $C(\Sigma,\uG)$. The isomorphism is defined by the relation:
\begin{align}
\label{eq:trivialbundlegauge}
f_{\lambda}(x,g) & = \alpha_{g^{-1}}(g_{\lambda}(x)),\ (x,g)\in\Sigma\times\uG.
\end{align}
\end{Remark}
\subsection{The algebras of loop quantum gravity $\mathfrak{P}_{\textup{LQG}}, \mathfrak{A}_{\textup{LQG}}$ \& the AIL representation}
\label{sec:prelqg}
In this subsection, we will stick to the semi-analytic category (cf. \cite{LewandowskiUniquenessOfDiffeomorphism, FleischhackRepresentationsOfThe} for the original utilization in the context of loop quantum gravity). \\[0.1cm]
Given a (right, semi-analytic) principal G-bundle $\uP\stackrel{\pi}{\rightarrow}\Sigma$ (G compact Lie group), as before, we consider the groupoid of (semi-analytic) paths $\cP_{\Sigma}$ in $\Sigma$. Fixing a system of reference points $\{p_{x}\}_{x\in\Sigma}$, we have the isomorphism
\begin{align}
\label{eq:genconhom}
\overline{\mathcal{A}}\cong\hom\left(\cP_{\Sigma},\uG\right)
\end{align}
by definition \ref{def:grouphol}.\\[0.1cm]
The construction of $\mathfrak{P}_{\textup{LQG}}$ and $\mathfrak{A}_{\textup{LQG}}$ is guided by the observation that $\overline{\mathcal{A}}$ may be endowed with a compact, Hausdorff topology, which makes it accessible to measure theoretic consideration (cf. \cite{AshtekarDifferentialGeometryOn, AshtekarRepresentationsOfThe, AshtekarRepresentationTheoryOf, BaezGeneralizedMeasuresIn, MarolfOnTheSupport, AshtekarProjectiveTechniquesAnd} for the original literature). This topology is induced by giving an isomorphism
\begin{align}
\label{eq:genconproj}
\hom(\cP_{\Sigma},\uG)\cong\varprojlim_{l\in\L}\hom\left(l,\uG\right)\subset\prod_{l\in\L}\hom(l,\uG),
\end{align}
where the projective limit is taken over subgroupoids $\L$ of $\cP_{\Sigma}$ generated by embedded, semi-analytic, compactly supported graphs $\gamma\in\Gamma^{\textup{sa}}_{0}$ in $\Sigma$. The projection $p_{l}:\hom(\cP_{\Sigma},\uG)\rightarrow\hom(l,\uG),\ l\in\L,$ are simply the restrictions of the homomorphisms. It follows that the projective limit is a closed subset of the product space $\prod_{l\in\L}\hom(l,\uG)$, where the latter carries the Tikhonov topology. The spaces \mbox{$p_{l}(\overline{\mathcal{A}})=:\overline{\mathcal{A}}_{|l}\cong\hom\left(l,\uG\right)$} acquire their compact topology by the map \eqref{eq:groupholmap}
\begin{align}
\label{eq:groupiso}
\hom(l,\uG)\cong\uG^{|\textup{E}(\gamma_{l})|},
\end{align}
which makes $\prod_{l\in\L}\hom(l,\uG)$ compact. Here, $|\textup{E}(\gamma_{l})|$ denotes the number of edges in $\gamma_{l}$. Furthermore, this allows for the definition of a smooth and an analytic structure on $\overline{\mathcal{A}}$, since these structures are left and right invariant, and thus are invariant under a change of reference points $\{p_{x}\}_{x\in\Sigma}$ \cite{AshtekarDifferentialGeometryOn, AbbatiOnDifferentialStructure}.\\[0.1cm]
Following this, let us introduce the basic building blocks of the algebra $\mathfrak{P}_{\textup{LQG}}$, which is constructed form certain (point-separating) functionals on $|\Lambda|^{1}T^{*}\mathcal{A}_{\uP}$. We loosely follow the notation of \cite{ThiemannModernCanonicalQuantum}.
\begin{Definition}[Cylindrical functions]
\label{def:cyl1}
The $C^*$-algebra $\overline{\Cyl}$ is the closure of the cylindrical functions $\Cyl=\bigcup_{l\in\L}C(\overline{\mathcal{A}}_{l})/\sim$ in the $\sup$-norm $\left\|\ .\ \right\|_{\infty}$. The equivalence is defined to be
\begin{align}
\label{eq:cylequivalence}
f_{l}\sim f_{l'}:\Leftrightarrow\exists l''\supseteq l,l': p_{l''l}^{*}f_{l}=p_{l''l'}^{*}f_{l'},
\end{align}
where $p_{l''l}:\overline{\mathcal{A}}_{l''}\rightarrow\overline{\mathcal{A}}_{l},\ l,l''\in\L,$ is the restriction map. Every $f\in\Cyl$ is given as a projective family of functions $\{f_{l}\}_{l\in\L}$. Explicitly, we have
\begin{align}
\label{eq:cyldef}
f(\bar{A}) & = p_{l}^{*}f_{l}(\bar{A}) = f_{l}(p_{l}(\bar{A})) = f_{l}(\{h^{\bar{A}}_{e}\}_{e\in E(\gamma_{l})}) \\ \nonumber
 & = F_{\gamma_{l}}(\{g(e,\bar{A},\{p_{x}\}_{x\in\Sigma})\}_{e\in E(\gamma_{l})}).
\end{align}
Here $F_{\gamma_{l}}\in C(\uG^{|E(\gamma_{l})|})$ is the function corresponding to $f_{l}\in C(\overline{\mathcal{A}}_{l})$ via \eqref{eq:groupiso}.
\end{Definition}
It is well known that the spectrum of $\overline{\Cyl}$ can be identified with the space of generalized connections $\overline{\mathcal{A}}$, thus leading to the isomorphism
\begin{align}
\label{eq:gelfandiso}
\overline{\Cyl}\cong C(\overline{\mathcal{A}}).
\end{align}
\begin{Definition}[Flux vector fields, cf. \cite{LewandowskiUniquenessOfDiffeomorphism}]
\label{def:fluxes}
The flux vector fields $\mathfrak{X}_{\Flux}$ on $\overline{\mathcal{A}}$ considered as derivations on $\Cyl^{1}$ are the (regularized\footnote{See \cite{ThiemannModernCanonicalQuantum} for a detailed account of the regularization of \eqref{eq:poisson}.}) Hamiltonian vector fields of the functions
\begin{align}
\label{eq:fluxes}
E_{n}(S)=\int_{S}*(E(n))
\end{align}
on $T^{*}\mathcal{A}$ defined by the pairing \eqref{eq:fluxpairing}, where $S$ is a face, i.e. an embedded, semi-analytic, connected hypersurface (without boundary) with oriented normal bundle $NS$, and $n\in\Gamma^{\textup{sa}}_{0}(\textup{Ad}(\uP_{|S}))$, a compactly supported, semi-analytic section of adjoint pullback bundle $\uP_{|S}=\iota^{*}_{S}\uP$. The action of the flux vector fields on $f\in\Cyl^{1}$ is obtained as follows:\\[0.1cm]
By proposition \ref{prop:horizontalforms}, we find a unique $\tilde{n}\in\overline{\Lambda}^{0}(\uP_{|S},\fg)_{\textup{Ad}}$, which gives rise to a $1$-parameter group of gauge transformation $\lambda_{t\tilde{n}}\in\cG_{\uP_{|S}}$ by theorem \ref{th:gaugealgexp}. These gauge transformations define generalized gauge transformations on $\overline{\mathcal{A}}$ in the following way:
\begin{align}
\label{eq:fluxflowdef}
h^{\lambda_{\frac{1}{2}t\tilde{n}}^{*}\bar{A}}_{e} & = h^{\bar{A}}_{e}\circ(\lambda_{\frac{1}{2}t\tilde{n}})^{\varepsilon(e,S)}\end{align}
where 
\begin{align}
\label{eq:indicatorfunctiondef}
\varepsilon(e,S) & =\left\{\begin{matrix*}[l] +1 & e\cap S=e(0)\wedge e\ \textup{is\ positively\ outgoing\ from\ }S\\ -1 & e\cap S=e(0)\wedge e\ \textup{is\ negatively\ outgoing\ from\ }S \\  0 & e\cap S=\emptyset\vee e\cap\bar{S}=e \end{matrix*}\right.
\end{align}
is the indicator function of $S$ w.r.t. to adapted edges\footnote{The factor $\frac{1}{2}$ in \eqref{eq:fluxflowdef} is a remnant of the regularization procedure for the Hamiltonian vector field of $E_{n}(S)$ (see \cite{ThiemannModernCanonicalQuantum} for further explanations).}. It is at this point, where semi-analyticity is crucial to ensure that an arbitrary edge $e'$ decomposes into a finite number of adapted edges $e$, which is necessary to get a well-defined action on $\Cyl$. On the group elements $g(e,\bar{A},\{p_{x}\}_{x\in\Sigma})\in\uG$ this leads to
\begin{align}
\label{eq:fluxflowgrouphol}
g(e,\lambda_{\frac{1}{2}t\tilde{n}}^{*}\bar{A},\{p_{x}\}_{x\in\Sigma}) & =g(e,\bar{A},\{p_{x}\}_{x\in\Sigma})\exp_{\uG}(\frac{1}{2}t\varepsilon(e,S)\tilde{n}_{|p_{e(0)}})
\end{align}
A flux vector field $E_{n}(S)$ is the generator of a generalized gauge transformation on $\Cyl^{1}$.
\begin{align}
\label{eq:fluxflowgenerator}
(E_{n}(S)\cdot F_{\gamma^{S}_{l}})(\{g(e,\bar{A},\{p_{x}\}_{x\in\Sigma})\}_{e\in E(\gamma_{l})}) & = \frac{\ud}{\ud t}_{|t=0}F_{\gamma^{S}_{l}}(\{g(e,\lambda_{\frac{1}{2}t\tilde{n}}^{*}\bar{A},\{p_{x}\}_{x\in\Sigma})\}_{e\in E(\gamma_{l})})\\ \nonumber
 &\ =\frac{1}{2}\sum_{e\in E(\gamma^{S}_{l})}\varepsilon(e,S)\tilde{n}^{i}_{|p_{e(0)}}(\uL^{e}_{i}F_{\gamma^{S}_{l}})(\{g(e,\bar{A},\{p_{x}\}_{x\in\Sigma})\}_{e\in E(\gamma^{S}_{l})}).
\end{align}
$F_{\gamma^{S}_{l}}$ denotes the representative of $f\in\Cyl^{1}$ w.r.t. an adapted decomposition $\gamma^{S}_{l}$ of an underlying graph $\gamma_{l}\in\Gamma^{\textup{sa}}_{0}$ and its associated groupoid $l\in\L$. In the following, we will always assume to work with an adapted decomposition of a graph, when we consider the action of a flux vector field.
\end{Definition}
\begin{Remark}
\label{rem:leftrightchange}
Note that we stick to the realization of the flux vector fields by left invariant vector fields on the structure group $\uG$. This is, again, due to the use of right principal bundles, and the requirement of an isomorphism $\overline{\mathcal{A}}\cong\hom(\mathcal{P}_{\Sigma},\uG)$ rather than an anti-isomorphism (cp. remark \ref{rem:gaugeactionhol}). Contrary, we could change the definition of adapted edges in such a way that the non-vanishing contributions would be due to edges ending at a face, i.e. $e\cap S=e(1)$, if we wanted to arrive at a formulation in terms of right invariant vector fields on $\uG$.
\end{Remark}
Our definition of the flux vector field appears to differ slightly from those existing in the literature (cf. especially \cite{LewandowskiUniquenessOfDiffeomorphism}), but is nevertheless equivalent by the following lemma.
\begin{Lemma}
\label{lem:fluxequivalence}
Instead of defining the flux vector field in terms of $S\subset\Sigma$ and $n\in\Gamma^{\textup{sa}}_{0}(\textup{Ad}(\uP_{|S}))$, we may equivalently define them by $S\subset\Sigma$ and $\tilde{X}\in\mathfrak{X}^{\textup{sa}}_{0}(\uP_{|S})^{\uG}_{\textup{vert}}$, a semi-analytic, compactly supported, right invariant, vertical vector field on $\uP_{|S}$ (cf. \cite{LewandowskiUniquenessOfDiffeomorphism}). \\[0.1cm]
More precisely, we consider the flow
\begin{align}
\label{eq:fluxequivalenceflow}
\phi^{\tilde{X}}_{t}:\uP_{|S}\longrightarrow\uP_{|S},\ t\in\R
\end{align}
generated by $\tilde{X}$, which is a gauge transformation of $\uP_{|S}$ by the right invariance of $\tilde{X}$. Then, we may replace $\lambda_{\pm\frac{1}{2}t\tilde{n}}$ in \eqref{eq:fluxflowdef} by $\phi^{\tilde{X}}_{\pm\frac{1}{2}t}$ and define flux vector fields $E_{\tilde{X}}(S)$ according to this relation.
\begin{Proof}
Note that for $\tilde{n}\in\overline{\Lambda}^{0}(\uP_{|S},\fg)_{\textup{Ad}}$ we have by theorem \ref{th:gaugealgexp}:
\begin{align}
\label{eq:fluxequivalenceproof1}
\frac{\ud}{\ud t}_{|t=0}\exp_{\cG_{\uP_{|S}}}(t\tilde{n})(p) & = \frac{\ud}{\ud t}_{|t=0}p\exp_{\uG}(t\tilde{n}_{|p})\\ \nonumber
 & = (\tilde{n}_{|p})^{*}_{|p},\ p\in\uP_{|S}.
\end{align}
Clearly, $(\tilde{n})^{*}\in\mathfrak{X}(\uP_{|S})$ is semi-analytic and compactly supported if and only if $\tilde{n}$ is. Moreover, due to the definition of $^{*}:\fg\rightarrow\mathfrak{X}(\uP_{|S})^{\uG}$ and the $\textup{Ad}$-equivariance of $\tilde{n}$, $(\tilde{n})^{*}$ is right invariant and vertical.\\[0.1cm]
Conversely, since $\phi^{\tilde{X}}_{t},\ t\in\R,$ is a $1$-parameter group (connected to the identity, $\phi^{\tilde{X}}_{t=0}=\id_{\uP_{|S}}$), we find a corresponding $1$-parameter group $f_{\tilde{X},t}\in C(\uP_{|S},\uG)_{\alpha},\ t\in\R,$ by lemma \ref{lem:gaugeauto}. Then, by theorem \ref{th:gaugealgexp}, we find $\tilde{n}_{\tilde{X}}$, s.t.
\begin{align}
\label{eq:fluxequivalenceproof2}
\phi^{\tilde{X}}_{t} & = \exp_{\cG_{\uP}}(t\tilde{n}_{\tilde{X}}),\ \forall t\in\R.
\end{align}
\end{Proof}
\end{Lemma}
In view of the calculations which will be performed in the following section of the article, we state a useful result about the flux vector fields.
\begin{Lemma}[cf. \cite{SahlmannOnTheSuperselection}]
\label{lem:fluxaction}
The action of the flux vector fields on $\Cyl^{1}$ can be computed as follows:
\begin{align}
\label{eq:fluxaction}
E_{n}(S)\cdot f=\sum_{x\in\Sigma}\sum_{[e]_{x}\in\mathcal{K}}\varepsilon([e]_{x},S)n^{i}_{|p_{x}}\uL^{[e]_{x}}_{i|x}f,
\end{align}
where $\varepsilon$ denotes the indicator functions of $S$ w.r.t. the edge germs $[e]_{x},\ x\in\Sigma$. The set of edge germs $\mathcal{K}_{x}$ does not depend on $x\in\Sigma$ in this setting. The action of the elementary vector fields $\uL^{i}_{x,[e]_{x}}$ is defined to be:
\begin{align}
\label{eq:pointfluxes}
\uL^{[e]_{x}}_{i|x}p^{*}_{l}f_{l}=p^{*}_{l}\left(\frac{1}{2}\sum_{\bar{e}\in E(\gamma_{l})}\delta_{x,\bar{e}(0)}\delta_{[e]_{x},[\bar{e}]_{\bar{e}(0)}}\uL^{\bar{e}}_{i}f_{l}\right),
\end{align} 
where an adapted representative $f_{l}$ of $f$ was chosen. The commutation relations between these vector fields are
\begin{align}
\label{eq:pointcom}
\left[\uL^{[e]_{x}}_{i|x},\uL^{[e']_{x'}}_{j|x'}\right]=\frac{1}{2}f_{ij}\!\ ^{k}\delta_{x,x'}\delta_{[e]_{x},[e']_{x}}\uL^{k}_{x,[e]_{x}},
\end{align}
where $[\tau_{i},\tau_{j}]=f_{ij}\!\ ^{k}\tau_{k}$ defines the structure constants of $\fg$.
\end{Lemma}
From the cylindrical functions $\Cyl$ and the flux vector fields (short: fluxes) we construct the *-algebra $\mathfrak{P}_{\textup{LQG}}$ and a certain Weyl form $\mathfrak{A}_{\textup{LQG}}$ of it. We denote by $\langle\mathfrak{X}_{\Flux}\rangle$ the Lie algebra span of $\mathfrak{X}_{\Flux}$.
\begin{Definition}[The holonomy-flux algebra, cf. \cite{LewandowskiUniquenessOfDiffeomorphism}]
\label{def:hfa}
The *-algebra $\mathfrak{P}_{\textup{LQG}}$ is the *-algebra given by the quotient $\mathfrak{F}/\mathfrak{I}$ of the tensor algebra $\mathfrak{F}$ generated by $\Cyl^{\infty}$ and $\langle\mathfrak{X}_{\Flux}\rangle\subset\mathfrak{X}(\overline{\mathcal{A}})$ by the two-sided -*-ideal $\mathfrak{I}$ defined by the elements:
\begin{align}
& Vf-fV-V\cdot f \\ \nonumber
& VV'-V'V-[V,V']_{\mathfrak{X}(\overline{\mathcal{A}})} \\ \nonumber
& ff'-f'f=0,\ \ \ \forall f,f'\in\Cyl^{\infty},\ V,V'\in\langle\mathfrak{X}_{\Flux}\rangle.
\end{align}
The tensor product is taken relative to the algebra structure of $\Cyl$ resp. $\Cyl^{\infty}$ to make $\mathfrak{F}$ a $\Cyl^{\infty}$-module. The involution $*$ is defined by complex conjugation on $\Cyl^{\infty}$, by $\overline{V\cdot f}=V\cdot\overline{f}$ on $\langle\mathfrak{X}_{\Flux}\rangle$, and extends to an anti-automorphism of $\mathfrak{F}$.\\ Note that the flux vector fields satisfy the reality condition \mbox{$E_{n}(S)=-E_{n}(S)^*, E_{n}(S)\in\mathfrak{X}_{\Flux}$.}
\end{Definition}
At this point, it is important to note, that there is a natural action by semi-analytic gauge transformations $\cG^{\textup{sa}}_{\uP}$ and, more generally, semi-analytic automorphisms $\aut^{\textup{sa}}(\uP)$ on the algebra $\mathfrak{P}_{\textup{LQG}}$. In general, the latter cover diffeomorphisms, $\textup{Diff}^{\textup{sa}}(\Sigma)$, different from the identity:
\begin{align}
\label{eq:bundleautomorphisms}
\xymatrix@=1.5cm{
\uP \ar[r]^{\chi}\ar[d]_{\pi} & \uP \ar[d]^{\pi} \\
\Sigma \ar[r]^{\phi_{\chi}} & \Sigma
}
\end{align}
with $\chi\in\aut^{\textup{sa}}(\uP),\ \phi_{\chi}\in\textup{Diff}^{\textup{sa}}(\Sigma)$. For general bundles, it is not necessarily the case that every diffeomorphism $\phi\in\textup{Diff}^{\textup{sa}}(\Sigma)$ is covered by an automorphism $\chi_{\phi}\in\aut^{\textup{sa}}(\uP)$, as this amounts to a non-trivial lifting problem (cf. \cite{AbbatiTheLieGroup}).
\begin{align}
\label{eq:nontrivlift}
\xymatrix@=1.5cm{
& \uP \ar[d]^{\pi} \\
\uP \ar@{.>}[ur]^{\chi_{\phi}} \ar[r]_{\phi\circ\pi} & \Sigma
}
\end{align}
In the smooth category, one finds a short exact sequence of NLF-manifolds \cite{AbbatiTheLieGroup}
\begin{align}
\label{eq:smoothexactsequence}
\xymatrix{
1 \ar[r] & \cG^{\infty}_{\uP} \ar[r] & \aut^{\infty}(\uP) \ar[r] & \textup{Diff}^{\infty}_{\natural}(\Sigma) \ar[r] & 1
}
\end{align}
with an open subgroup $\textup{Diff}^{\infty}_{\natural}(\Sigma)$ of $\textup{Diff}^{\infty}(\Sigma)$ containing the connected component of the identity. This issue does not arise for the Ashtekar-Barbero variables, since then the bundle $\uP$ comes from the natural bundle $\uP_{SO}(\Sigma)$ \cite{KolarNaturalOperationsIn}. The actions of both groups of transformations on the basic elements, i.e. the cylindrical functions and the fluxes, look as follows:
\begin{Definition}
\label{def:gaugetrafo}
The transformations $\cG^{\textup{sa}}_{\uP}$ and automorphisms $\aut^{\textup{sa}}(\uP)$ have natural (right) actions on $\Cyl$ and $\mathfrak{X}_{\Flux}$ induced by those of corollary \ref{cor:gaugeaction} and lemma \ref{lem:gaugeactiontensorial}.
\begin{align}
\label{eq:gaugeactionalgebra}
\alpha_{\lambda}(f)(\bar{A}) & = p_{l}^{*}f_{l}((\lambda^{-1})^{*}\bar{A})=F_{\gamma_{l}}(\{f_{\lambda}(p_{e(1)})g(e,\bar{A},\{p_{x}\}_{x\in\Sigma})(f_{\lambda}(p_{e(0)}))^{-1}\}_{e\in E(\gamma(l))}) \\[0.1cm] \nonumber
\alpha_{\lambda}(E_{n}(S)) & = (\lambda\triangleright E)_{n}(S) = E_{\lambda\triangleright n}(S) \\[0.25cm]
\label{eq:diffaction}
\alpha_{\chi}(f)(\bar{A}) & =p^{*}_{l}f_{l}((\chi^{-1})^{*}\bar{A})= f_{l}(\{\chi\circ h^{A}_{\phi_{\chi}^{-1}(e)}\circ\chi^{-1}\}_{e\in E(\gamma_{l})}) \\ \nonumber
 & \ = F_{\gamma_{l}}(\{g(e,(\chi^{-1})^{*}A,\{p_{x}\}_{x\in\Sigma})\}_{e\in E(\gamma_{l})}) \\ \nonumber
 & \ = F_{\phi^{-1}_{\chi}(\gamma_{l})}(\{g_{\chi}(\bar{e}(1))g(\bar{e},A,\{p_{x}\}_{x\in\Sigma})g_{\chi}(\bar{e}(0))^{-1}\}_{\bar{e}\in E(\phi^{-1}_{\chi}(\gamma_{l}))}), \\[0.1cm] \nonumber
\alpha_{\chi}(E_{n}(S)) & = (\chi_{*}E)_{n}(S) = E_{\chi^{*}n}(\phi_{\chi}^{-1}(S)),\ \ f\in\Cyl, E_{n}(S)\in\mathfrak{X}_{\Flux}, \lambda\in\cG^{\textup{sa}}_{\uP}, \chi\in\aut^{\textup{sa}}(\uP),
\end{align}
where $g_{\chi}:\Sigma\rightarrow\uG$, s.t. $\chi(p_{x})=R_{g_{\chi}(x)}(p_{\phi_{\chi}(x)})$\footnote{This implies: $g_{\chi^{-1}}\circ\phi_{\chi}=g^{-1}_{\chi}$. Furthermore, this definition has the necessary equivariance properties w.r.t. a change of reference system $\{p_{x}\}_{x\in\Sigma}\mapsto\{p'_{x}\}_{x\in\Sigma}$.}, and $\chi^{*}n=\chi^{-1}\circ n\circ\phi_{\chi}$. These actions extend to *-automorphic actions on $\mathfrak{P}_{\textup{LQG}}$.
\end{Definition}
Since the algebra $\mathfrak{P}_{\textup{LQG}}$ is supposed to be generated by the cylindrical functions and the fluxes, it is necessary to allow only semi-analytic gauge transformations or automorphisms, as otherwise the action of the transformations groups would not preserve the elementary operators of the algebra. Nevertheless, (distributional) extensions of these transformations groups have been discussed in the literature \cite{VelhinhoFunctorialAspectsOf, BahrAutomorphismsInLoop}, and can be shown to have a well-defined action on $\Cyl^{\infty}$, but which do not preserve $\mathfrak{X}_{\Flux}$. As an example, we show that, in case of a trivial bundle $\uP\cong\Sigma\times\uG$, the extension of $\cG^{\textup{sa}}_{\uP}\cong C^{\textup{sa}}(\Sigma,\uG)$ to $G^{\Sigma}=\left\{g:\Sigma\rightarrow\uG\right\}$ leads to elements that are not generated from finite linear combinations of fluxes (unless $\textup{Ad}:\uG\rightarrow\aut(\fg)$ is trivial):\\[0.1cm]
Let us consider a flux $E_{n}(S)$ and, without loss of generality, a generalized gauge transformation $\{g_{x}\}_{x\in S}$, s.t. \mbox{$\textup{Ad}_{g^{-1}}(n)(x)=m(x)\nparallel n(x)$} and $\forall y\neq x: \textup{Ad}_{g^{-1}}(n)(y)=n(y)$. Then the element
\begin{align}
\label{eq:singularflux}
\alpha_{g}(E_{n}(S)) & = E_{\textup{Ad}_{g^{-1}}(n)}(S)
\end{align}
is not of the form required of a flux. Furthermore, this leads to
\begin{align}
\label{eq:nonclosure}
\left[E_{n}(S),\alpha_{g}(E_{n}(S))\right]=E_{[n,\textup{Ad}_{g^{-1}}(n)]}(|S|)=\tfrac{1}{2}\sum_{[e]_{x}\in\mathcal{K}}\varepsilon([e]_{x},S)^{2}[n,m]_{j}(x)\uL^{[e]_{x}}_{j|x},
\end{align}
which is a point-localized vector field on $\Cyl^{\infty}$. Although it is not a point-localized flux, as it contains the squared type indicator function of $S$. Clearly, such a point-localized object cannot be obtained from fluxes, as these are defined w.r.t. to open, semi-anlaytic surfaces S and compactly supported, semi-analytic functions $n\in C^{\textup{sa}}_{0}(S,\fg)$ thereon.\\[0.1cm]
This subtlety indicates that the known proof of uniqueness of the Ashtekar-Isham-Lewandowski representation \cite{LewandowskiUniquenessOfDiffeomorphism}, which requires the algebra $\mathfrak{P}_{\textup{LQG}}$ to be generated by finite linear combinations of products of the cylindrical functions and the fluxes, strictly speaking only holds without considering the action of $\uG^{\Sigma}$. On the other hand, this subtlety poses no problem for the proof of uniqueness given in \cite{FleischhackRepresentationsOfThe} involving a generalized Weyl form of $\mathfrak{P}_{\textup{LQG}}$.\\
From a practical point of view the extension of $\cG^{\textup{sa}}_{\uP}$ to $\uG^{\Sigma}$ appears to be unnecessary, because the action of the semi-analytic gauge transformations is sufficiently localizable due to the existence of semi-analytic partitions of unity (cf. \cite{LewandowskiUniquenessOfDiffeomorphism} and references therein). Additionally, the use of $\cG^{\textup{sa}}_{\uP}$ entails the occurrence of large gauge transformation, i.e. gauge transformation not homotopic to the identity, which might be useful in the discussion of chiral symmetry breaking in loop quantum gravity (see below). \\[0.1cm]
The algebra $\mathfrak{A}_{\textup{LQG}}$ is obtained by partially extending and exponentiating the generators of $\mathfrak{P}_{\textup{LQG}}$, and providing it with the formal commutation relations induced by the Lie bracket on $\mathfrak{X}(\overline{\mathcal{A}})$. The reason for not exponentiating the cylindrical functions is due to the fact, that they are essentially continuous functions of holonomies, the latter being already a sort of exponential of the connection $1$-form $A$. In contrast, the flux vector fields are not exponentiated up to this point, being essentially Hamiltonian vector fields of the (smeared) vector densities $E$.
\begin{Definition}[The *-algebra in Weyl form, cf. \cite{SahlmannOnTheSuperselection}]
\label{def:lqgweyl}
The *-algebra $\mathfrak{A}_{\textup{LQG}}$ is generated by the elements of $\Cyl$ and the Weyl elements $W_{S}(tn)=e^{tE_{n}(S)}=\alpha_{\lambda_{\frac{1}{2}t\tilde{n}}^{*}}$ subject to the following relations (cp. \eqref{eq:fluxflowdef} \& \eqref{eq:gaugeactionalgebra}):
\begin{align}
\label{eq:lqgweyl}
f^{*} & =\overline{f},& ff'  & =f\cdot_{\Cyl}f', \\ \nonumber
W_{S}(tn)^{*} & =W_{S}(tn)^{-1}=W_{S}(-tn),& W_{S}(tn)W_{S}(t'n) & = W_{S}((t+t')n), \\ \nonumber
W_{S}(tn)fW_{S}(tn)^{-1} & = W_{S}(tn)\cdot f = \alpha_{\lambda_{\frac{1}{2}t\tilde{n}}^{*}}(f), & W_{S}(0) & = \mathds{1},
\end{align}
\begin{align}
\label{eq:lqgweylcomp} \nonumber
W_{S}(tn)W_{S'}(t'n')W_{S}(tn)^{-1}W_{S'}(t'n')^{-1} & =\alpha_{\lambda_{\frac{1}{2}t\tilde{n}}^{*}}\!\!\circ\alpha_{\lambda_{\frac{1}{2}t'\tilde{n}'}^{*}}\!\!\circ\alpha^{-1}_{\lambda_{\frac{1}{2}t\tilde{n}}^{*}}\!\!\circ\alpha^{-1}_{\lambda_{\frac{1}{2}t'\tilde{n}'}^{*}},
\end{align}
where $f,f'\in\Cyl$ and $\lambda_{\frac{1}{2}t\tilde{n}}^{*},\lambda_{\frac{1}{2}t'\tilde{n}'}^{*}$ are  as in definition \ref{def:fluxes}. The action of the Weyl elements on $\Cyl$ implements the formal identity $W_{S}(tn)\cdot f=\sum_{k=0}^{\infty}\frac{t^{k}}{k!}E_{n}(S)^{k}\cdot f$ on $\Cyl^{\omega}$. The set of Weyl elements will be denoted by $\mathcal{W}$, and the group generated by this set by $\langle\mathcal{W}\rangle$.
\end{Definition}
\begin{Remark}
\label{rem:lqgweyl}
This definition of the algebra $\mathfrak{A}_{\textup{LQG}}$ is not equivalent to the definition in \cite{FleischhackRepresentationsOfThe}, because we do not regard $\mathfrak{A}_{\textup{LQG}}$ as a (closed) subalgebra of $B(L^{2}(\overline{\mathcal{A}},d\mu_{0}))$ (see below, \eqref{eq:AILstate}), and thus do not require all relations among the generating elements that would follow from such a definition. We will further explain the consequences of this difference in the next section.
\end{Remark}
Additionally, we consider the extended algebra $\mathfrak{A}^{\textup{ext}}_{\textup{LQG}}$ of $\mathfrak{A}_{\textup{LQG}}$ generated by elementary, point-localised fluxes \eqref{eq:pointfluxes} and the cylindrical functions. The extended algebra allows us to obtain an explicit expression for the commutator between the fluxes
\begin{align}
\label{eq:fluxcom}
[E_{n}(S),E_{n'}(S')] & =\tfrac{1}{2}\sum_{x\in S\cap S'}\sum_{[e]_{x}\in\mathcal{K}}\varepsilon([e]_{x},S)\varepsilon([e]_{x},S')[n,n']_{k}(x)\uL^{[e]_{x}}_{k|x},
\end{align}
which will be important in the following sections. It is interesting to note that the commutator does not close among the fluxes in the non-Abelian case\footnote{In the Abelian case the relation \eqref{eq:fluxcom} is trivially closed, i.e. $[E_{n}(S),E_{n'}(S')]=0$.} precisely because of the indicator function $\varepsilon$, i.e. the product $\varepsilon([e]_{x},S)\varepsilon([e]_{x},S')$ is in general not of the form $\varepsilon([e]_{x},S'')$ for a suitable surface $S''$. Although, there are certain special cases where iterated commutators lead to fluxes again, e.g.
\begin{align}
\label{eq:higherfluxcom}
[E_{n}(S),[E_{n'}(S),E_{n''}(S)]] & =\tfrac{1}{4}\sum_{x\in S\cap S'}\sum_{[e]_{x}\in\mathcal{K}}\varepsilon([e]_{x},S)[n,[n',n'']_{k}(x)\uL^{[e]_{x}}_{k|x}=\tfrac{1}{4}E_{[n,[n',n'']]}(S),
\end{align}
since $\varepsilon([e]_{x},S)^{3}=\varepsilon([e]_{x},S)$. This is a feature that is missed by a restriction to Abelian groups $G$, e.g. $G=U(1)^{3}$(``Abelian artifact'', cf. \cite{BojowaldMathematicalStructureOf}).\\[0.1cm]
Typically, the Ashtekar-Isham-Lewandowski representation is invoked as a Hilbert space representation of $\mathfrak{P}_{\textup{LQG}}$ resp. $\mathfrak{A}_{\textup{LQG}}$, defined by the irregular (algebraic) state
\begin{align}
\label{eq:AILstate}
\omega_{0}(fE_{n_{1}}(S_{1})...E_{n_{j}}(S_{j})) & = \left\{\begin{matrix*}[l] \mu_{0}(f) & \textup{if}\ \{1,..,j\}=\emptyset \\[0.1cm] 0 & \textup{else} \end{matrix*}\right. ,\ \forall f\in\Cyl^{\infty},\ E_{n_{1}}(S_{1})...E_{n_{j}}(S_{j})\in\mathfrak{X}_{\Flux}, \\[0.25cm] \nonumber
\omega_{0}(fW_{S_{1}}(n_{1})...W_{S_{j}}(n_{j})) & =\mu_{0}(f),\ \forall f\in\Cyl,\ W_{S_{1}}(n_{1})...W_{S_{j}}(n_{j})\in\mathfrak{A}_{\textup{LQG}},
\end{align}
where $\mu_{0}$ denotes the Ashtekar-Isham-Lewandowski measure induced by the Haar measure on $G$. In terms of (gauge-variant) spin network functions $T_{s},\ s\in S$, which form a special orthonormal basis in $\mathfrak{H}_{\omega_{0}}\cong L^{2}(\overline{\mathcal{A}},d\mu_{0})$, \eqref{eq:AILstate} reads
\begin{align}
\label{eq:SNstate}
\omega_{0}(T_{s}E_{n_{1}}(S_{1})...E_{n_{j}}(S_{j})) & = \left\{\begin{matrix*}[l] \delta_{s,0} & \textup{if}\ \{1,..,j\}=\emptyset \\[0.1cm] 0 & \textup{else} \end{matrix*}\right.,\ \forall f\in\Cyl,\ E_{n_{1}}(S_{1})...E_{n_{j}}(S_{j})\in\mathfrak{X}_{\Flux}, \\[0.25cm] \nonumber
\omega_{0}(T_{s}W_{S_{1}}(n_{1})...W_{S_{j}}(n_{j})) & =\delta_{s,0},\ \forall s\in S,\ W_{S_{1}}(n_{1})...W_{S_{j}}(n_{j})\in\mathfrak{A}_{\textup{LQG}},
\end{align}
where $s=0$ denotes the spin network label corresponding to the empty graph $\gamma=\emptyset$. This representation enjoys a uniqueness property under certain natural assumptions \cite{LewandowskiUniquenessOfDiffeomorphism}.\\[0.1cm]
At this point, we want to state a short lemma regarding the regularity and gauge invariance of states on $\mathfrak{P}_{\textup{LQG}}$ and $\mathfrak{A}_{\textup{LQG}}$ for compact, connected $\uG$.
\begin{Lemma}
\label{lem:gaugeinvariantstates}
Let $\omega$ be a gauge invariant state, i.e. $\omega\circ\alpha_{\lambda}=\omega,\ \lambda\in\cG^{\textup{sa}}_{\uP}$, on $\mathfrak{P}_{\textup{LQG}}$ or $\mathfrak{A}_{\textup{LQG}}$. Then, $\omega$ is irregular w.r.t. the gauge variant spin network functions (cf. \cite{ThiemannModernCanonicalQuantum}),
\begin{align}
\label{eq:gaugevariantSNfunctions}
T_{\gamma,\vec{\pi},\vec{m},\vec{n}}(\bar{A}) & = \prod_{e\in E(\gamma)}\sqrt{\dim(\pi_{e})}\ \pi_{e}(g(e,\bar{A},\{p_{x}\}_{x\in\Sigma}))_{m_{e},n_{e}},
\end{align}
with $\gamma\in\Gamma^{\textup{sa}}_{0},\ \{[\pi_{e}]\}_{e\in E(\gamma)}\in(\hat{G}\setminus\{[\pi_{\textup{triv}}]\})^{|E(\gamma)|},\ m_{e},n_{e}=1,...,\dim(\pi_{e})$. Here, irregularity is understood in the sense that for any $\pi_{e}\neq\pi_{\textup{triv}}$ there exist $m_{e},n_{e}=1,...,\dim(\pi_{e})$, s.t.
\begin{align}
\label{eq:SNirregularity}
[0,1]\ni s\longmapsto\omega(T_{e_{s},\pi_{e},m_{e},n_{e}}),\ e\in\cP_{\Sigma},\ e_{s}(t)=e(st),\ t\in[0,1]
\end{align}
is not continuous from the right in $[0,1]$ at $s=0$.
\begin{Proof}
The action of the gauge transformations $\lambda\in\cG^{\textup{sa}}_{\uP}$ on the gauge variant spin network functions looks as follows:
\begin{align}
\label{eq:gaugeactionSNfunctions}
\alpha_{\lambda}(T_{\gamma,\vec{\pi},\vec{m},\vec{n}})(\bar{A}) & = \prod_{e\in E(\gamma)}\sqrt{\dim(\pi_{e})}\ \pi_{e}(f_{\lambda}(p_{e(1)})g(e,\bar{A},\{p_{x}\}_{x\in\Sigma})f_{\lambda}(p_{e(0)})^{-1})_{m_{e},n_{e}} \\ \nonumber
 & = \prod_{e\in E(\gamma)}\sqrt{\dim(\pi_{e})}\ \sum_{k_{e},l_{e}=1}^{\dim(\pi_{e})}\pi_{e}(f_{\lambda}(p_{e(1)}))_{m_{e},k_{e}}\pi_{e}(g(e,\bar{A},\{p_{x}\}_{x\in\Sigma}))_{k_{e},l_{e}}\pi_{e}(f_{\lambda}(p_{e(0)})^{-1})_{l_{e},n_{e}}.
\end{align}
Now, let us choose maximal torus $\textup{T}\subset\uG$ and consider spin network functions $T_{e_{s},\pi_{e},m_{e},n_{e}}$ defined on single edges $\{e_{s}\}_{s\in[0,1]}\subset\cP_{\Sigma}$, and gauge transformations $\lambda_{e_{s}(1)}$ localized at the vertex $e_{s}(1)$ of $e_{s}$, s.t. $\forall\ 1\geq s>0$: $f_{\lambda}(p_{e_{s}(1)})=t\neq1_{\uG}\in\textup{T}\subset\uG$ and $f_{\lambda}(p_{e_{s}(0)})=1_{G}$. Such gauge transformations exist because of the existence of semi-analytic partitions of unity \cite{LewandowskiUniquenessOfDiffeomorphism, FleischhackRepresentationsOfThe}. Next, we notice that $[\pi_{e}]\neq[\pi_{\textup{triv}}]$ implies the non-triviality of
\begin{align}
\label{eq:nontrivialtorusrep}
\pi_{e|\textup{T}}:\textup{T}\rightarrow\aut(V_{\pi_{e}}),\ \exists t\in\textup{T}:\ \pi_{e|\textup{T}}(t)\neq\mathds{1}_{V_{\pi_{e}}},
\end{align}
since every $g\in\uG$ is conjugate to some $t_{g}\in\textup{T}$ \cite{BroeckerRepresentationsOfCompact}. Thus, we obtain, by diagonalizing the representation of $\textup{T}$, a non-trivial decomposition 
\begin{align}
\label{eq:nontrivialtorusdecomp}
V_{\pi_{e}}\cong\bigoplus_{\rho_{e}}V_{\rho_{e}},\ &\ \pi_{e|\textup{T}}\cong\bigoplus_{\rho_{e}}\rho_{e},
\end{align}
where $\rho_{e}:\textup{T}\rightarrow\mathds{T},\ \dim(V_{\rho_{e}})=1,$ are irreducible representations of $\textup{T}$, i.e. characters of the maximal torus, $\rho_{e}\in\hat{\textup{T}}$. From \eqref{eq:gaugeactionSNfunctions}, \eqref{eq:nontrivialtorusrep} and \eqref{eq:nontrivialtorusdecomp}, we conclude that we find an element $t\in\textup{T}$, s.t.
\begin{align}
\omega(T_{e_{s},\pi_{e},m_{e},n_{e}}) & = \omega(\alpha_{\lambda_{s}}(T_{e_{s},\pi_{e},m_{e},n_{e}})) \\ \nonumber
 & = \underbrace{\rho_{e}(t)}_{\neq1}\omega(T_{e_{s},\pi_{e},m_{e},n_{e}}),\ s\in[0,1],
\end{align}
for some $m_{e},n_{e}$, and we have $\forall 1\geq s>0:\ \omega(T_{e_{s},\pi_{e},m_{e},n_{e}})=0$. We may even choose $m_{e}=n_{e}$. But, $\omega(T_{e_{0},\pi_{e},m_{e},n_{e}})=\sqrt{\dim(\pi_{e})}\delta_{m_{e},n_{e}}$, because $g(e_{0},A,\{p_{x}\}_{x\in\Sigma})=1$. Thus, discontinuity follows for diagonal expectation value functions $[0,1]\in s\mapsto\omega(T_{e_{s},\pi_{e},m_{e},m_{e}}),\ m_{e}=1,...,\dim(\pi_{e})$.
\end{Proof}
\end{Lemma}
This result is inspired by a similar statement in the algebraic formulation of quantum gauge field theories \cite{LoeffelholzMathematicalStructureOf}. Interestingly, in quantum field theory the only way to avoid irregular representations of the gauge field variable $A\in\mathcal{A}_{\uP}$, seems to be the use of indefinite inner product (Krein) spaces (cf. \cite{LoeffelholzMathematicalStructureOf, StrocchiNonPerturbativeFoundations}). 
\subsection{The algebra of loop quantum cosmology $\mathfrak{A}_{\textup{LQC}}$ \& the Bohr representation}
\label{sec:prelqc}
The algebra $\mathfrak{A}_{\textup{LQC}}$ of (homogeneous, isotropic) loop quantum cosmology is given by the Weyl algebra associated with the space $\mathds{R}^{2}=\{(\lambda,\theta)\ |\ \lambda,\theta\in\R\}$ with the (canonical) symplectic structure (cf. \cite{StrocchiAnIntroductionTo}):
\begin{align}
\label{eq:r2sympl}
\sigma((\lambda_{1},\theta_{1}),(\lambda_{2},\theta_{2}))=\lambda_{1}\theta_{2}-\lambda_{2}\theta_{1}.
\end{align}
\begin{Definition}
\label{def:lqcweyl}
The algebra $\mathfrak{A}_{\textup{LQC}}$ is the *-algebra generated by the elements $U(\lambda)=e^{i\lambda b},\ \lambda\in\R$, and $V(\theta)=e^{i\theta\nu},\ \theta\in\R$, subject to the relations
\begin{align}
\label{eq:lqcweyl}
& U(\lambda)^{*}=U(-\lambda)=U(\lambda)^{-1},\ U(0)=\mathds{1} & & V(\theta)^{*}=V(-\theta)=V(\theta)^{-1},\ V(0)=\mathds{1} \\ \nonumber
& U(\lambda_{1})U(\lambda_{2})=U(\lambda_{1}+\lambda_{2}),\ V(\theta_{1})V(\theta_{2})=V(\theta_{1}+\theta_{2}) , & & U(\lambda)V(\theta)=e^{-i\lambda\theta}V(\theta)U(\lambda).
\end{align}
$\mathfrak{A}_{\textup{LQC}}$ can be made a $C^{*}$-algebra by completing it w.r.t. the maximal $C^{*}$-norm (cf. \cite{BaerWaveEquationsOn})
\begin{align}
\label{eq:cstarmax}
||W||_{\textup{max}}=\sup\{||W||\ |\ ||\ .\ ||\ \textup{is\ a}\ C^{*}-\textup{norm\ on}\ \mathfrak{A}_{\textup{LQC}}\},\ W\in\mathfrak{A}_{\textup{LQC}}.
\end{align}
\end{Definition}
The generators $b,\ \nu$ defined w.r.t. a regular representations are related to the Hubble parameter and the oriented volume respectively. These elementary variables are related to those in standard treatments of LQC, where $\{b,v\}=2\gamma$, by rescaling the volume $\nu=\frac{1}{2\gamma}v$. Alternatively, this algebra is written in terms of the combined operators $W(\lambda,\theta)=e^{i(\lambda b+\nu\theta)}=e^{i\frac{\lambda\theta}{2}}U(\lambda)V(\theta),\ (\lambda,\theta)\in\R^{2}$.
\begin{align}
\label{eq:lqcweylalternative}
& W(\lambda,\theta)^{*}=W(-\lambda,-\theta)=W(\lambda,\theta)^{-1},\ W(0,0)=\mathds{1} \\ \nonumber
& W(\lambda_{1},\theta_{1})W(\lambda_{2},\theta_{2})=e^{-\frac{i}{2}\sigma((\lambda_{1},\theta_{1}),(\lambda_{2},\theta_{2}))}W(\lambda_{1}+\lambda_{2},\theta_{1}+\theta_{2})=e^{-i\sigma((\lambda_{1},\theta_{1}),(\lambda_{2},\theta_{2}))}W(\lambda_{2},\theta_{2})W(\lambda_{1},\theta_{1}).
\end{align}
This algebra is obtained by restricting the holonomies to a cubic graph and the fluxes to surfaces dual to this graph, and exploiting isotropy to reduce from $SU(2)$ to $U(1)$, followed by a ``decompactification'' to $\mathds{R}_{\textup{Bohr}}$ (cf. \cite{AshtekarLoopQuantumCosmology, BojowaldMathematicalStructureOf}).\\[0.1cm]
In analogy with the Hilbert space representation typically chosen for $\mathfrak{A}_{\textup{LQG}}$, one selects a preferred (irregular) representation induced by the (algebraic) state
\begin{align}
\label{eq:Bohrstate}
\omega_{0}(W(\lambda,\theta))=\delta_{\lambda,0},\ \forall\lambda,\theta\in\mathds{R}.
\end{align}
The representation of this state can be understood in terms of Besicovitch’s almost-periodic functions, i.e. $\fH_{\omega_{0}}\cong L^{2}(\R_{\textup{Bohr}},d\mu_{\textup{Bohr}})$ (cf. \cite{LoeffelholzSpectralStochasticProcesses, StrocchiAnIntroductionTo}). The uniqueness of this state was recently justified \cite{AshtekarOnTheUniqueness} along the same lines as the uniqueness of the Ashtekar-Isham-Lewandowski representation for $\mathfrak{A}_{\textup{LQG}}$ \cite{LewandowskiUniquenessOfDiffeomorphism}. In contrast, the usual Fock or Schr\"odinger representation is obtained from the (regular) state
\begin{align}
\label{eq:Fockstate}
\omega_{F}(W(\lambda,\theta))=e^{-\frac{\lambda^{2}+\theta^{2}}{4}},\ \forall\lambda,\theta\in\mathds{R}.
\end{align}
\subsubsection{Dynamically induced ``superselection'' sectors in LQC}
\label{sec:dynsectors}
The quantization of the (gravitational) Hamiltonian constraint $\mathcal{H}$ in the spatially flat case ($k=0$) that is derived w.r.t. the GNS representation $(\fH_{\omega_{0}},\pi_{\omega_{0}},\Omega_{\omega_{0}})$ of the state \eqref{eq:Bohrstate} takes the form (up to numerical constants) \cite{AshtekarLoopQuantumCosmology}:
\begin{align}
\label{eq:hamiltonconstraint}
\mathcal{H} & \sim\nu\frac{\sin(\lambda_{0}b)}{\lambda_{0}}\nu\frac{\sin(\lambda_{0}b)}{\lambda_{0}} \sim \nu\Im(U(\lambda_{0}))\nu\Im(U(\lambda_{0})),\ \lambda_{0}\in\R,
\end{align}
where $\Im(U(\lambda_{0}))$ denotes the imaginary part, $\lambda_{0}$ is a minimal length scale connected to the (kinematical) minimal area eigenvalue of loop quantum gravity\footnote{The presence of the minimal length scale $\lambda_{0}$ serves as an argument for the use of the irregular representation, as the limit $\lambda_{0}\rightarrow0$ is supposed to be forbidden in the quantum theory.}, and $\nu$ is the (densely defined) generator of the $1$-parameter group $\{\pi_{\omega_{0}}(V(\theta))\}_{\theta\in\R}$, which exists by the continuity of the state w.r.t. $\theta$ (cf. \cite{KaminskiTheFlatFRW} for details regarding the domain $D(\mathcal{H})$). It is easy to see that $\mathcal{H}$ commutes with $\pi_{\omega_{0}}(V(\theta=\frac{\pi}{\lambda_{0}}))$, and the representation admits a direct sum decomposition w.r.t. the spectrum of the latter ($\sigma(\pi_{\omega_{0}}(V(\theta=\frac{\pi}{\lambda_{0}})))=S^{1}=\{e^{i\vartheta}\ |\ \vartheta\in[0,2\pi)\}$):
\begin{align}
\label{eq:Bohrdirectsum}
\fH_{\omega_{0}}\cong\bigoplus_{\vartheta\in[0,2\pi)}\fH_{\vartheta},\ \pi_{\omega_{0}}\cong\bigoplus_{\vartheta\in[0,2\pi)}\pi_{\vartheta},
\end{align}
where the summands $(\fH_{\vartheta},\pi_{\vartheta}),\ \vartheta\in[0,2\pi),$ are preserved by the subalgebra $\mathfrak{A}^{U(1)}_{\textup{LQC}}\subset\mathfrak{A}_{\textup{LQC}}$
\begin{align}
\label{eq:circlealgebra}
\mathfrak{A}^{\lambda_{0}}_{\textup{LQC}}=\langle\{W(2\lambda_{0}n,\theta)\ |\ (n,\theta)\in\mathds{Z}\times\R\subset\R^{2}\}\rangle,
\end{align}
i.e. the Weyl algebra associated with the cotangent bundle $T^{*}U(1)\cong S^{1}\times\R$. Thus, the parameter $\lambda_{0}$ is one half of the inverse radius of the $S^{1}$-factor, and plays the role of a ``compactification scale''. In the literature, it is argued that this gives rise to a ``superselection'' structure induced by $\mathcal{H}$, and Dirac observables are computed w.r.t. one of the $\vartheta$-sectors (cf. \cite{AshtekarQuantumNatureOf1, AshtekarLoopQuantumCosmology}).\\[0.1cm]
In the following sections we will explain further similarities between this structure in loop quantum cosmology and the Koslowki-Sahlmann representations \cite{KoslowskiLoopQuantumGravity}. 
\section{Central operators and the structure group}
\label{sec:centers}
The algebras $\mathfrak{A}_{\textup{LQG}}$ and $\mathfrak{A}^{\lambda_{0}}_{\textup{LQC}}$ have a common feature that will be in focus of this section. Namely, both algebras, as defined in \ref{def:lqcweyl} and \ref{def:lqgweyl}, have non-trivial centers, which implies that they have no irreducible, faithful representations. Furthermore, there is a common cause for the appearance of non-trivial central elements in these algebras, as both can be related to quantizations of the cotangent bundle of a compact group, i.e. the structure group $\uG$ and the dual $U(1)$ of the invariance group $\mathds{Z}$ of $\mathcal{H}$ respectively. This feature affects the representations theory of both algebras, since irreducible representation require that elements of the center are represented by multiples of the identity (superselection structure). In the subsequent discussion of these aspects, we will repeatedly encounter a unifying algebraic structure consisting of the following data (cf. \cite{StrocchiAnIntroductionTo}):
\begin{itemize}
	\item[1.] An algebra $\mathfrak{A}$ of ``observables'' with a non-trivial center $\mathfrak{Z}$.
	\item[2.] An extended algebra $\mathfrak{F}\supset\mathfrak{A}$, which is called the ``field algebra''.
	\item[3.] A group of automorphisms $G$ representing the adjoint action of $\mathfrak{Z}$ on $\mathfrak{F}$, s.t. $\mathfrak{A}\subseteq\mathfrak{F}^{G}$ is contained in the fix-point algebra of $\mathfrak{F}$ w.r.t. this action. $G$ is called ``global gauge group''.
	\item[4.] A group of automorphisms of $\mathfrak{A}$, which does not leave the center $\mathfrak{Z}$ pointwise invariant, i.e. $\rho(Z)\neq Z$ for $\rho\in\mathcal{C}, Z\in\mathfrak{Z}$. Elements of $\mathcal{C}$ are called ``charge automorphisms''. 
\end{itemize}
Let us briefly explain the nomenclature (see also remark \ref{rem:gausslaw} below): The algebra $\mathfrak{A}$ is the algebra of which we intend to understand the representation theory. In many cases this will be the algebra of observables of a given quantum system. The center $\mathfrak{Z}$ of $\mathfrak{A}$ reflects the superselection structure of the quantum system. Moreover, if we are dealing with a gauge theory, we will have to deal with the question of gauge invariance, and the issue of charged fields (charged with respect to the global gauge group $G$) from which we construct observables. As the observables should be invariant under the global gauge group $G$, and the charged fields cannot be part of $\mathfrak{A}$, but should belong to an (extended) field algebra $\mathfrak{F}$, we need to require that $\mathfrak{A}\subset\mathfrak{F}^{G}$. On the other hand, the center $\mathfrak{Z}$ embeds into the gauge group $G$ via the adjoint action, and is, therefore, part of the global gauge group, while the adjoint action of a (unitary) charged field can lead to an automorphism of $\mathfrak{A}$ (charge and conjugate charge combine to zero charge) that moves the elements of center among each other, which amounts to a shift in the superselection structure. \\[0.1cm]
Thus, we see that the construction of a field algebra $\mathfrak{F}$ from a given algebra $\mathfrak{A}$ with center $\mathfrak{Z}$ provides candidates of charged automorphisms, $\rho\in\mathcal{C}$, which can be combined with known representations $\pi$ of $\mathfrak{A}$ to give new, inequivalent representations $\pi\circ\rho$.\\[0.1cm]
Furthermore, the inequivalence of the representations $\pi$ and $\pi\circ\rho$ can be understood as an instance of spontaneous symmetry breaking (in the algebraic sense, cf. \cite{BratteliOperatorAlgebrasAnd1}). Namely, in spite of the fact that $\rho$ is an automorphisms (a symmetry) of $\mathfrak{A}$, a pure (or primary) algebraic state $\omega$ on $\mathfrak{A}$ cannot be invariant w.r.t. $\rho$, i.e. $\omega\circ\rho\neq\omega$. To see this, we recall that a state $\omega$ is pure (or primary) if and only if the associated GNS representation $(\pi_{\omega},\fH_{\omega},\Omega_{\omega})$ has a 1-dimensional commutant $\pi_{\omega}(\mathfrak{A})'=\C\cdot\mathds{1}_{\fH_{\omega}}$ (or center $\mathfrak{Z}_{\omega}=\pi_{\omega}(\mathfrak{A})'\cap\pi_{\omega}(\mathfrak{A})''=\C\cdot\mathds{1}_{\fH_{\omega}}$). But, this implies non-invariance of $\omega$, because $\pi_{\omega}(\mathfrak{Z})\subset\pi_{\omega}(\mathfrak{A})'=\C\cdot\mathds{1}_{\fH_{\omega}}$, and $\rho$ acts non-trivially on $\mathfrak{Z}$, which are incompatible requirements.\\
If we intended to consider an invariant state $\omega$ on $\mathfrak{A}$, we will have an extremal (or central) decomposition into pure (or primary) states $\omega_{x},\ x\in X,$:
\begin{align}
\label{eq:generaldecomposition}
\omega & = \int_{X}\omega_{x}d\mu(x).
\end{align} 
The central decomposition of $\omega$ is related to the central decomposition of the von Neumann algebra $\pi_{\omega}(\mathfrak{A})''$ with respect to its center $\mathfrak{Z}_{\omega}=\pi_{\omega}(\mathfrak{A})'\cap\pi_{\omega}(\mathfrak{A})''$, which is especially important in physics, because elements of the center describe invariants of the quantum system, which assume specific values in the states $\omega_{x},\ x\in X$ \cite{BratteliOperatorAlgebrasAnd1}.
This point of view will be important in the discussion of chiral symmetry breaking in section \ref{sec:theta}.\\
The algebraic formulation of spontaneous symmetry breaking connects with the standard formulation that a symmetry of the Hamiltonian of a quantum system does not entail a symmetric (ground) state of the latter in the following way: Symmetries $\rho$ of the algebra $\mathfrak{A}$ of observables, to which the Hamiltonian is affiliated, are not necessarily symmetries of a state $\omega$ on $\mathfrak{A}$.
\subsection{Central operators in $\mathfrak{A}^{\lambda_{0}}_{\textup{LQC}}$}
\label{sec:lqcweyl}
Our analysis of the algebra $\mathfrak{A}^{\lambda_{0}}_{\textup{LQC}}$, and how the structure of its center $\mathfrak{Z}^{\lambda_{0}}_{\textup{LQC}}$ reflects the decomposition \eqref{eq:Bohrdirectsum}, follows closely the analysis of the Weyl algebra for a quantum particle on a circle as given in \cite{StrocchiAnIntroductionTo}.\\[0.1cm]
The center $\mathfrak{Z}^{\lambda_{0}}_{\textup{LQC}}$ is generated by the element $W(0,\frac{\pi}{\lambda_{0}})$, and we may regard the algebra $\mathfrak{A}^{\lambda_{0}}_{\textup{LQC}}$ as the fix-point algebra $\mathfrak{A}_{\textup{LQC}}^{\mathds{Z}}$ w.r.t the $\mathds{Z}$-action
\begin{align}
\label{eq:zaction}
\alpha_{m}(W(\lambda,\theta)) & =e^{i\frac{\pi}{\lambda_{0}} m\lambda}W(\lambda,\theta),\ m\in\mathds{Z}.
\end{align}
In this setting, $\mathds{Z}$ is called the global gauge group, and $\mathfrak{A}_{\textup{LQC}}$ the field algebra. The extension of $\mathfrak{A}^{\lambda_{0}}_{\textup{LQC}}$ to $\mathfrak{A}_{\textup{LQC}}$ is minimal in a precise sense (cf. \cite{AcerbiInfraredSingularFields}). Clearly, the action of the global gauge group is implemented by the adjoint action generator $W(0,\frac{\pi}{\lambda_{0}})$ of the center $\mathfrak{Z}^{\lambda_{0}}_{\textup{LQC}}$ on $\mathfrak{A}_{\textup{LQC}}$:
\begin{align}
\label{eq:adjointzaction}
\alpha_{m}(W(\lambda,\theta)) & = W(0,\tfrac{\pi}{\lambda_{0}})^{m}W(\lambda,\theta)W(0,-\tfrac{\pi}{\lambda_{0}})^{m}.
\end{align}
We observe that the irregular state $\omega_{0}$ \eqref{eq:Bohrstate} is gauge invariant, in contrast to the regular Fock state $\omega_{F}$ \eqref{eq:Fockstate}. On the other hand, requiring a gauge invariant state $\omega\circ\alpha_{m}=\omega$ immediately leads to $\omega(W(\lambda,\theta))=0\ \textup{if}\ \lambda\notin2\lambda_{0}\mathds{Z}$. Additionally, we have the so-called charged automorphisms of $\mathfrak{A}^{\lambda_{0}}_{\textup{LQC}}$:
\begin{align}
\label{eq:chargedauto}
\rho_{\vartheta}(W(2\lambda_{0}n,\theta)) & = e^{-i\vartheta\frac{\lambda_{0}}{\pi}\theta}W(2\lambda_{0}n,\theta),\ \vartheta\in[0,2\pi),
\end{align}
which are necessarily outer automorphisms, as they do not the leave the center $\mathfrak{Z}^{\lambda_{0}}_{\textup{LQC}}$ pointwise invariant. These automorphisms are inner in the larger algebra $\mathfrak{A}_{\textup{LQC}}$
\begin{align}
\label{eq:adjointchargeauto}
\rho_{\vartheta}(W(2\lambda_{0}n,\theta)) & = W(\vartheta\tfrac{\lambda_{0}}{\pi},0)W(2\lambda_{0}n,\theta)W(-\vartheta\tfrac{\lambda_{0}}{\pi},0),
\end{align}
and they intertwine inequivalent irreducible representations of $\mathfrak{A}^{\lambda_{0}}_{\textup{LQC}}$. The latter follows, because every irreducible representation $\pi$ requires that we have for the generator of the center $\pi(W(0,\frac{\pi}{\lambda_{0}}))=e^{i\vartheta},\ \vartheta\in[0,2\pi)$. Thus, an irreducible representation $\pi_{\vartheta}$ is labeled by an ``angle'' $\vartheta\in[0,2\pi)$, and we find that
\begin{align}
\label{eq:inequivalentrepslqcweyl}
\pi_{0} & = \pi_{\vartheta}\circ\rho_{\vartheta}
\end{align}
is a representation satisfying
\begin{align}
\label{eq:inequivalentrepsuniquenesslqc}
\pi_{0}(W(0,\tfrac{\pi}{\lambda_{0}}))=1.
\end{align}
One can show that any two irreducible representations of $\mathfrak{A}^{\lambda_{0}}_{\textup{LQC}}$ supplemented by \eqref{eq:inequivalentrepsuniquenesslqc}, which are regular w.r.t. the $1$-parameter group $\{W(0,\theta)\}_{\theta\in\R}$, are unitarily equivalent. The representations $\pi_{\vartheta}$ can be realized by the GNS representation of the state
\begin{align}
\label{eq:thetastateslqc}
\omega_{\vartheta}(W(2\lambda_{0}n,\theta)) & = e^{i\vartheta\frac{\lambda_{0}}{\pi}\theta}\delta_{n,0},\ \forall n\in\mathds{Z}, \theta\in\R.
\end{align}
The difference between the representations with distinct values of $\vartheta$ are also seen on the level of the generator $\nu_{\vartheta}$ of $\{\pi_{\vartheta}(V(\theta))\}_{\theta\in\R}$, i.e. we have $\fH_{\vartheta}\cong L^{2}([0,\frac{\pi}{\lambda_{0}}),db)$ and $\nu_{\theta}$ is the self-adjoint extension of $-i\frac{\partial}{\partial b}$ subject to the boundary condition $\psi(\frac{\pi}{\lambda_{0}})=e^{i\vartheta}\psi(0)$. \\[0.1cm]
The occurrence of these structures can be related to the topology of group $U(1)$, which is the dual of the invariance group $\mathds{Z}$ acting according to \eqref{eq:zaction} (cf. \cite{MorchioQuantumMechanicsOn}). The $\mathds{Z}$-action corresponds geometrically to translations of the variable $b$, i.e. $b\mapsto b+\frac{\pi}{\lambda_{0}}m,\ m\in\mathds{Z}$. In the restricted setting of the algebra $\mathfrak{A}^{\lambda_{0}}_{\textup{LQC}}$, it can be interpreted as the action of the large gauge transformations with winding number $m$, which are the rotations by $2\pi m$ of the underlying circle group $U(1)\subset T^{*}U(1)$. As argued in \cite{AcerbiThetaVacuaCharge, LoeffelholzMathematicalStructureOf, StrocchiNonPerturbativeFoundations}, there is a strong analogy between these algebraic structures and those present in the context of chiral symmetry breaking and the vacuum structure of QCD (see section \ref{sec:theta}). To this end the following remark is in order:\\[0.1cm]
Although the charge automorphisms $\rho_{\vartheta}$, which play the role of the chiral automorphisms of QCD, and gauge automorphisms $\alpha_{m}$ commute, i.e.
\begin{align}
\label{eq:chargegaugecommutation}
\rho_{\vartheta}\circ\alpha_{m} & = \alpha_{m}\circ\rho_{\vartheta},\ \vartheta\in[0,2\pi), m\in\mathds{Z},
\end{align}
the implementers of the gauge transformations $W(0,m\frac{\pi}{\lambda_{0}})\in\mathfrak{Z}^{\lambda_{0}}_{\textup{LQC}}$ are not invariant under the charge automorphisms $\rho_{\vartheta}$ by \eqref{eq:chargedauto}. Thus, the charge symmetry is necessarily spontaneously broken in any irreducible representation $\pi_{\vartheta}$ of $\mathfrak{A}^{\lambda_{0}}_{\textup{LQC}}$. \\[0.1cm]
Interestingly, there is also way to relate the $\vartheta$-sectors to a purely imaginary topological term contributing to the action of a free particle on the circle via the functional integral point of view (cf. \cite{StrocchiAnIntroductionTo}):
\begin{align}
\label{eq:topologicaltermlqc}
S=\frac{m}{2}\int\dot{x}(\tau)^{2}d\tau + i\vartheta\frac{\lambda_{0}}{\pi}\int\dot{x}(\tau)d\tau.
\end{align}
We conclude this subsection by pointing out how the appearance of these structures differs in the GNS representation of the Fock state \eqref{eq:Fockstate} from that in the Bohr state \eqref{eq:Bohrstate}. The Fock state leads to a representation $(\fH_{F},\pi_{F},\Omega_{F})$ of $\mathfrak{A}_{\textup{LQC}}$ that is unitarily equivalent to the Schr{\"o}dinger representation by von Neumann's uniqueness theorem, but this representation is reducible for $\mathfrak{A}^{\lambda_{0}}_{\textup{LQC}}$. In fact, we have a central decomposition of the representation over the spectrum of $W(0,\frac{\pi}{\lambda_{0}})$:
\begin{align}
\label{eq:Fockdirectintegral}
\fH_{F}\cong\int_{\vartheta\in[0,2\pi)}\fH_{\vartheta}d\vartheta,\ \pi_{F}\cong\int_{\vartheta\in[0,2\pi)}\pi_{\vartheta}d\vartheta.
\end{align}
In comparison with \eqref{eq:Bohrdirectsum}, which reflects that $\pi_{\omega_{0}}(W(0,\frac{\pi}{\lambda_{0}}))$ has only pure point spectrum, the spectrum of $\pi_{F}(W(0,\frac{\pi}{\lambda_{0}}))$ is purely absolutely continuous, and the GNS vectors $\Omega_{\vartheta}$ are its improper eigenvectors.
\subsection{Central operators in $\mathfrak{A}_{\textup{LQG}}$}
\label{sec:lqgweyl}
In the construction of $\mathfrak{A}_{\textup{LQG}}$, it is assumed that the structure group of the principal bundle $\uP$ is a compact Lie group $\uG$. By compactness, $\uG$ is the finite extension of its (connected) identity component $\uG_{0}$ by $\uG/\uG_{0}\cong\pi_{0}(\uG)$. On the other hand, it is well-known \cite{BroeckerRepresentationsOfCompact} that $\uG_{0}$ is isomorphic to the quotient of the product of a $n$-torus $U(1)^{n}$ and a compact, connected, simply connected Lie group $\uK$ by a central, finite, Abelian subgroup $\uA$. Furthermore, $\uK$ is isomorphic to a finite product of compact, connected, simply connected, simple Lie groups.
\begin{align}
\label{eq:compactliegroup}
\uG_{0}\cong(\uK\times U(1)^{n})/\uA
\end{align}
Therefore, we will give separate discussions of the structure of $\mathfrak{A}_{\textup{LQG}}$ in the two cases:
\begin{itemize}
	\item[1.] $\uG\cong U(1)^{n}$ for some $n\in\mathds{N}$.
	\item[2.] $\uG\cong\uK$ is compact, connected, simply connected and simple.
\end{itemize}
In the second case, we will also comment on the case $\uG\cong\uK/\uA,\ \uA\subset Z(\uK)\ \&\ \textup{finite}$, i.e. $\pi_{1}(G)\neq\{1\}$.
\subsubsection{$\uG\cong U(1)^{n}$}
\label{sec:abelianlqgweyl}
If $\uG\cong U(1)^{n}$, we notice that the only non-trivial relation among the generators of $\mathfrak{A}_{\textup{LQG}}$ is (cp. \eqref{eq:lqgweyl})
\begin{align}
\label{eq:abelianlqgweylrelation}
W_{S}(tn)f & = \alpha_{\lambda^{*}_{\frac{1}{2}t\tilde{n}}}(f)W_{S}(tn).
\end{align}
Furthermore, the generators $\{\tau_{i}\}_{i=1,...,n}\subset\underline{u(1)}^{\oplus n}$ ($\underline{u(1)}=\textup{Lie}(U(1))=i\R$) can be treated independently, because $U(1)^{n}$ is Abelian. Thus, it is sufficient to discuss the relation \eqref{eq:abelianlqgweylrelation} for $n=1$.\\[0.1cm]
Since the $C^{*}$-algebra $C(U(1))$ is generated by the characters $(\ .\ )^{n}:U(1)\rightarrow\C,\ g\mapsto g^{n}$, let us consider \eqref{eq:abelianlqgweylrelation} for the spin (or charge) network functions
\begin{align}
\label{eq:CNfunctions}
T_{\gamma,\vec{m}}(\bar{A}) & = \prod_{e\in E(\gamma)}g(e,\bar{A},\{p_{x}\}_{x\in\Sigma})^{m_{e}},\ \gamma\in\Gamma^{\textup{sa}}_{0}, \bar{A}\in\overline{\mathcal{A}},
\end{align}
where $\vec{m}=(m_{e})_{e\in E(\gamma)}\in\mathds{Z}^{|E(\gamma)|}_{\neq0}$:
\begin{align}
\label{eq:CNweylrelation}
W_{S}(tn)T_{\gamma,\vec{m}} & = \left(\prod_{e\in E(\gamma^{S})}e^{m^{S}_{e}\frac{1}{2}t\varepsilon(e,S)\tilde{n}_{|p_{e(0)}}}\right)T_{\gamma,\vec{m}}W_{S}(tn).
\end{align}
The labels $\{m^{S}_{e}\}_{e\in E(\gamma^{S})}$ are those defined by $T_{\gamma,\vec{m}}$ for the adapted graph $\gamma^{S}$. This relation basically resembles the commutation relations of $\mathfrak{A}^{\lambda_{0}}_{\textup{LQC}}$ (cp. \eqref{eq:lqcweyl}), apart from the complication due to the intersection properties of $\gamma$ and $S$. Therefore, the center $\mathfrak{Z}_{\textup{LQG}}$ of $\mathfrak{A}_{LQG}$ is generated by elements $W_{S}(tn)$ with $t\tilde{n}_{|p_{x}}=4\pi i\ \forall x\in S$. But by definition \ref{def:fluxes}, this is only possible if $S$ is closed and compact, as otherwise $n\in\Gamma(\textup{Ad}(\uP))$ is not allowed to be constant on $S$. Examples of such closed and compact $S$ are given by embedded compact Riemann surfaces, e.g. $S=S^{2}$ or $\mathds{T}^{2}$. Thus, we have:
\begin{align}
\label{eq:centerabelianlqg}
\mathfrak{Z}_{\textup{LQG}} & = \langle W_{S}(4\pi i)\rangle,\ S\ \textup{closed\ and\ compact}.
\end{align}
As in the previous subsection, we conclude that in any irreducible representation $\pi$ of $\mathfrak{A}_{\textup{LQG}}$ the generators of $\mathfrak{Z}_{\textup{LQG}}$ are represented by multiples of the identity, i.e.
\begin{align}
\label{eq:abelianlqgirrepcondition}
\pi(W_{S}(4\pi i)) & = e^{i\vartheta_{S}},\ \vartheta_{S}\in[0,2\pi) \ \forall S\ \textup{closed\ and\ compact},
\end{align}
which implies that any irreducible representation $\pi=\pi_{\vartheta}$ is labeled by family of ``angles'' $\vartheta=\{\vartheta_{S}\}_{S}$, where we set $\vartheta_{S}=0$ if $S$ is not closed and compact.\\[0.1cm]
If we define a type of (charged) automorphisms $\rho_{\vartheta,E^{(0)}}=\{\rho_{\vartheta_{S},E}\}_{S}$ by
\begin{align}
\label{eq:abelianlqgchargeauto}
\rho_{\vartheta_{S},E^{(0)}}(W_{S}(tn)) & = \left\{\begin{matrix*}[l] e^{i\frac{\vartheta_{S}}{4\pi\vol_{*E^{(0)}(i)}(S)}t\int_{S}*E^{(0)}(n)}W_{S}(tn) & S\ \textup{closed\ and\ compact} \\ W_{S}(tn) & \textup{otherwise} \end{matrix*}\right.,
\end{align}
for some $E^{(0)}\in\Gamma(T\Sigma\otimes\textup{Ad}^{*}(\uP)\otimes|\Lambda|^{1}(\Sigma))$\footnote{$\vol_{*E^{(0)}(i)}(S)=\int_{S}*E^{(0)}(i)$ is the volume of $S$ relative to the pairing of $E^{(0)}$ and the generator $i$ of $\underline{u(1)}$, and serves as a normalization factor.}, we find a relation analogous to \eqref{eq:inequivalentrepslqcweyl}:
\begin{align}
\label{eq:iequivalentrepsabelianlqg}
\pi_{0} & = \pi_{\vartheta}\circ\rho_{\vartheta, E^{(0)}}.
\end{align}
Clearly, the Ashtekar-Isham-Lewandowski representation \eqref{eq:AILstate} is a representation with $\vartheta=\{0\}_{S}$, and is singled out by automorphism invariance or diffeomorphism and gauge invariance (cf. \cite{LewandowskiUniquenessOfDiffeomorphism, FleischhackRepresentationsOfThe}). The question, if this is the only representation with $\vartheta=\{0\}_{S}$, is more subtle, and will be discussed elsewhere. Inspecting \eqref{eq:abelianlqgchargeauto} more closely, we may even choose $\vartheta_{S}\neq0$ for arbitrary faces $S$, and set $\vartheta_{S}=\vartheta_{0}\ \forall S$
\begin{align}
\label{eq:abelianlqgchargeautoext}
\rho_{\vartheta_{0},E^{(0)}}(W_{S}(tn)) & = e^{i\frac{\vartheta_{0}}{4\pi\vol_{*E^{(0)}(i)}(S)}t\int_{S}*E^{(0)}(n)}W_{S}(tn)\ \forall S,
\end{align}
which would lead us to the Koslowksi-Sahlmann representations $\pi_{\vartheta_{0},E^{(0)}}=\pi_{\omega_{0}}\circ\rho^{-1}_{\vartheta_{0},E^{(0)}}$ \cite{KoslowskiLoopQuantumGravity} (see below). \\[0.1cm]
Following the discussion of the previous subsection, we can also ask, whether we can regard $\mathfrak{A}_{\textup{LQG}}$ as the fix-point algebra of a larger field algebra $\mathfrak{F}_{\textup{LQG}}$ under the adjoint action of the generators of the center $\mathfrak{Z}_{\textup{LQG}}$. To this end, we exploit the similarity of \eqref{eq:CNweylrelation} and \eqref{eq:lqcweyl}:\\[0.1cm]
First, we use the covering homomorphism $\pi_{0}:\R\rightarrow U(1),\ \varphi\mapsto e^{i\varphi}$, which coincides with the exponential map $\exp:\underline{u(1)}\rightarrow U(1)$, to lift the functions $F_{\gamma_{l}}$ on $U(1)^{|E(\gamma_{l})|}\cong\hom(l,U(1))$ to functions $\tilde{F}_{\gamma_{l}}=F_{\gamma_{l}}\circ\pi_{0}^{\times|E(\gamma_{l})|}$ to $\R^{|E(\gamma)|}\cong\hom(l,\R)$. Clearly, the lifting is isometric w.r.t the sup-norm and compatible with the projective structure of $\hom(\cP_{\Sigma},\R)$. Thus, we are allowed to consider $\tilde{F}_{\gamma_{l}}$ as defining a cylindrical function on the latter via $p_{l}:\hom(\cP_{\Sigma},\R)\rightarrow\hom(l,\R)$. This is possible, because the construction of $\overline{\mathcal{A}}$ does not require the compactness of $\uG$. Only the construction of the Ashtekar-Isham-Lewandowski measure requires a compact structure group. Especially, we may lift the spin network functions $\tilde{T}_{\gamma,\vec{m}}$, which form a subset of the Fourier network functions on $\hom(\cP_{\Sigma},\R)$:
\begin{align}
\label{eq:FNfunctions}
\tilde{T}_{\gamma,\vec{\beta}}(\{\varphi_{e}\}_{e\in E(\gamma)}) & = \prod_{e\in E(\gamma)}e^{i\alpha_{e}\varphi_{e}},\ \gamma\in\Gamma^{\textup{sa}}_{0},\vec{\beta}=(\beta_{e})_{e\in E(\gamma)}\in\R^{|E(\gamma)|}_{\neq0}.
\end{align}
Second, we note that the action of the Weyl elements $W_{S}(tn)$ on the cylindrical functions, which defines the commutation relation \eqref{eq:abelianlqgweylrelation}, is compatible with lift through $\xi_{0}:\R\rightarrow U(1)$, as well.
\begin{align}
\label{eq:abelianlqgweylactionlift}
(W_{S}(tn)\cdot\tilde{F}_{\gamma^{S}_{l}})(\{\varphi_{e}\}_{e\in E(\gamma^{S}_{l})}) & = \tilde{F}_{\gamma^{S}_{l}}(\{\varphi_{e}-i\tfrac{1}{2}t\varepsilon(e,S)\tilde{n}_{|p_{e(0)}}\}_{e\in E(\gamma^{S}_{l})})=F_{\gamma^{S}_{l}}(\{e^{i\varphi_{e}}e^{\frac{1}{2}t\varepsilon(e,S)\tilde{n}_{|p_{e(0)}}}\}_{e\in E(\gamma^{S}_{l})}).
\end{align}
Third, we define the field algebra $\mathfrak{F}_{\textup{LQG}}$ to be generated by the Fourier network functions $\tilde{T}_{\gamma,\vec{\beta}}$ and the Weyl elements $W_{S}(tn)$ subject to the equivalent set of relations as in \eqref{eq:lqgweyl}, but involving the lifted action \eqref{eq:abelianlqgweylactionlift}.
\begin{Remark}
\label{rem:liftingclassicalcorrespondence}
The lifting of the structure group $U(1)$ to $\R$ by the covering homomorphism $\pi$, requires on the classical level, i.e. for the construction to be related to structures in principal $\uG$-bundles, the existence of a non-trivial covering of the principal $U(1)$-bundle $\uP$ by a principal $\R$-bundle $\uP_{\R}$
\begin{align}
\label{eq:abliancoveringhom}
\xymatrix@=1.5cm{
\ar@{}[dr] |{\circlearrowright} \uP_{\R} \ar[r]^{\xi} \ar[d]_{\pi_{\R}} & \uP \ar[d]^{\pi} \\
\Sigma \ar[r]^{\id_{\Sigma}} & \Sigma
} &\hspace{0.5cm} 
\xymatrix@=1.5cm{
\ar@{}[dr] |{\circlearrowright} \uP_{\R} \ar[r]^{\xi} \ar[d]_{R_{\varphi}} & \uP \ar[d]^{R_{\xi_{0}(\varphi)}} \\
\uP_{\R} \ar[r]^{\xi} & \uP
}
\end{align}
with a diagram of fibrations:
\begin{align}
\label{eq:abelianfibration}
\xymatrix@R=0.25cm{ 
& \R \ar[dd] \ar[rr]^{\xi_{0}} & & U(1) \ar[dd] \\
\mathds{Z} \ar[ur] \ar[dr] & & & \\
& \uP_{\R} \ar[rr]^{\xi} \ar[dr]_{\pi_{\R}} & & \uP \ar[dl]^{\pi} \\
& & \Sigma & 
}
\end{align}
\end{Remark}
By construction, the adjoint action of the generators of $\mathfrak{Z}_{\textup{LQG}}$ on $\mathfrak{F}_{\textup{LQG}}$ fixes the algebra $\mathfrak{A}_{\textup{LQG}}$\footnote{Strictly speaking, it fixes an algebra containing the lift of $\mathfrak{A}_{\textup{LQG}}$ in $\mathfrak{F}_{\textup{LQG}}$.}, which we infer from:
\begin{align}
\label{eq:abelianlqgfixpointrelation}
\alpha^{S}_{m}(T_{\gamma,\vec{\beta}}) & =W_{S}(4\pi i)^{m}\tilde{T}_{\gamma,\vec{\beta}}W_{S}(-4\pi i)^{m} = \left(\prod_{e\in E(\gamma^{S})}e^{2\pi i\beta^{S}_{e}\varepsilon(e,S)m}\right)\tilde{T}_{\gamma,\vec{\beta}}, \\ \nonumber
\alpha^{S}_{m}(W_{S'}(tn)) & = W_{S}(4\pi i)^{m}W_{S'}(tn)W_{S}(-4\pi i)^{m} = W_{S'}(tn),
\end{align}
where we defined $\mathds{Z}$-actions $\alpha^{S}:\mathds{Z}\rightarrow\aut(\mathfrak{F}_{\textup{LQG}})$ (the ``global'' gauge group) for every closed, compact face $S$. On the other hand, we get (charged) automorphisms by the adjoint action of the Fourier network functions $\tilde{T}_{\gamma,\vec{\beta}},\ \vec{\beta}=(\frac{\vartheta_{e}}{2\pi})_{e\in E(\gamma)}\in[0,1]^{|E(\gamma)|}$ on $\mathcal{A}_{\textup{LQG}}$:
\begin{align}
\label{eq:abelianlqgchargeautoadjoint}
\rho^{\gamma}_{\vec{\beta}}(T_{\gamma',\vec{m}}) & = \tilde{T}_{\gamma,\vec{\beta}}T_{\gamma',\vec{m}}\tilde{T}_{\gamma,-\vec{\beta}} = T_{\gamma',\vec{m}}, \\ \nonumber
\rho^{\gamma}_{\vec{\beta}}(W_{S}(tn)) & = \tilde{T}_{\gamma,\vec{\beta}}W_{S}(tn)\tilde{T}_{\gamma,-\vec{\beta}} = \left(\prod_{e\in E(\gamma^{S})}e^{-\beta^{S}_{e}\frac{1}{2}t\varepsilon(e,S)\tilde{n}_{|p_{e(0)}}}\right)W_{S}(tn).
\end{align}
On the generators of the center $\mathfrak{Z}_{\textup{LQG}}$ these automorphisms lead to
\begin{align}
\label{eq:abelianlqgchargeautoadjointcenter}
\rho^{\gamma}_{\vec{\beta}}(W_{S}(4\pi i)) & = e^{-i\sum_{e\in E(\gamma^{S})}\varepsilon(e,S)\vartheta^{S}_{e}}W_{S}(4\pi i) = e^{-i\vartheta_{S}}W_{S}(4\pi i),
\end{align}
where we defined $\vartheta_{S}=\sum_{e\in E(\gamma)}\varepsilon(e,S)\vartheta^{S}_{e}$. Thus, we arrive at a second type of (charged) automorphisms (cp. \eqref{eq:abelianlqgchargeauto} \& \eqref{eq:abelianlqgchargeautoext}) labeled by a graph $\gamma\in\Gamma^{\textup{sa}}_{0}$ and an associated set of angles $\{\vartheta_{e}\}_{e\in E(\gamma)}$. As in the previous section, we have (cp. \eqref{eq:chargegaugecommutation})
\begin{align}
\label{eq:chargegaugecommutationabelianlqg}
\rho_{\vartheta_{S},E^{(0)}}\circ\alpha^{S}_{m} & = \alpha^{S}_{m}\circ\rho_{\vartheta_{S},E^{(0)}}, \\[0.25cm] \nonumber
\rho^{\gamma}_{\vec{\beta}}\circ\alpha^{S}_{m} & = \alpha^{S}_{m}\circ\rho^{\gamma}_{\vec{\beta}}.
\end{align}
Similar to the discussion of $\vartheta$-representations of $\mathfrak{A}^{\lambda_{0}}_{\textup{LQC}}$, the difference between representations of $\mathfrak{A}_{\textup{LQG}}$ with distinct labels $\{\vartheta_{S}\}_{S}$, which are regular w.r.t. the Weyl elements $W_{S}(tn)$, can be seen on the level of the fluxes, e.g. in the GNS representation of $\omega_{0}$ \eqref{eq:AILstate}:
\begin{align}
\label{eq:abelianlqgchargedfluxes}
E^{\vartheta, E^{(0)}}_{S}(n) & = E_{S}(n) + i\frac{\vartheta_{0}}{4\pi\vol_{*E^{(0)}(i)}(S)}\int_{S}*E^{(0)}(n) \\ \nonumber
E^{\gamma, \vartheta}_{S}(n) & = E_{S}(n) + \tfrac{1}{4\pi}\sum_{e\in E(\gamma^{S})}\varepsilon(e,S)\vartheta^{S}_{e}\tilde{n}_{|p_{e(0)}}.
\end{align}
Actually, this is the starting point for the construction of Koslowski-Sahlmann representations (see below).
\subsubsection{$\uG\cong\uK$ is compact, connected, simply connected and simple}
\label{sec:simplelqg}
In this subsection, we assume that $\uG\cong\uK$ is a compact, connected, simply connected and simple Lie group, which is the most important case for loop quantum gravity, because in a version of the theory in the Ashtekar-Barbero variables $\uK=SU(2)$. A variant of loop quantum gravity w.r.t. the new variables has $\uG=\textup{Spin}_{4}$ \cite{BodendorferNewVariablesFor1}, which is compact, connected, simply connected and semi-simple, because $\textup{Spin}_{4}\cong SU(2)\times SU(2)$ \cite{LawsonSpinGeometry}, and thus can be reduced to the simple case.\\[0.1cm]
In view of the previous subsection, we have additional non-trivial relations among the generators of $\mathfrak{A}_{\textup{LQG}}$ (cp. \eqref{eq:lqgweyl}):
\begin{align}
\label{eq:nonabelianlqgweylrelations}
W_{S}(tn)fW_{S}(tn)^{-1} & = W_{S}(tn)\cdot f = \alpha_{\lambda_{\frac{1}{2}t\tilde{n}}^{*}}(f) \\ \nonumber
W_{S}(tn)W_{S'}(t'n')W_{S}(tn)^{-1}W_{S'}(t'n')^{-1} & =\alpha_{\lambda_{\frac{1}{2}t\tilde{n}}^{*}}\circ\alpha_{\lambda_{\frac{1}{2}t'\tilde{n}'}^{*}}\circ\alpha^{-1}_{\lambda_{\frac{1}{2}t\tilde{n}}^{*}}\circ\alpha^{-1}_{\lambda_{\frac{1}{2}t'\tilde{n}'}^{*}}.
\end{align}
As in the case of the Weyl algebra associated with a linear symplectic space (cp. \eqref{eq:lqcweylalternative}), the second relation sets up a strong relation between the product of Weyl elements $W_{S}(tn)$ and the composition of the maps $\alpha_{\lambda^{*}_{\frac{1}{2}t\tilde{n}}}$, and thus the group product of $\uK$. But, the relation leaves room for the existence of non-trivial central elements. \\
To be more precise, the existence of the central elements is due to the relations \eqref{eq:nonabelianlqgweylrelations} and the fact that for a compact Lie group we can find $0\neq X\in\fk$ s.t. $\exp_{\uK}(X)=1_{\uK}$, because there exist maximal tori in $\uK$. Moreover, since we assume $\uK$ to be simple, it has a non-degenerate, negative definite (by compactness) Killing form $(X,Y)_{\fk}=\tr_{\mathfrak{gl}(\fk)}(\textup{ad}_{X}\circ\textup{ad}_{Y})$, which is $\textup{Ad}$-invariant, i.e. $\textup{Ad}_{\uK}\subset SO(\fk)$. This implies, that all elements in the adjoint orbit of $X\in\fk$, s.t. $\exp_{\uK}=1_{\uK}$, are mapped to $1_{\uK}$:
\begin{align}
\label{eq:adjointorbits}
\exp_{\uK}(\textup{Ad}_{g}(X)) & = \alpha_{g}(\exp_{\uK}(X)) = \alpha_{g}(1_{\uK}) = 1_{\uK},\ \forall g\in\uK.
\end{align}
In general, a similar observation can be made for all elements $g\in Z(\uK)$, since then $\textup{Stab(g)}=\uK$. But by the first line of \eqref{eq:nonabelianlqgweylrelations}, only the cut locus $\exp^{-1}_{\uK}(\{1_{\uK}\})$ of $1_{\uK}$ in $\fk$ will define central elements of $\mathfrak{A}_{\textup{LQG}}$:
\begin{align}
\label{eq:nonabeliancenter}
\mathfrak{Z}_{\textup{LQG}}=\langle W_{S}(tn)\rangle,\ \frac{1}{2}t\tilde{n}_{|p_{x}}=X\in\exp^{-1}_{\uK}(\{1_{\uK}\})\ \textup{for \ all\ reference\ points}\ p_{x},\ S\ \textup{closed\ and\ compact}.
\end{align}
The restriction to closed and compact faces $S$ is again necessary, because of the support properties of $n$. By \eqref{eq:adjointorbits}, the $\textup{Ad}$-equivariance of $\tilde{n}$ implies that $\frac{1}{2}t\tilde{n}_{|p}\in\exp_{\uK}^{-1}(\{1_{\uK}\})\ \forall p\in\uP$. In the case $\uK=SU(2)$, we have:
\begin{align}
\label{eq:su2cutlocus}
\exp^{-1}_{\uK}(\{1_{\uK}\}) & =\left\{4\pi(\vec{e}\cdot\vec{\tau})\ |\ \vec{e}\in S^{2}\subset\R^{3},\ \tau_{i}=-\tfrac{i}{2}\sigma_{i}\right\},
\end{align}
where $\{\sigma_{i}\}_{i=1,2,3}$ are the Pauli matrices. As above, we conclude that $\mathfrak{Z}_{\textup{LQG}}$ is non-trivial, and that in any irreducible representation $\pi$ the identities
\begin{align}
\label{eq:nonabelianlqgirrepcondition}
\pi(W_{S}(2X)) & = e^{i\vartheta_{S}(2X)},\ \vartheta_{S}(2X)\in[0,2\pi) \ \forall X\in\exp^{-1}_{\uK}(\{1_{\uK}\}),\ S\ \textup{closed\ and\ compact},
\end{align}
hold, with $\vartheta_{S}(2X+2X')=\vartheta_{S}(2X)+\vartheta_{S}(2X')\ \textup{mod}\ 2\pi$ if $X=\mu X'$ for some $\mu\in\R$. Unfortunately, we cannot define (charged) isomorphisms by the analog of \eqref{eq:abelianlqgchargeauto}, because we have non-trivial relations among Weyl elements.\\[0.1cm]
For example, since we have that $\uK$ is simply connected, we know that $\textup{Ad}(\uP)$ is spin \cite{LawsonSpinGeometry}. Therefore, we find that $\textup{Ad}(\uP_{|S})\cong S\times\fk$ \cite{LawsonSpinGeometry}, which gives the identification $\Gamma^{\textup{sa}}_{0}(\textup{Ad}(\uP_{|S}))\cong C^{\textup{sa}}_{0}(S,\fk)$. Now, we specialized to $\uK=SU(2)$ choose two constant, ortho-normalized functions $f^{i}_{n}\in C^{\textup{sa}}_{0}(S,\fk),\ i=1,2$, i.e. $(f^{i}_{n},f^{j}_{n})_{\fk}=-\delta_{ij}$ and $f^{i}_{n}(x)=X_{i}\in\fk\forall x\in S$, and consider the associated Weyl element $W_{S}(n_{i}),\ i=1,2$. By the second line of \eqref{eq:nonabelianlqgweylrelations} and the $\textup{Ad}$-equivariance of $\tilde{n}_{i},\ i=1,2$, we obtain
\begin{align}
\label{eq:compactweylperfectness}
W_{S}(4\pi n_{1})W_{S}(-n_{2})W_{S}(4\pi n_{1})^{-1}W_{S}(-n_{2})^{-1} & = W_{S}(2n_{2}).
\end{align}
On the level of the holonomy-flux algebra, the presence of non-trivial relations is exemplified by \eqref{eq:higherfluxcom}. Clearly, an analogue of this construction works for $\uG=\textup{Spin}_{4}$ by the isomorphism $\textup{Spin}_{4}=SU(2)\times SU(2)$. We summarize this observation in the following proposition.
\begin{Proposition}
\label{prop:compactweylperfectness}
For $\uG$ compact, connected and simply connected, assume that $\forall X\in\fg\exists g_{X}:\textup{Ad}_{g_{X}}(X)=-X$. Then, the subgroup $\langle\mathcal{W}_{0}\rangle\subset\langle\mathcal{W}\rangle$, generated by the Weyl elements
\begin{align}
\label{eq:perfectweylsubgroup}
W_{S}(tn),\ S\ \textup{closed\ and\ compact},\ n\in\Gamma^{\textup{sa}}_{0}(\textup{Ad}(\uP_{|S}))\ \textup{constant\ w.r.t.\ some\ triv.\ of\ }\textup{Ad}(\uP_{|S}),
\end{align}
is perfect, i.e. $\langle[\langle\mathcal{W}_{0}\rangle,\langle\mathcal{W}_{0}\rangle]\rangle=\langle\mathcal{W}_{0}\rangle$.
\begin{Proof}
From the simply connectedness of $\uG$, we deduce the triviality of $\textup{Ad}(\uP_{|S})\cong S\times\fg$, as above. Thus, $W_{S}(tn)\in\langle\mathcal{W}_{0}\rangle$ is determined by a constant function $f_{n}:S\rightarrow\fg$. By assumption, we are allowed to choose an element $g_{n}\in\uG$, s.t. $\textup{Ad}_{g_{n}}(f_{n})=-f_{n}$, and by compactness and connectedness of $\uG$, we find $0\neq X_{n}\in\fg$, s.t. $\exp_{\uG}(X_{n})=g_{n}$. If we define a constant section $s_{n}\in\Gamma^{\textup{sa}}_{0}(\textup{Ad}(\uP_{|S}))$ by $S\ni x\mapsto X_{n}\in\fg$, we have
\begin{align}
\label{eq:perfectweylsubgrouprelation}
W_{S}(s_{n})W_{S}(-\tfrac{1}{2}tn)W_{S}(s_{n})^{-1}W_{S}(-\tfrac{1}{2}tn)^{-1} & = W_{S}(tn).
\end{align}
This implies the proposition.
\end{Proof}
\end{Proposition}
\begin{Corollary}
\label{cor:nocharacterautomorphisms}
By proposition \ref{prop:compactweylperfectness}, representations $\pi$ of $\mathfrak{A}_{\textup{LQG}}$ with $\vartheta_{S}\neq0$, for some closed and compact $S$, cannot be induced by character automorphisms
\begin{align}
\label{eq:characterauto}
\rho_{\chi}(W_{S}(tn)) & =\chi(W_{S}(tn))W_{S}(tn)
\end{align}
of the Weyl group $\langle\mathcal{W}\rangle$, where $\chi:\langle\mathcal{W}\rangle\rightarrow U(1)$ is a character.
\end{Corollary}
\begin{Proof}
The generators of the center $\mathfrak{Z}_{\textup{LQG}}$ are contained in the subgroup $\langle\mathcal{W}_{0}\rangle$, which is perfect. Therefore, the restriction $\chi_{|\langle\mathcal{W}_{0}\rangle}$ is trivial.
\end{Proof}
\begin{Remark}
\label{rem:perfectweyl}
The proof of the proposition \ref{prop:compactweylperfectness} clearly fails in this form for $\uG=SU(2)$, if we remove the condition that $n$ is constant. To see this, choose $S\cong S^{2}$ subordinate to a coordinate chart of $\Sigma$, and define $n:S^{2}\rightarrow\fg\cong\R^{3}$ to be the (outward) unit normal vector field on $S^{2}$. This implies that $n^{\perp}_{|x}\cong T_{x}S^{2}$. But, $TS^{2}$ admits no nowhere vanishing, continuous section $m:S^{2}\rightarrow TS^{2}$, since the Euler characteristic is positive, $\chi(S^{2})=2$ \cite{SteenrodTheTopologyOf}.
\end{Remark}
A similar result can formulated for the algebra of flux vector fields $\langle\mathfrak{X}_{\Flux}\rangle$ (cp. \eqref{eq:higherfluxcom}).
\begin{Proposition}
\label{prop:compactfluxperfectness}
Assume that $\fg$ is perfect, which will be the case, if $\fg$ is simple or semi-simple, i.e. $[\fg,\fg]=\fg$. Then, the subalgebra $\langle\mathfrak{X}_{\Flux,0}\rangle\subset\langle\mathfrak{X}_{\Flux}\rangle$ generated by the elements
\begin{align}
\label{eq:perfectfluxsubalgebra}
E_{n}(S),\ S\ \textup{closed\ and\ compact},\ n\in\Gamma^{\textup{sa}}_{0}(\textup{Ad}(\uP_{|S}))\ \textup{constant\ w.r.t.\ some\ triv.\ of\ }\textup{Ad}(\uP_{|S}),
\end{align}
is perfect, i.e. $\langle[\langle\mathfrak{X}_{\Flux,0}\rangle,\langle\mathfrak{X}_{\Flux,0}\rangle]\rangle=\langle\mathfrak{X}_{\Flux,0}\rangle$.
\begin{Proof}
This follows immediately from equation \eqref{eq:higherfluxcom} and the perfectness of $\fg$.
\end{Proof}
\end{Proposition}
\begin{Remark}
\label{rem:gausslaw}
The subalgebra $\langle\mathfrak{X}_{\Flux,0}\rangle$ and subgroup $\langle\mathcal{W}_{0}\rangle$ also exist in the Abelian case, where they admit a natural, heuristic interpretation in terms of Gau{\ss}' law. Formally, we have $W_{S}(tn)=e^{tE_{n}(S)}$, and for $S\ \textup{closed\ and\ compact},\ n\in\Gamma^{\textup{sa}}_{0}(\textup{Ad}(\uP_{|S}))\ \textup{constant}$, we find from the (classical) formula \eqref{eq:fluxes} and the Gau{\ss}' theorem:
\begin{align}
\label{eq:gausslaw}
\int_{V}(\div_{T\Sigma}(E))(n) & = \int_{S}*E(n),\ \partial V = S,
\end{align}
where $n$ is extended constantly to the region $V$ bounded by $S$, and the adjoint bundle is assumed to trivialize over $V$, i.e. $\textup{Ad}(\uP_{|V})\cong V\times\fg$. Thus, $\langle\mathfrak{X}_{\Flux,0}\rangle$ and subgroup $\langle\mathcal{W}_{0}\rangle$ are quantizations of the smeared Gau{\ss}' constraints $(\div_{T\Sigma}(E))_{n}(V) = \int_{V}(\div_{T\Sigma}(E))(n)$, and can serve as implementers of the gauge transformations generated by $n\in\Gamma(\textup{Ad}(\uP_{|V})),\ n=\textup{constant}$. This justifies the terminology ``global'' gauge group for the $\mathds{Z}$-automorphisms $\alpha^{S}$ defined by the adjoint action of the center $\mathfrak{Z}_{\textup{LQG}}$ on the field algebra $\mathfrak{F}_{\textup{LQG}}$:
\begin{align}
\label{eq:gaussgroup}
\mathfrak{Z}_{\textup{LQG}}\subset\cG_{\textup{LQG}}=\langle\mathcal{W}_{0}\rangle=\alpha_{\cG^{\textup{sa},0}_{\uP}}.
\end{align}
The relations \eqref{eq:chargegaugecommutationabelianlqg} generalize accordingly
\begin{align}
\label{eq:spontaneousshiftbreakdown}
\rho_{\vartheta_{S},E^{(0)}}\circ\alpha_{W_{S}(n)} & = \alpha_{W_{S}(n)}\circ\rho_{\vartheta_{S},E^{(0)}}, \\[0.25cm] \nonumber
\rho^{\gamma}_{\vec{\beta}}\circ\alpha_{W_{S}(n)} & = \alpha_{W_{S}(n)}\circ\rho^{\gamma}_{\vec{\beta}}.
\end{align}
We conclude that the charged automorphisms are spontaneously broken w.r.t. gauge invariant, pure states $\omega$.\\[0.1cm]
An interpretation along these lines is not available in the non-Abelian setting, because the Gau{\ss}' law is given by the vanishing of the smeared horizontal (or covariant) divergences:
\begin{align}
\label{eq:nonabeliangausslaw}
\int_{V}(\div_{T\Sigma}^{A}(E))(n)=-\int_{V}\underline{\tilde{E}(d_{A}\tilde{n})}=0,\ (A,\tilde{E})\in|\Lambda|^{1}T^{*}\mathcal{A}_{\uP},
\end{align} 
which spoils the applicability of the Gau{\ss}' theorem. Here, $\underline{\tilde{E}(d_{A}\tilde{n})}$ denotes the projection of the right invariant density $\tilde{E}(d_{A}\tilde{n})$ on $\uP$ to $\Sigma$ (cp. \eqref{eq:compatibleduality}).
\end{Remark}
\begin{Corollary}
\label{cor:nocentralterm}
There are no representations $\pi$ of $\mathfrak{P}_{\textup{LQG}}$ satisfying
\begin{align}
\label{eq:centraltermreps}
\pi(E_{n}(S)) & = \pi_{\omega_{0}}(E_{n}(S))+c_{n}(S)\cdot\mathds{1}_{\fH_{\omega_{0}}},\ c_{n}(S)\in\C,
\end{align}
with $c_{n}(S)\neq0$ for closed and compact $S,\ n\in\Gamma^{\textup{sa}}_{0}(\textup{Ad}(\uP_{|S}))\ \textup{constant\ w.r.t.\ some\ triv.\ of\ }\textup{Ad}(\uP_{|S})$. $\omega_{0}$ is the Ashtekar-Isham-Lewandowski state \eqref{eq:AILstate}
\begin{Proof}
Let $E_{n}(S)\in\langle\mathfrak{X}_{\Flux,0}\rangle$. Then, we have by proposition \ref{prop:compactfluxperfectness} and \eqref{eq:higherfluxcom}:
\begin{align}
\label{eq:centraltermproblem}
\pi(E_{n}(S)) & = \pi([E_{n'},[E_{n''}(S),E_{n'''}(S)]]) = [\pi(E_{n'}(S)),[\pi(E_{n''}(S)),\pi(E_{n'''}(S))]] \\ \nonumber
 & = [\pi_{\omega_{0}}(E_{n'}(S)),[\pi_{\omega_{0}}(E_{n''}(S)),\pi_{\omega_{0}}(E_{n'''}(S))]] = \pi_{\omega_{0}}([E_{n'}(S),[E_{n''}(S),E_{n'''}(S)]]) \\ \nonumber
 & = \pi_{\omega_{0}}(E_{n}(S)),
\end{align}
where we chose $n',n'',n'''$, s.t. $[n',[n'',n''']]=n$ by the perfectness of $\fg$. Thus, $c_{n}(S)=0$.
\end{Proof}
\end{Corollary}
Applying the same reasoning to general $E_{n}(S)\in\langle\mathfrak{X}\rangle$, we conclude, that for arbitrary faces $S$, we are forced to set $c_{n}(S)=0\ \forall\ n\in\Gamma^{\textup{sa}}_{0}(\textup{Ad}(\uP_{|S}))$, s.t. $\tilde{n}\in[\fG^{\textup{sa},0}_{\uP_{|S}},[\fG^{\textup{sa},0}_{\uP_{|S}},\fG^{\textup{sa},0}_{\uP_{|S}}]]$, where $\fG^{\textup{sa},0}_{\uP_{|S}}$ denotes the semi-analytic, compactly supported gauge algebra of $\uP_{|S}$ (see definition \ref{def:gaugealg}). \\[0.25cm]
Finally, we want to consider the case $\uG\cong\uK/\uA$, for some Abelian, finite group $\uA\subset Z(\uK)$. With minor modifications, similar results holds for semi-simple $\uK$, e.g. $\textup{Spin}_{4}$. In the same way, as in the discussion of $\uG=U(1)$, we use the covering homomorphism $\xi_{\uA}:\uK\rightarrow\uG$ to construct a lift of the algebra $\mathfrak{A}_{\textup{LQG}}$ to an extended algebra $\mathfrak{F}_{\textup{LQG}}$. Because $\uG$ is compact, the spin network functions of $\uG$,
\begin{align}
\label{eq:GSNfunctions}
T_{\gamma,\vec{\pi},\vec{m},\vec{n}}(\bar{A}) & = \prod_{e\in E(\gamma)}\sqrt{\dim(\pi_{e})}\pi_{e}(g(e,\bar{A},\{p_{x}\}_{x\in\Sigma}))_{m_{e},n_{e}},
\end{align}
generate the algebra $\Cyl_{\uG}$ by the Peter-Weyl theorem, where we introduced the notation $\Cyl_{\uG}$ to indicate the Lie group the cylindrical functions are based on. Here, we denote by $\pi_{e}(\ .\ )_{m_{e},n_{e}},\ e\in E(\gamma),$ a matrix entry of a non-trivial, unitary, irreducible representation of $\uG$. Therefore, we only need to define the lifts of these functions and the Weyl elements. The lift of a spin network function is defined via pullback:
\begin{align}
\label{eq:GSNfunctionslift}
\tilde{T}_{\gamma,\vec{\pi},\vec{m},\vec{n}}(\{g_{e}\}_{e\in E(\gamma)}) & = \prod_{e\in E(\gamma)}\sqrt{\dim(\pi_{e})}\pi_{e}(\xi_{\uA}(k_{e}))_{m_{e},n_{e}},\ \{k_{e}\}_{e\in E(\gamma)}\in\uK^{|E(\gamma)|},
\end{align}
which embeds these function, isometrically w.r.t. sup-norm, into the spin network functions of $\uK$, and is compatible with the projective structure of $\hom(\cP_{\Sigma},\uK)$:
\begin{align}
\label{eq:KSNfunctions}
\tilde{T}_{\gamma,\vec{\eta},\vec{i},\vec{j}}(\{k_{e}\}_{e\in E(\gamma)}) & = \prod_{e\in E(\gamma)}\sqrt{\dim(\eta_{e})}\eta_{e}(k_{e})_{i_{e},j_{e}},\ \{[\eta_{e}]\}_{e\in E(\gamma)}\in(\hat{\uK}\setminus\{[\eta_{\textup{triv}}]\})^{|E(\gamma)|},\ \{k_{e}\}_{e\in E(\gamma)}\in\uK^{|E(\gamma)|}.
\end{align}
The naturalness of the exponential maps, $\exp_{\uG}\circ\ d\xi_{\uA|1_{\uK}}=\xi_{\uA}\circ\exp_{\uK}$, and the fact that $d\xi_{\uA|1_{\uK}}:\fk\rightarrow\fg$ is an isomorphism, gives rise to a compatible action of the Weyl elements:
\begin{align}
\label{eq:nonabelianweyllift}
(W_{S}(tn)\cdot\tilde{T}_{\gamma^{S},\vec{\pi},\vec{m},\vec{n}})(\{k_{e}\}_{e\in E(\gamma^{S})}) & = \tilde{T}_{\gamma^{S},\vec{\pi},\vec{m},\vec{n}}(\{k_{e}\exp_{\uK}(\tfrac{1}{2}t\varepsilon(e,S)(d\xi_{\uA|1_{\uK}})^{-1}(\tilde{n}_{|p_{e(0)}}))\}_{e\in E(\gamma^{S})}) \\ \nonumber
& = T_{\gamma^{S},\vec{\pi},\vec{m},\vec{n}}(\{\xi_{\uA}(k_{e})\exp_{\uG}(\tfrac{1}{2}t\varepsilon(e,S)\tilde{n}_{|p_{e(0)}})\}_{e\in E(\gamma^{S})}).
\end{align}
This action respects the $\textup{Ad}$-equivariance of $\tilde{n}$, because $\textup{Ad}_{k}=(d\xi_{\uA|1_{\uK}})^{-1}\circ\textup{Ad}_{\xi_{A}(g)}\circ\ d\xi_{\uA|1_{\uK}},\ k\in\uK$. The field algebra $\mathfrak{F}_{\textup{LQG}}$ is defined as the algebra generated by the cylindrical functions on $\hom(\cP_{\Sigma},\uK)$ and the Weyl elements of $\mathfrak{A}_{\textup{LQG}}$ subject to the relations \eqref{eq:lqgweyl} and the compatible action \eqref{eq:nonabelianweyllift}.\\[0.1cm]
In view of remark \ref{rem:liftingclassicalcorrespondence}, the construction requires on the level of the principal $\uG$-bundle $\uP$ the existence of a non-trivial covering
\begin{align}
\label{eq:nonabeliancoveringhom}
\xymatrix@=1.5cm{
\ar@{}[dr] |{\circlearrowright} \uP_{\uK} \ar[r]^{\xi} \ar[d]_{\pi_{\uK}} & \uP \ar[d]^{\pi} \\
\Sigma \ar[r]^{\id_{\Sigma}} & \Sigma
} &\hspace{0.5cm} 
\xymatrix@=1.5cm{
\ar@{}[dr] |{\circlearrowright} \uP_{\uK} \ar[r]^{\xi} \ar[d]_{R_{k}} & \uP \ar[d]^{R_{\xi_{\uA}(k)}} \\
\uP_{\uK} \ar[r]^{\xi} & \uP
}
\end{align}
with a diagram of fibrations:
\begin{align}
\label{eq:nonabelianfibration}
\xymatrix@R=0.25cm{ 
& \uK \ar[dd] \ar[rr]^{\xi_{A}} & & \uG \ar[dd] \\
\uA \ar[ur] \ar[dr] & & & \\
& \uP_{\uK} \ar[rr]^{\xi} \ar[dr]_{\pi_{\uK}} & & \uP \ar[dl]^{\pi} \\
& & \Sigma & 
}
\end{align}
The fact that $\textup{Ad}:\uK\rightarrow\aut(\uK)$ descends to the central quotient $\uG\cong\uK/\uA$, since $\ker(\textup{Ad})=Z(\uK)$, implies the equivalence of the adjoint bundles of $\uP$ and $\uP_{\uK}$:
\begin{align}
\label{eq:adjointbundleequivalence}
\textup{Ad}(\uP) & \cong\textup{Ad}(\uP_{\uK}).
\end{align}
Thus, we can regard the field algebra $\mathfrak{F}_{\textup{LQG}}$ as the a Weyl algebra of $\uP_{\uK}$. The algebra $\mathfrak{A}_{\textup{LQG}}$ embeds into $\mathfrak{F}_{\textup{LQG}}$ via the lifting procedure, and its image is contained in fix-point algebra under the adjoint action of the generators of $\mathfrak{Z}_{\textup{LQG}}$, i.e. $W_{S}(tn),\ \frac{1}{2}t\tilde{n}=X\in\exp^{-1}_{\uG}(\{1_{\uG}\}),$ $S$ closed and compact:
\begin{align}
\label{eq:nonabelianfixpoint}
\alpha^{S}_{m}(T_{\gamma,\vec{\eta},\vec{i},\vec{j}}) & =W_{S}(tn)^{m}\tilde{T}_{\gamma,\vec{\eta},\vec{i},\vec{j}}W_{S}(-tn)^{m} = W_{S}(tn)^{m}\cdot\tilde{T}_{\gamma,\vec{\eta},\vec{i},\vec{j}}, \\[0.25cm] \nonumber
\alpha^{S}_{m}(W_{S'}(t'n')) & = W_{S}(tn)^{m}W_{S'}(t'n')W_{S}(-tn)^{m} = W_{S'}(t'n'),\ W_{S}(tn)\in\mathfrak{Z}_{\textup{LQG}},\ m\in\mathds{Z}.
\end{align}
In fact, the first line of \eqref{eq:nonabelianfixpoint} is trivial for those irreducible representations $\eta_{e},\ e\in E(\gamma),$ of $\uK$ that are trivial on $\uA$, which are precisely the irreducible representation of $\uG$. Moreover, the algebra $\mathfrak{F}_{\textup{LQG}}$ is not equal to the fix-point algebra, which can be seen by considering the action \eqref{eq:nonabelianfixpoint} on spin network functions $\tilde{T}_{e,\eta_{e},i_{e},j_{e}}$ defined on single edges $e\in\cP_{\Sigma}$. The actions of the gauge transformations $\cG^{\textup{sa}}_{\uP_{\uK}}$ and $\cG^{\textup{sa}}_{\uP}$ are compatible with the covering, as we will show next.
\begin{Lemma}
\label{lem:coveredgaugetrafos}
Given a bundle covering $\xi:\uP_{\uK}\rightarrow\uP$ as in \eqref{eq:nonabeliancoveringhom} \& \eqref{eq:nonabelianfibration}, every $\lambda_{\uK}\in\cG^{\textup{sa}}_{\uP_{\uK}}$ induces a $(\lambda_{\uK})_{\uG}\in\cG^{\textup{sa}}_{\uP}$ by
\begin{align}
\label{eq:coveredgaugetrafos}
(\lambda_{\uK})_{\uG}(p) & =\xi(\lambda_{\uK}(q))
\end{align}
for some $q\in\uP_{\uK}$, s.t. $\xi(q)=p$. The map $(\ .\ )_{\uG}:\cG^{\textup{sa}}_{\uP_{\uK}}\rightarrow\cG^{\textup{sa}}_{\uP}$ is a homomorphism.
\begin{Proof}
Clearly, \eqref{eq:coveredgaugetrafos} is well-defined: If $q'\in\uP_{\uK}$ is another element, s.t. $\xi(q')=p$, we know by \eqref{eq:nonabelianfibration} that \mbox{$q'=qa,\ a\in\uA$.} This implies:
\begin{align}
\label{eq:coveredgaugetrafoswelldefined}
\xi(\lambda_{\uK}(q')) & = \xi(\lambda_{\uK}(qa)) = \xi(\lambda_{\uK}(q)a) = \xi(\lambda_{\uK}(q))\xi_{\uA}(a) \\ \nonumber
 & = \xi(\lambda_{\uK}(q)).
\end{align}
Semi-analyticity follows from the semi-analyticity of the involved maps, and we have $(\lambda_{\uK})^{-1}_{\uG}=(\lambda^{-1}_{\uK})_{\uG},\ (\lambda_{\uK}\circ\lambda'_{\uK})_{\uG}=(\lambda_{\uK})_{\uG}\circ(\lambda'_{\uK})_{\uG}$. Now, we only need to verify that $(\lambda_{\uK})_{\uG}$ is a right equivariant bundle map covering the identity.
\begin{align}
\label{eq:coveredgaugetrafosbundlemap}
\pi((\lambda_{\uK})_{\uG}(p)) & = \pi(\xi(\lambda_{\uK}(q))) = \pi_{\uK}(\lambda_{\uK}(q)) = \pi_{\uK}(q) = \pi(\xi(q)) \\ \nonumber
 & = \pi(p),\ q\in\uP_{\uK}:\xi(q)=p \\[0.25cm]
(\lambda_{\uK})_{\uG}(pg) & = \xi(\lambda_{\uK}(qk)) = \xi(\lambda_{\uK}(q))\xi_{\uA}(k) \\ \nonumber
 & = (\lambda_{\uK})_{\uG}(p)g,\ q\in\uP_{\uK}:\xi(q)=p,\ k\in\uK:\xi_{A}(k)=g.
\end{align}
\end{Proof}
\end{Lemma}
The lemma tells us that the action of $\cG^{\textup{sa}}_{\uP_{\uK}}$ on $\mathfrak{A}_{\textup{LQG}}\subset\mathfrak{F}_{\textup{LQG}}$ descends to the action of induced gauge transformations in $\cG^{\textup{sa}}_{\uP}$. If we assume that $\uP_{\uK}$ and $\uP$ are path-connected, we may conclude that $(\ .\ )_{\uG}:\cG^{\textup{sa}}_{\uP_{\uK}}\rightarrow\cG^{\textup{sa}}_{\uP}$ is onto.
\begin{Proposition}[Lifting of gauge transformations]
\label{prop:gaugetrafoslift}
Assume that $\uP_{\uK}$ and $\uP$ are path-connected. Given $\lambda_{\uG}\in\cG^{\textup{sa}}_{\uP}$ and two points $q,q'\in\uP_{\uK}$, s.t. $\lambda_{\uG}(\xi(q))=\xi(q')$, there exist a unique lift $\tilde{\lambda}_{\uG}\in\cG^{\textup{sa}}_{\uP_{\uK}}$, s.t. $\lambda_{\uG}\circ\xi=\xi\circ\tilde{\lambda}_{G}$ and $\tilde{\lambda}_{\uG}(q)=q'$. The diagram of pointed spaces is
\begin{align}
\label{eq:basepointdiagram}
\xymatrix{
& (\uP_{\uK},q') \ar[d]^{\xi} \\
(\uP_{\uK},q) \ar@{.>}[ur]^{\tilde{\lambda}_{\uG}} \ar[r]_{\lambda_{\uG}\circ\xi} & (\uP,p)
}
\end{align}
\begin{Proof}
Form the assumptions, we deduce the existence of a lift $\tilde{\lambda}_{\uG}\in\textup{Diff}^{\textup{sa}}((\uP_{\uK},q),(\uP_{\uK},q'))$ by the lifting theorem for covering spaces \cite{BredonTopologyAndGeometry}, which applies, because $[\lambda_{\uG}\circ\xi](\pi_{1}(\uP_{\uK},q))=[\xi](\pi_{1}(\uP_{\uK},q'))$ since $\lambda_{\uG}$ is a diffeomorphisms. That $\tilde{\lambda}_{\uG}$ covers the identity, is evident from the definition of the lift:
\begin{align}
\label{eq:gaugetrafoslift}
\pi_{\uK}\circ\tilde{\lambda}_{\uG} & = \pi\circ\xi\circ\tilde{\lambda}_{\uG} = \pi\circ\lambda_{\uG}\circ\xi = \pi\circ\xi \\ \nonumber
 & = \pi_{\uK}.
\end{align}
Finally, we need to check that $\tilde{\lambda}_{\uG}$ is right equivariant. We know that $\forall\ k\in\uK:\ R_{k}\circ\tilde{\lambda}_{\uG}$ and $\tilde{\lambda}_{\uG}\circ R_{k}$ are lifts of $\lambda_{\uG}\circ R_{\xi_{\uA}(k)}$ by the equivariance of $\lambda_{\uG}$. Furthermore, we find $R_{k}(\tilde{\lambda}_{\uG}(q))=q'k$ and $\tilde{\lambda}_{\uG}(R_{k}(q))=q'ka(q)$ for some continuous $a:\uP_{\uK}\rightarrow\uA$. But, $\uA$ is discrete by assumption, which implies $\forall\ q\in\uP_{\uK}:\ a(q)=a_{0}\in\uA$. Thus, we have $\forall k\in\uK:\ \tilde{\lambda}_{\uG}\circ R_{k}=R_{a_{0}}\circ R_{k}\circ\tilde{\lambda}_{\uG}$, which leads to $a_{0}=1_{\uK}$ for $k=1_{\uK}$ and the free action of $\uK$ on $\uP_{\uK}$.
\end{Proof}
\end{Proposition}
A similar reasoning applies to the actions of the automorphisms $\aut^{\textup{sa}}(\uP_{\uK})$ and $\aut^{\textup{sa}}(\uP)$.\\[0.25cm]
Candidates for (charged) automorphisms of $\mathfrak{A}_{\textup{LQG}}$ can be defined with the help of unitary cylindrical functions $f$ on $\hom(\cP_{\Sigma},\uK)$ that do not descend through $\uA$, i.e. $f=p_{l}^{*}f_{l}\in\Cyl_{K}$ s.t. $f_{l}\bar{f}_{l}=1$ and $f_{l}$ does not define a function on $\uG^{|E(\gamma_{l})|}$:
\begin{align}
\label{eq:nonabelianchargeauto}
\rho_{f}(T_{\gamma,\vec{\pi},\vec{m},\vec{n}}) & = f\ T_{\gamma,\vec{\pi},\vec{m},\vec{n}}\ f^{*} = T_{\gamma,\vec{\pi},\vec{m},\vec{n}}, \\[0.25cm] \nonumber
\rho_{f}(W_{S}(tn)) & = f\ W_{S}(tn)\ f^{*}= f(W_{S}(tn)\cdot f^{*})W_{S}(tn).
\end{align}
Two examples of such functions are:
\begin{align}
\label{eq:nonabelianunitaryfunctions}
f^{\Re}_{\gamma,\vec{\eta},\vec{i},\vec{j}} & = e^{i\Re(T_{\gamma,\vec{\eta},\vec{i},\vec{j}})}, & f^{\Im}_{\gamma,\vec{\eta},\vec{i},\vec{j}}& = e^{i\Im(T_{\gamma,\vec{\eta},\vec{i},\vec{j}})},
\end{align} 
for irreducible representations $\eta_{e},\ e\in E(\gamma),$ of $\uK$ that do not reduce to $\uG$. To arrive at a true automorphism of $\mathfrak{A}_{\textup{LQG}}$, we need to ensure that $\forall\ W_{S}(tn):\ f(W_{S}(tn)\cdot f^{*})$ gives a cylindrical function on $\hom(\cP_{\Sigma},\uG)$. We call functions $f\in\Cyl_{K}$ satisfying these requirements $(\uK,\uA)$-admissible, and denote them by $U(\Cyl_{\uK})_{\uA}$. Examples of $(K,A)$-admissible functions could be generated from 1-dimensional, unitary representations $\chi:\uK\rightarrow\mathds{T}$, s.t. $\chi_{|\uA}\neq1$. But, unfortunately compact, connected, semi-simple Lie groups are (topologically) perfect, and thus do not posses non-trivial 1-dimensional, unitary representations \cite{GotoATheoremOn}. Therefore it seems possible that there are no $(K,A)$-admissible functions, i.e. $U(\Cyl_{\uK})_{\uA}=\emptyset$. Nevertheless, we observe that $U(\Cyl_{\uK})_{\uA}$ is preserved by gauge transformations and automorphisms (see lemma \ref{lem:coveredgaugetrafos}).\\[0.25cm]
Thus, we conclude the section with the observation that for structure groups $\uG$ admitting non-trivial coverings $\xi:\uK\rightarrow\uG$, together with a bundle covering \eqref{eq:nonabeliancoveringhom} \& \eqref{eq:nonabelianfibration}, we can construct an embedding of algebras $\mathfrak{A}_{\textup{LQG}}\subset\mathfrak{F}_{\textup{LQG}}$, which allows to construct candidates for (charged) automorphisms $\rho_{f},\ f\in\Cyl_{K}$ as in \eqref{eq:nonabelianchargeauto}. If we find among the latter a true automorphism of $\mathfrak{A}_{\textup{LQG}}$ that acts non-trivially on the center $\mathfrak{Z}_{\textup{LQG}}$, we will obtain a new irreducible representations of $\mathfrak{A}_{\textup{LQG}}$ from the state (cp. \eqref{eq:AILstate}):
\begin{align}
\label{eq:nonabelianchargedstate}
\omega_{f}=\omega_{0}\circ\rho_{f}.
\end{align}
Let us also shortly comment on the issue of gauge and automorphism invariance of the state $\omega_{f}$. From the invariance of $\omega_{0}$, we find:
 \begin{align}
 \label{eq:nonabelianchargedstateinv}
 \omega_{f}\circ\alpha_{\lambda} & = \omega_{\alpha_{\lambda^{-1}}(f)}, & \omega_{f}\circ\alpha_{\phi} & = \omega_{\alpha_{\phi^{-1}}(f)} 
 \end{align}
 for $\lambda\in\cG^{\textup{sa}}_{P_{\uK}},\phi\in\textup{Diff}^{\textup{sa}}(\Sigma)$. Thus, gauge invariance could be achieved by the additional requirement $\alpha_{\lambda}(f)=f\ \forall\lambda\in\cG^{\textup{sa}}_{P_{\uK}}$, although it is not obvious that this condition can be satisfied non-trivially in combination with the additional constraints on $f$. Requiring automorphism invariance poses a much more severe constraint, because the analogous requirement $\alpha_{\chi}(f)=f\ \forall\chi\in\aut^{\textup{sa}}(\uP)$ can probably not be satisfied non-trivially \cite{MouraoPhysicalPropertiesOf}, i.e. it leads to $f\equiv1$.\\[0.1cm]
In general, we could follow the same strategy as in \cite{KoslowskiLoopQuantumGravity} to obtain a unitary implementation of the gauge transformations and automorphisms in a representation constructed from states of the form \eqref{eq:nonabelianchargedstate}. That is, we make use of the Ashtekar-Isham-Lewandowski representation $(\fH_{\omega_{0}},\pi_{\omega_{0}},\Omega_{\omega_{0}})$ of the field algebra $\mathfrak{F}_{\textup{LQG}}$, w.r.t. which the (charged) automorphisms $\rho_{f},\ f\in U(\Cyl_{\uK})_{\uA},$ of $\mathfrak{A}_{\textup{LQG}}$ are unitarily implemented, because they are inner automorphisms of $\mathfrak{F}_{\textup{LQG}}$. Therefore, we find $\mathfrak{A}_{\textup{LQG}}$-invariant subspaces
\begin{align}
\label{eq:nonabelianinvariantsubspaces}
\fH_{f} & = \pi_{\omega_{0}}(f^{*})(\overline{\pi_{\omega_{0}}(\mathfrak{A}_{\textup{LQG}})\Omega_{\omega_{0}}}), & \pi_{f} & = \pi_{\omega_{0}|\fH_{f}}.
\end{align}
The implementers of the gauge transformations and automorphisms $U(\lambda)$, $\lambda\in\cG^{\textup{sa}}_{P_{\uK}},$ and $U(\chi)$, $\chi\in\aut^{\textup{sa}}(\uP_{\uK})$ map these subspaces into each other according to \eqref{eq:nonabelianchargedstateinv}:
\begin{align}
\label{eq:nonabelianchargedautoaction}
U(\lambda)\fH_{f} & = U(\lambda)\pi_{\omega_{0}}(f^{*})(\overline{\pi_{\omega_{0}}(\mathfrak{A}_{\textup{LQG}})\Omega_{\omega_{0}}}) \\ \nonumber
& = \pi_{\omega_{0}}(\alpha_{\lambda}(f^{*}))(\overline{\pi_{\omega_{0}}(\alpha_{\lambda}(\mathfrak{A}_{\textup{LQG}}))\Omega_{\omega_{0}}}) \\ \nonumber 
& = \pi_{\omega_{0}}(\alpha_{\lambda}(f)^{*})(\overline{\pi_{\omega_{0}}(\mathfrak{A}_{\textup{LQG}})\Omega_{\omega_{0}}}) \\ \nonumber
& = \fH_{\alpha_{\lambda}(f)} \\[0.25cm] \nonumber
U(\chi)\fH_{f} & = U(\chi)\pi_{\omega_{0}}(f^{*})(\overline{\pi_{\omega_{0}}(\mathfrak{A}_{\textup{LQG}})\Omega_{\omega_{0}}}) \\ \nonumber
& = \pi_{\omega_{0}}(\alpha_{\chi}(f^{*}))(\overline{\pi_{\omega_{0}}(\alpha_{\chi}(\mathfrak{A}_{\textup{LQG}}))\Omega_{\omega_{0}}}) \\ \nonumber 
& = \pi_{\omega_{0}}(\alpha_{\chi}(f)^{*})(\overline{\pi_{\omega_{0}}(\mathfrak{A}_{\textup{LQG}})\Omega_{\omega_{0}}}) \\ \nonumber
& = \fH_{\alpha_{\chi}(f)}.
\end{align}
If we denote by $[f]$ the equivalence class of $f\in U(\Cyl_{\uK})_{\uA}$ under the actions of $\cG^{\textup{sa}}_{\uP_{\uK}}$ and $\aut^{\textup{sa}}(\uP_{\uK})$, we can form the direct sum
\begin{align}
\label{eq:nonabelianlargedirectsum}
\fH_{[f]} = \bigoplus_{f'\in[f]}\fH_{f'},&\ \pi_{[f]} = \bigoplus_{f'\in[f]}\pi_{f'}.
\end{align}
This gives us a (reducible) representation of $\mathfrak{A}_{\textup{LQG}}$ with a unitary implementation of $\cG^{\textup{sa}}_{\uP_{\uK}}$ and $\aut^{\textup{sa}}(\uP_{\uK})$. On it, we can apply the usual group averaging procedure to obtain gauge or automorphism invariant spaces $(\fH^{\cG}),\ (\fH)^{\aut}$ (cf. \cite{AshtekarQuantumTheoryOf1, ThiemannModernCanonicalQuantum}).
\section{The Koslowski-Sahlmann representations}
\label{sec:KSreps}
Now, we turn to the discussion of the Koslowski-Sahlmann representations, mainly for non-Abelian structure group $\uG$, \cite{KoslowskiLoopQuantumGravity} in view of the results of the previous section. The discussion will be split into two parts related to a similar division in \cite{KoslowskiLoopQuantumGravity}:
\begin{itemize}
	\item[1.] ``Central extensions'' of holonomy-flux algebras an non-degenerate backgrounds,
	\item[2.] Weyl forms of the holonomy-flux algebra and non-degenerate backgrounds.
\end{itemize}
\subsection{``Central extensions'' of holonomy-flux algebras and non-degenerate backgrounds}
\label{sec:HFKSreps}
The holonomy-flux algebras considered in \cite{KoslowskiLoopQuantumGravity} are essentially of the form, we defined in subsection \ref{sec:prelqg} (see \ref{def:hfa}). That is, the algebras are generated by elements $f\in\Cyl^{\infty}$, $Y_{n}(S)$, $S$ a face, $n\in\Gamma^{\textup{sa}}_{0}(\textup{Ad}(\uP_{|S}))$ together with the commutation relation
\begin{align}
\label{eq:holfluxcommutator}
[Y_{n}(S),f] & = E_{n}(S)\cdot f, &  [f,f'] & = 0
\end{align}
and the reality and linearity conditions
\begin{align}
\label{eq:holfluxreality}
f^{*} & =\bar{f}, &  Y_{n}(S)^{*} & =-Y_{n}(S),
\end{align}
\begin{align}
\label{eq:holfluxlin}
Y_{n+n'}(S) & =Y_{n}(S)+Y_{n'}(S).
\end{align}
But, in contrast to definition \ref{def:hfa}, the higher commutation relations for the elements $Y_{n}(S)$ are not specified, but only required to satisfy the Jacobi identity, i.e.
\begin{align}
\label{eq:Jacobiidentity}
[[Y_{n}(S),Y_{n'}(S')],f] & = [Y_{n}(S),[Y_{n'}(S'),f]]-[Y_{n'}(S'),[Y_{n}(S),f]] \\ \nonumber
 & = [E_{n}(S),E_{n'}(S)]_{\mathfrak{X}(\overline{\mathcal{A}})}\cdot f,
\end{align}
and similar higher order relations. While, in the case of an Abelian structure group, e.g. $\uG=U(1)$, this poses no specific constraints\footnote{Although, we are allowed to consider modifications, e.g. a central extension $[Y_{n}(S),Y_{n'}(S')]=c_{[n,n']}(S,S')$.} on the algebraic relations for the elements $Y_{n}(S)$ to make the Koslowski-Sahlmann representations well-defined, this is not the case for a non-Abelian structure group, e.g. $\uG=SU(2)$. In the latter case, we find (cp. \eqref{eq:higherfluxcom}):
\begin{align}
\label{eq:higherJacobiidentities}
[[Y_{n}(S),[Y_{n'}(S),Y_{n''}(S)]],f] & = [E_{n}(S),[E_{n'}(S),E_{n''}(S)]_{\mathfrak{X}(\overline{\mathcal{A}})}]_{\mathfrak{X}(\overline{\mathcal{A}})}\cdot f \\ \nonumber
 & = \tfrac{1}{4}E_{[n,[n',n'']]}(S)\cdot f \\ \nonumber
 & = \tfrac{1}{4}[Y_{[n,[n',n'']]}(S),f].
\end{align}
Thus, we are forced to require the additional relation
\begin{align}
\label{eq:quasicentralextension}
[Y_{n}(S),[Y_{n'}(S),Y_{n''}(S)]] & = \tfrac{1}{4}Y_{[n,[n',n'']]}(S) + c_{[n,[n',n'']]}(S),
\end{align}
where $c_{[n,[n',n'']]}(S)$ is an element of the algebra that commutes with the subalgebra $\Cyl^{\infty}$. If we, additionally, assume that $c_{[n,[n',n'']]}(S)$ commutes with the generators $Y_{m}(S)$, the Jacobi identity will give us a ``co-cycle condition'':
\begin{align}
\label{eq:fluxcocycle}
c_{[n,[n',n'']]}(S) + c_{[n',[n'',n]]}(S) + c_{[n'',[n,n']]}(S) = 0.
\end{align}
Clearly, $c_{[n,[n',n'']]}(S)$ needs to satisfy linearity conditions related to \eqref{eq:holfluxlin} as well.\\
This said, we return to the Koslowski-Sahlmann representations, which are proposed to be defined by an \mbox{$E^{(0)}\in\Gamma(T\Sigma\otimes\textup{Ad}^{*}(\uP)\otimes|\Lambda|^{1}(\Sigma))$:}
\begin{align}
\label{eq:KSreps}
\pi_{E^{(0)}}(Y_{n}(S)) & = \pi_{\omega_{0}}(E_{n}(S))+i\int_{S}*E^{(0)}(n)\cdot\mathds{1}_{\fH_{\omega_{0}}}, & \pi_{E^{(0)}}(f) & = \pi_{\omega_{0}}(f),
\end{align}
w.r.t. to the Ashtekar-Isham-Lewandowski representation $(\pi_{\omega_{0}},\fH_{\omega_{0}},\Omega_{\omega_{0}})$. A similar construction applies in the temporal gauge to algebraic formulation of quantum electrodynamics \cite{LoeffelholzMathematicalStructureOf}. The interpretation of these representation is obtained from the consideration of the limit
\begin{align}
\label{eq:boundaryvalues}
\lim_{R\rightarrow\infty}(\omega_{0}\circ\rho_{E^{(0)}})(W_{S_{R}}(n)) = e^{i\theta(n)}
\end{align}
in the Abelian case (cp. \eqref{eq:abelianlqgchargeauto}) for $\Sigma=\R^{3}$, where we chose $S_{R}=S^{2}_{R}$ (2-sphere), $n=n(\vartheta,\varphi)$, $(\tilde{n},\tilde{n})_{\fg}=1$, \mbox{$E^{(0)}(r,\vartheta,\varphi)\sim\theta(\vartheta,\varphi)r^{-2}$.} Thus, the choice of $E^{(0)}$ affects the asymptotic flux configuration (cp. \cite{BuchholzThePhysicalState}).\\
We will analyze the Koslowski-Sahlmann representations of the holonomy-flux algebras, in the above sense, from two different, though related, points of view. First, we will argue that the Koslowski-Sahlmann representations require a modification of the commutation relations by a non-trivial ``central term'', if we want the $Y_{n}(S)$ to correspond to the fluxes $E_{n}(S)$. Second, we will show, that we can interpret the $Y_{n}(S)$ as shifted fluxes $E_{n}(S) + i\int_{S}*E^{(0)}(n)$, which leads to the conclusion that the Koslowski-Sahlmann representations are the Ashtekar-Isham-Lewandowski representations w.r.t. the shifted fluxes. The two points are related by the observation that the shift transformation
\begin{align}
\label{eq:affineshift}
\rho_{E^{(0)}}:E_{n}(S)\mapsto E_{n}(S) + i\int_{S}*E^{(0)}(n)
\end{align}
is not a *-automorphism of $\mathfrak{P}_{\textup{LQG}}$ but only an affine transformation. Thus, in contrast to section \ref{sec:centers} the charge transformations $\rho_{E^{(0)}}$ are already broken on the level of the algebra $\mathfrak{P}_{\textup{LQG}}$, and not on the level of a state or representation.\\[0.2cm]
As we discussed in subsection \ref{sec:abelianlqgweyl}, the Koslowski-Sahlmann representations can be understood in terms of charged automorphisms of the Weyl algebra $\mathfrak{A}_{\textup{LQG}}$ ($c\equiv0$) in the Abelian case (cp. \eqref{eq:abelianlqgchargeautoext} \& \eqref{eq:abelianlqgchargedfluxes}). In the non-Abelian setting, the question, whether \eqref{eq:KSreps} defines representations of a holonomy-flux algebra is more subtle, because of \eqref{eq:quasicentralextension}:
\begin{align}
\label{eq:KSrepscondition}
\pi_{E^{(0)}}([Y_{n}(S),[Y_{n'}(S),Y_{n''}(S)]]) & = [\pi_{E^{(0)}}(Y_{n}(S),)[\pi_{E^{(0)}}(Y_{n'}(S)),\pi_{E^{(0)}}(Y_{n''}(S))]] \\ \nonumber
& = \pi_{\omega_{0}}([E_{n}(S),[E_{n'}(S),E_{n''}(S)]_{\mathfrak{X}(\overline{\mathcal{A}})}]_{\mathfrak{X}(\overline{\mathcal{A}})}) \\ \nonumber
& = \tfrac{1}{4}\pi_{\omega_{0}}(E_{[n,[n',n'']]}(S)), \\[0.25cm]
\pi_{E^{(0)}}([Y_{n}(S),[Y_{n'}(S),Y_{n''}(S)]]) & = \pi_{E^{(0)}}(Y_{[n,[n',n'']]}(S)) + \pi_{E^{(0)}}(c_{[n,[n',n'']]}(S)) \\ \nonumber
& = \tfrac{1}{4}\pi_{\omega_{0}}(E_{[n,[n',n'']]}(S))+\tfrac{i}{4}\int_{S}*E^{(0)}([n,[n',n'']])\cdot\mathds{1}_{\fH_{\omega_{0}}} \\ \nonumber
 &\hspace{0.25cm} + \pi_{E^{(0)}}(c_{[n,[n',n'']]}(S)).
\end{align}
Therefore, we find, that the Koslowski-Sahlmann representations require the presence of a non-trivial ``central term'' in the higher commutation relations (cp. corollary \ref{cor:nocentralterm}):
\begin{align}
\label{eq:centraltermrep}
\pi_{E^{(0)}}(c_{[n,[n',n'']]}(S)) & = -\tfrac{i}{4}\int_{S}*E^{(0)}([n,[n',n'']])\cdot\mathds{1}_{\fH_{\omega_{0}}}.
\end{align}
This relation could be easily satisfied by
\begin{align}
\label{eq:centralterm}
c^{E^{(0)}}_{[n,[n',n'']]}(S) & = -\tfrac{i}{4}\int_{S}*E^{(0)}([n,[n',n'']])\cdot1,\ \ \pi_{E^{(0)}}(c_{[n,[n',n'']]}(S))=\pi_{\omega_{0}}(c^{E^{(0)}}_{[n,[n',n'']]}(S)),
\end{align}
which satisfies the ``co-cycle condition'' due to the linearity of the integral and the Jacobi identity of $[\ .\ ,\ .\ ]:\fg\times\fg\rightarrow\fg$. But, we would still have to check that there is compatible definition for $[Y_{n}(S),Y_{n'}(S)]$, and that there are no other higher order relations in conflict with it. Moreover, we have to extend the actions of the gauge transformation $\cG^{\textup{sa}}_{\uP}$ and automorphisms $\aut^{\textup{sa}}(\uP)$ to account for the ``central term''.
\begin{align}
\label{eq:gaugeanddiffeoextension}
\alpha_{\lambda}(c_{[n,[n',n'']]}(S)) & = c_{\lambda\triangleright[n,[n',n'']]}(S) \\[0.25cm] \nonumber
\alpha_{\chi}(c_{[n,[n',n'']]}(S)) & = c_{\chi^{*}[n,[n',n'']]}(\phi^{-1}_{\chi}(S)),\ \lambda\in\cG^{\textup{sa}}_{\uP},\ \chi\in\aut^{\textup{sa}}(\uP),
\end{align}
where we assumed that the actions are natural w.r.t. to the generators $Y_{n}(S)$, i.e. identical to those on the flux vector fields $E_{n}(S)$ (see definition \ref{def:gaugetrafo}). Unfortunately, this leads to the conclusion that \eqref{eq:centralterm} and \eqref{eq:gaugeanddiffeoextension} are incompatible, i.e. the ``central term'' cannot be proportional to the unit element. Moreover, the unitary implementers of the gauge transformations and automorphisms in the Ashtekar-Isham-Lewandowski representation are not compatible with the choice \eqref{eq:centraltermrep}, and an extension like \eqref{eq:gaugeanddiffeoextension} for generic $E^{(0)}$:
\begin{align}
\label{eq:AILunitaryincompatible}
\pi_{E^{(0)}}(\alpha_{\lambda}(c_{[n,[n',n'']]}(S))) & = -\tfrac{i}{4}\int_{S}*E^{(0)}(\lambda\triangleright[n,[n',n'']])\cdot\mathds{1}_{\fH_{\omega_{0}}} \\ \nonumber
& = -\tfrac{i}{4}\int_{S}*(\lambda\triangleright E^{(0)})([n,[n',n'']])\cdot\mathds{1}_{\fH_{\omega_{0}}} \\ \nonumber
& \neq U_{\omega_{0}}(\lambda)\left(-\tfrac{i}{4}\int_{S}*E^{(0)}([n,[n',n'']])\cdot\mathds{1}_{\fH_{\omega_{0}}}\right)U_{\omega_{0}}(\lambda)^{*} \\ \nonumber
& = U_{\omega_{0}}(\lambda)\pi_{E^{(0)}}(c_{[n,[n',n'']]}(S))U_{\omega_{0}}(\lambda)^{*}. \\[0.25cm]
\pi_{E^{(0)}}(\alpha_{\chi}(c_{[n,[n',n'']]}(S))) & = -\tfrac{i}{4}\int_{\phi^{-1}_{\chi}(S)}\!\!\!\!*E^{(0)}(\chi^{*}[n,[n',n'']])\cdot\mathds{1}_{\fH_{\omega_{0}}} \\ \nonumber
& = -\tfrac{i}{4}\int_{S}*(\chi_{*}E^{(0)})([n,[n',n'']])\cdot\mathds{1}_{\fH_{\omega_{0}}} \\ \nonumber
& \neq U_{\omega_{0}}(\chi)\left(-\tfrac{i}{4}\int_{S}*E^{(0)}([n,[n',n'']])\cdot\mathds{1}_{\fH_{\omega_{0}}}\right)U_{\omega_{0}}(\chi)^{*}\\ \nonumber
& = U_{\omega_{0}}(\chi)\pi_{E^{(0)}}(c_{[n,[n',n'']]}(S))U_{\omega_{0}}(\chi)^{*}.
\end{align}
The latter issue can be fixed in the same way a proposed in \cite{KoslowskiLoopQuantumGravity}, i.e. the unitary implementers intertwine between representations with different background.
\begin{align}
\label{eq:AILunitaryintertwiners}
\pi_{E^{(0)}}(\alpha_{\lambda}(Y_{n}(S))) & = U_{\omega_{0}}(\lambda)\pi_{\lambda_{*}E^{(0)}}(Y_{n}(S))U_{\omega_{0}}(\lambda)^{*}, \\[0.1cm] \nonumber
\pi_{E^{(0)}}(\alpha_{\lambda}(c_{[n,[n',n'']]}(S))) & = U_{\omega_{0}}(\lambda)\pi_{\lambda_{*}E^{(0)}}(c_{[n,[n',n'']]}(S))U_{\omega_{0}}(\lambda)^{*}, \\[0.25cm]
\pi_{E^{(0)}}(\alpha_{\chi}(Y_{n}(S))) & = U_{\omega_{0}}(\chi)\pi_{\chi_{*}E^{(0)}}(Y_{n}(S))U_{\omega_{0}}(\chi)^{*}, \\[0.1cm] \nonumber
\pi_{E^{(0)}}(\alpha_{\chi}(c_{[n,[n',n'']]}(S))) & = U_{\omega_{0}}(\chi)\pi_{\chi_{*}E^{(0)}}(c_{[n,[n',n'']]}(S))U_{\omega_{0}}(\chi)^{*}.
\end{align}
The second way of thinking about the Koslowski-Sahlmann representations, which is more along the lines of section \ref{sec:centers}, is offered by the following observation:\\[0.1cm]
There is a shift transformation of the holonomy-flux algebra $\mathfrak{P}_{\textup{LQG}}$ defined by:  
\begin{align}
\label{eq:affineiso}
\rho_{E^{(0)}}(E_{n}(S)) & = E_{n}(S)+i\int_{S}E^{(0)}(n)\cdot 1, & \rho_{E^{(0)}}(f) & = f.
\end{align}
It resembles the (classical) moment map problem, i.e. the association of a phase space function with a Hamiltonian vector field is only unique up to constant terms. By the same argument as before, this transformation is not a *-automorphism of the holonomy-flux algebra $\mathfrak{P}_{\textup{LQG}}$, but only an affine transformation:
\begin{align}
\label{eq:noalgebraisomorphism}
\rho_{E^{(0)}}([E_{n}(S),[E_{n'}(S),E_{n''}(S)]]) & \neq [\rho_{E^{(0)}}(E_{n}(S)),[\rho_{E^{(0)}}(E_{n'}(S)),\rho_{E^{(0)}}(E_{n''}(S))]],
\end{align}
and we will be forced to introduce ``central'' elements, if we want it to be a *-isomorphism. Thus, the shift transformations $\rho_{E^{(0)}}$ can be considered as charge transformations that are already broken on the level of the algebra $\mathfrak{P}_{\textup{LQG}}$. The Koslowski-Sahlmann representations arise by the identification $Y_{n}(S)=\rho_{E^{(0)}}(E_{n}(S))$ and the use of the Ashtekar-Isham-Lewandowski representation $(\pi_{\omega_{0}},\fH_{\omega_{0}},\Omega_{\omega_{0}})$. In terms of an algebraic state $\omega$, we have:
\begin{align}
\label{eq:shiftstate}
\omega(fY_{n_{1}}(S_{1})...Y_{n_{j}}(S_{j})) & = \left\{\begin{matrix*}[l] \mu_{0}(f)\left(i\int_{S_{1}}E^{(0)}(n_{1})\right)...\left(i\int_{S_{j}}E^{(0)}(n_{j})\right) & \textup{if}\ \{1,..,j\}=\emptyset \\[0.1cm] 0 & \textup{else} \end{matrix*}\right. ,
\end{align}
for every $f\in\Cyl^{\infty},\ Y_{n_{1}}(S_{1})...Y_{n_{j}}(S_{j})\in\mathfrak{X}_{\Flux}$.\\[0.25cm]
Let us summarize our findings in this subsection:
\begin{itemize}
	\item[1.] The Koslowski-Sahlmann representations require, at least, the modification of the commutation relations of the standard holonomy-flux algebra (see definition \ref{def:hfa}) by a ``central term'' \eqref{eq:quasicentralextension} to be well-defined (cp. corollary \ref{cor:nocentralterm}) w.r.t. the identification $Y_{n}(S)=E_{n}(S)$. But, it is so far unclear, whether the addition of a ``central term'' suffices to satisfy all relation imposed by higher order commutators.
	\item[2.] If the extension exists, and the ``central term'' commutes with the generators $Y_{n}(S)$, it has to satisfy the ``co-cycle condition'' \eqref{eq:fluxcocycle}. The actions of the gauge transformations and automorphisms have to be modified to account for the presence of the ``central term''. If the gauge transformations and automorphisms are supposed to act naturally on the generators $Y_{n}(S)$, i.e. the actions are identical to those on the flux vector fields $E_{n}(S)$, the standard unitary implementers of both groups of transformations in the Koslowski-Sahlmann representations do not implement the modified actions, but intertwine between different backgrounds.
	\item[3.] The Koslowski-Sahlmann representations can be interpreted as the Ashtekar-Isham-Lewandowski representation after shifting the generators of $\mathfrak{P}_{\textup{LQG}}$ by $\rho_{E^{(0)}}$. But, the shift transformation is not a *-automorphism, and thus broken on the level of the algebra.
	\item[4.] An important difference between the first and the second point of view is the relation to gauge and automorphism invariance, because the second perspective does not allow to treat different choices of $E^{(0)}$ as representations of the same generators, i.e. $\rho_{E^{(0)}}(E_{n}(S))\neq\rho_{E'^{(0)}}(E_{n}(S))$ for generic $E^{(0)}\neq E'^{(0)}$. Although, the various generators are realized in the Ashtekar-Isham-Lewandowski representation, there is only the standard vacuum $\Omega_{\omega_{0}}$. Thus, a treatment along the lines of \cite{KoslowskiLoopQuantumGravity} requires the first attitude towards the Koslowski-Sahlmann representations. 
\end{itemize}
\subsection{Weyl form of the holonomy-flux algebras and non-degenerate backgrounds}
\label{sec:weylKSreps}
Regarding the Weyl form of the holonomy-flux algebra in relation to non-degenerate backgrounds, we will only comment on the version defined by Fleischhack in \cite{FleischhackRepresentationsOfThe}. Koslowski and Sahlmann also consider a different version generated by ``exponentials of area operators'', which we will not discuss in this article (cf. \cite{KoslowskiLoopQuantumGravity}).\\[0.1cm]
The $C^{*}$-Weyl algebra defined by Fleischhack is similar to the concrete realization of the algebra $\mathfrak{A}_{\textup{LQG}}$ for $\uG=SU(2)$ via the Ashtekar-Isham-Lewandowski representation, i.e.
\begin{align}
\label{eq:concreteweylalgebra}
\overline{\pi_{\omega_{0}}(\mathfrak{A}_{\textup{LQG}})}^{||\ .\ ||_{\mathfrak{B}(\fH_{\omega_{0}})}}\subset\mathfrak{B}(\fH_{\omega_{0}}).
\end{align}
Especially, both algebras contain the perfect subgroup of Weyl elements $\pi_{\omega_{0}}(\langle\mathcal{W}_{0}\rangle)$ (see proposition \ref{prop:compactweylperfectness}). Thus, by corollary \ref{cor:nocharacterautomorphisms}, there is a severe constraint on the definition of new representations via character automorphisms \eqref{eq:characterauto} as suggested in \cite{KoslowskiLoopQuantumGravity}. This observation is in accordance with the result of the previous subsection that the Koslowski-Sahlmann representations for holonomy-flux algebras cannot be defined for the algebra $\mathfrak{P}_{\textup{LQG}}$. Therefore, it appears to be necessary to look for Weyl forms of the (possibly) modified holonomy-flux algebras proposed above.
On the other hand, there is the second possibility, in analogy with the preceding discussion, to consider the representation of Weyl form $\mathfrak{A}_{\textup{LQG}}$ defined by the state
\begin{align}
\label{eq:shiftedweylstate}
\omega_{0}(fV_{S_{1}}(n_{1})...V_{S_{j}}(n_{j})) & =\mu_{0}(f)\ e^{i\int_{S_{1}}*E^{(0)}(n_{1})}\cdot...\cdot e^{i\int_{S_{j}}*E^{(0)}(n_{j})},\ \forall f\in\Cyl,\ V_{S_{1}}(n_{1})...V_{S_{j}}(n_{j})\in\mathfrak{A}_{\textup{LQG}}
\end{align}
w.r.t. the shifted generators $V_{n}(S)=e^{i\int_{S}*E^{(0)}(n)}W_{n}(S)$. The GNS representation realizes the Koslowski-Sahlmann representation with $E^{(0)}$ for $\mathfrak{A}_{\textup{LQG}}$:
\begin{align}
\label{eq:weylKSrep}
\pi_{\omega}(V_{n}(S)) & = e^{i\int_{S}*E^{(0)}(n)}\pi_{\omega_{0}}(W_{n}(S)), & \pi_{\omega}(f) & = \pi_{\omega_{0}}(f),
\end{align}
but again the shift transformation $W_{n}(S)\mapsto V_{n}(S)$ is not a *-automorphism of $\mathfrak{A}_{\textup{LQG}}$.
\section{Chiral symmetry breaking and $\theta$-vacua in loop quantum gravity}
\label{sec:theta}
In the last section of this article, we would like to present another application of the relation between central operators and representation theory (see section \ref{sec:centers}), and outline a setup for the discussion of chiral symmetry breaking and occurrence of $\theta$-vacua in the framework of loop quantum gravity. This setup is inspired by and strongly resembles a discussion of these topics in the setting of algebraic quantum field theory, which was given by Morchio and Strocchi in \cite{MorchioChiralSymmetryBreaking} (see also \cite{JackiwTopologicalInvestigationsOf} for the original account on the ideas involving the topology of the gauge group without the use of semi-classical approximations).\\[0.1cm]
Let us briefly, recall the problem of chiral symmetry breaking and the $\theta$-vacuum structure in quantum field theory. If we consider a field theory on Minkowski space $\mathds{M}$ in the temporal gauge given in terms of gauge field variables $(A,E)$ chirally coupled to fermion field variables $(\Psi,\bar{\Psi})$ (notably the standard model), we will have a chiral symmetry associated with the transformation
\begin{align}
\label{eq:chiralsym}
\rho_{\zeta}(\Psi)= e^{\zeta\gamma_{5}}\Psi, &\ \rho_{\zeta}(A)=A, & \rho_{\zeta}(\bar{\Psi})=\bar{\Psi}e^{\zeta\gamma_{5}}, &\ \rho_{\zeta}(E)=E,
\end{align}
where $\gamma_{5}^{*}=-\gamma_{5}$. If this symmetry were preserved in the quantization of the field theory, we would expect the presence of associated parity doublets. In the case of the standard model, such parity doublets are missing, and the chiral symmetry is said to be broken. Since the standard model is also missing Goldstone bosons related to breaking of the chiral symmetry, we arrive at the so-called axial $U(1)$-problem \cite{WeinbergTheU(1)Problem}, the solution of which is argued to be the chiral anomaly and its relation to the large gauge transformations \cite{tHooftSymmetryBreakingThrough, JackiwTopologicalInvestigationsOf} in standard treatments. The arguments goes, loosely speaking, as follows \cite{JackiwTopologicalInvestigationsOf}:\\[0.1cm]
The regularized expression for the symmetry generating axial current $j^{5}_{\mu}=i\bar{\Psi}\gamma^{5}\gamma_{\mu}\Psi$ acquires the famous gauge dependent axial anomaly, which is crucial for the theoretical explanation of the $\pi_{0}\rightarrow\gamma\gamma$ decay:
\begin{align}
\label{eq:axialanomaly}
\partial^{\mu}j^{5}_{\mu}& = -2P = -2\partial^{\mu}C_{\mu},
\end{align}
where $P$ denotes the Pontryagin density, which equals the divergence of the Chern-Simons form
\begin{align}
\label{eq:chernsimons}
 C^{\mu}=-\frac{1}{16\pi^{2}}\varepsilon^{\mu\nu\rho\sigma}\tr(F^{A}_{\nu\rho}A_{\sigma}-\frac{2}{3}A_{\nu}A_{\rho}A_{\sigma})
\end{align} 
Thus, the conserved current $J^{5}_{\mu}=j^{5}_{\mu}+2C_{\mu}$ gives rise to a gauge dependent symmetry generator\footnote{The expression for $Q^{5}$ is heuristic, but there are known strategies to regularize such expressions \cite{SchroerCurrentCommutationRelations, MorchioChargeDensityAnd}.}
\begin{align}
\label{eq:gaugedependentcurrent}
Q^{5}=\int_{\Sigma}J^{5}_{0}d^{3}x\stackrel{\lambda}{\longmapsto}\int_{\Sigma}J^{5}_{0}+2n_{[\lambda]},\ \lambda\in\cG_{0}
\end{align}
and is therefore rejected. Here, $n_{[\lambda]}$ is the winding number of the extension of $\lambda$ to the 1-point compactification $\dot{\R}\ \!\!^{3}=\R\ \!\!\!^{3}\cup\{\infty\}=S^{3}$, i.e. $\lambda:S^{3}=\dot{\R}\ \!\!^{3}\rightarrow\uG$, which is defined by $\lambda(\infty)=1_{\uG}$, since $\lambda$ differs from $1_{\uG}$ only on a compact set. \\[0.1cm]
On the other hand, it is argued in \cite{MorchioChiralSymmetryBreaking} that this line of thought is incomplete in view of the results of Bardeen \cite{BardeenAnomalousCurrentsIn}, who showed that $J^{5}_{\mu}$ gives rise to a well-defined symmetry on the observable algebra in perturbation theory in local gauges. Furthermore, in \cite{MorchioChiralSymmetryBreaking} Morchio and Strocchi put forward a way to close this gap, which we will argue could apply in the framework of loop quantum gravity, as well. This is of particular interest in the setting of deparametrizing models, which provide an arena for the discussion of the standard model and related theories in the context of loop quantum gravity (see \cite{GieselScalarMaterialReference} for a review), and for which the Ashtekar-Isham-Lewandowski representation can provide the physical Hilbert space.\\[0.1cm]
The main ingredients necessary for a discussion of chiral symmetry breaking along the lines of \cite{MorchioChiralSymmetryBreaking}, are an algebra of (localized) observables $\mathfrak{A}$, containing unitary elements $U(\lambda)$ implementing the (localized) gauge transformations $\cG_{0}$, and a $1$-parameter group of chiral automorphisms $\rho_{\zeta}:\mathfrak{A}\rightarrow\mathfrak{A}$, interacting non-trivially with elements associated with large gauge transformation $\lambda,\ 0\neq[\lambda]\in\pi_{3}(\uG)\textup{(}\cong\mathds{Z}$ in many relevant cases, e.g. $SU(n),\ n\geq2$):
\begin{align}
\label{eq:largegaugeanomaly}
\rho_{\zeta}(U(\lambda)) & = e^{-i2\zeta n_{[\lambda]}}U(\lambda).
\end{align}
In the following, we will argue that, if we assume the existence of a $1$-parameter group of automorphisms of the form \eqref{eq:largegaugeanomaly} for the algebra $\mathfrak{A}_{\textup{LQG}}$ (or a slightly extended version of it), we will have all ingredients at our disposal. A discussion of the possibility to obtain a chiral symmetry \eqref{eq:largegaugeanomaly} in loop quantum gravity will be given elsewhere.
\subsection{An extension of the algebra $\mathfrak{A}_{\textup{LQG}}$}
\label{sec:extendedlqg}
We start our discussion with the observation that the algebra $\mathfrak{A}_{\textup{LQG}}$ admits an extension by operators $U(\lambda),\ \lambda\in\cG^{\textup{sa},0}_{\uP}$, representing the semi-analytic, compactly supported gauge transformations, in the following way:
\begin{Definition}
\label{def:gaugeextendedalgebra}
The extension $\cG^{\textup{sa},0}_{\uP}\rtimes\mathfrak{A}_{\textup{LQG}}$ of $\mathfrak{A}_{\textup{LQG}}$ is given along the lines of definition \ref{def:lqgweyl}, but with the additional elements $U(\lambda),\ \lambda\in\cG^{\textup{sa},0}_{\uP}$ and relations
\begin{align}
\label{eq:extendedlqg}
U(\lambda)^{*} & =U(\lambda^{-1}), &\ U(\lambda\circ\lambda') & =U(\lambda')U(\lambda), \\ \nonumber
U(\lambda)f & =\alpha_{\lambda}(f)U(\lambda), &\ U(\lambda)W_{S}(tn) & =\alpha_{\lambda}(W_{S}(tn))U(\lambda),
\end{align}
for any $f, W_{S}(tn)\in\mathfrak{A}_{\textup{LQG}}$. The action of the automorphims $\aut^{\textup{sa}}(\uP)$ extends to this algebra by conjugation on the gauge transformations $\cG^{\textup{sa},0}_{\uP}$, i.e.
\begin{align}
\alpha_{\chi}(U(\lambda)) & = U(\chi^{-1}\circ\lambda\circ\chi).
\end{align}
\end{Definition}
Evidently, there is an analogous construction on the basis of the holonomy-flux algebra $\mathfrak{P}_{\textup{LQG}}$,  and it is possible to extend by the automorphisms $\aut^{\textup{sa}}(\uP)$ in a similar way. As a simple corollary we have:
\begin{Corollary}
\label{cor:extendedrep}
The Ashtekar-Isham-Lewandowski representation $(\pi_{\omega_{0}},\fH_{\omega_{0}},\Omega_{\omega_{0}})$ extends to a representation of $\cG^{\textup{sa},0}_{\uP}\rtimes\mathfrak{A}_{\textup{LQG}}$ of $\mathfrak{A}_{\textup{LQG}}$, which can be defined by the (algebraic) state:
\begin{align}
\label{eq:extendedAILstate}
\omega^{\textup{ext}}_{0}(fW_{S}(n)...W_{S'}(n')U(\lambda)) & =\mu_{0}(f),\ \forall f\in\Cyl,\ W_{S}(n)...W_{S'}(n'),\ U(\lambda)\in\cG^{\textup{sa},0}_{\uP}\rtimes\mathfrak{A}_{\textup{LQG}}.
\end{align}
\begin{Proof}
This is immediate from the invariance properties of $\omega_{0}$.
\end{Proof}
\end{Corollary}
Interestingly, if we extend the algebra $\mathfrak{A}_{\textup{LQG}}$ only by the subgroup of gauge transformations close to the identity $\textup{Gauss}_{\uP}=\exp_{\cG_{\uP}}(\fG_{\uP})$, this will correspond to the inclusion of (smeared) generators of the gauge transformation into $\mathfrak{P}_{\textup{LQG}}$, and fits with their separate quantization (cf. \cite{ThiemannModernCanonicalQuantum}, cp. also \eqref{eq:infgaugeactionhol2}):
\begin{align}
\label{eq:quantumgauss}
G_{V}(\Lambda) & = \int_{V}\div^{A}_{T\Sigma}(E)(\Lambda),\ \Lambda\in\Gamma^{\textup{sa}}_{0}(\textup{Ad}(\uP_{|V})),\ V\subset\Sigma\ \textup{open\ and\ semi-analytic}.
\end{align}
Thus, the extension $\textup{Gauss}_{\uP}\rtimes\mathfrak{A}_{\textup{LQG}}$ appears to be natural from the point of view that the algebra $\mathfrak{A}_{\textup{LQG}}$ contains (smeared) functions of the (classical) variables $(A,\tilde{E})\in|\Lambda|^{1}T^{*}\mathcal{A}_{\uP}$.
\subsection{Chiral symmetry breaking and $\theta$-vacua for $\Sigma=\R^{3}$ \& $\uP=\R^{3}\times\uG$}
\label{sec:chiraltrivialbundle}
For the discussion of chiral symmetry breaking in the context of loop quantum gravity it is important to note, that the formalism, recalled here, is capable of treating gravitational and Yang-Mills degrees of freedom at the same time. At the given structural level, this is reflected in the choice of structure group $\uG$. Further differences would arise at the level of dynamics and the associated Hamiltonian constraints. In the following, we will not distinguish between the different types of degrees of freedom, and therefore in principle allow for chiral symmetry breaking w.r.t. the gravitational degrees of freedom. Furthermore, it is possible to include fermions into the treatment (cf. \cite{ThiemannKinematicalHilbertSpaces}), which points out a potential direction to investigate the existence of a gauge dependent chiral symmetry \eqref{eq:largegaugeanomaly}. Interestingly, the anomaly \eqref{eq:axialanomaly} can be generated by a gauge invariant regularization procedure by point-split objects like
\begin{align}
\label{eq:pointsplithol}
j^{5}_{\mu}(e(1),e(0)) & = i\bar{\Psi}(e(1))\gamma^{5}\gamma_{\mu}\hol^{A}_{e}\Psi(e(0)),
\end{align}
which have natural analogs in the loop quantum gravity framework (cf. \cite{MoralesTecotlLoopSpaceRepresentation, MoralesTecotlFermionsInQuantum, BaezQuantizationOfDiffeomorphism}). But, these objects behave complicated w.r.t. general automorphisms $\aut^{\textup{sa}}(\uP)$ \cite{ThiemannKinematicalHilbertSpaces}, which might restrict their applicability to deparametrized models.\\[0.25cm]
Let us now turn to the mechanism for chiral symmetry breaking in the loop quantum gravity framework. To simplify the discussion, we will restrict to a spatial manifold $\Sigma=\R^{3}$ and a trivial bundle $\uP=\R^{3}\times\uG$. The quantum field algebra will be $\textup{Gauss}_{\uP}\rtimes\mathfrak{A}_{\textup{LQG}}$ or $\cG^{\textup{sa},0}_{\uP}\rtimes\mathfrak{A}_{\textup{LQG}}$, and the existence of a chiral symmetry $\{\rho_{\zeta}\}_{\zeta\in\R}$ with the property \eqref{eq:largegaugeanomaly} will be assumed in the latter case. \\[0.1cm]
This has the important implication, that $\cG^{\textup{sa},0}_{\uP}\cong C^{\textup{sa}}_{0}(\R^{3},\uG)$. Thus every $\lambda\in\cG^{\textup{sa},0}_{\uP}$ determines uniquely a map (see above)
\begin{align}
\label{eq:gaugeaxtensionhomotopy}
g_{\lambda}:\dot{\R}\ \!\!^{3}=S^{3}\rightarrow\uG,
\end{align}
and a homotopy class $[\lambda]=[g_{\lambda}]\in\pi_{3}(\uG)$. From this point on, let us assume that $\pi_{3}(\uG)\cong\mathds{Z}$, which holds for $\uG=SU(n),\ n\geq2$ or $\uG=SO(n),\ n\geq3,\ n\neq4$\footnote{Note that this excludes the case $\uG=\textup{Spin}_{4},\ \pi_{3}(\textup{Spin}_{4})\cong\mathds{Z}\times\mathds{Z}$, which is important in the treatment of the new variables \cite{BodendorferNewVariablesFor1}.}. Then, $[\lambda]$ is uniquely determined by the winding number or instanton number \cite{ChruscinskiGeometricPhasesIn}
\begin{align}
\label{eq:windingnumberdef}
n_{[\lambda]}=\frac{1}{24\pi^{2}}\int_{\R^{3}}\tr(g^{-1}_{\lambda}dg_{\lambda}\wedge g^{-1}_{\lambda}dg_{\lambda}\wedge g^{-1}_{\lambda}dg_{\lambda}).
\end{align}
Gauge transformations $\lambda$ with $n_{[\lambda]}\neq0$ are called large gauge transformations.\\[0.1cm] 
Next, we analyze the difference between $\textup{Gauss}_{\uP}$-invariance and gauge invariance for $\textup{Gauss}_{\uP}\rtimes\mathfrak{A}_{\textup{LQG}}$. Again, the argument follows Morchio and Strocchi \cite{MorchioChiralSymmetryBreaking}, who exploit the localization properties of operators in the quantum algebra, which is also possible for the algebra $\textup{Gauss}_{\uP}\rtimes\mathfrak{A}_{\textup{LQG}}$.
\begin{Lemma}
\label{lem:gaussgaugeinv}
Any $\textup{Gauss}_{\uP}$-invariant state $\omega$ on $\textup{Gauss}_{\uP}\rtimes\mathfrak{A}_{\textup{LQG}}$ is also gauge invariant, and the large gauge transformations are unitarily implemented in the GNS representation $(\pi_{\omega},\fH_{\omega},\Omega_{\omega})$. Furthermore, any $\textup{Gauss}_{\uP}$-invariant operator in $\textup{Gauss}_{\uP}\rtimes\mathfrak{A}_{\textup{LQG}}$ is also gauge invariant.
\begin{Proof}
Let $\lambda\in C^{\textup{sa}}_{0}(\R^{3},\uG)$ be a large gauge transformation, and define $\lambda_{a}(x)=\lambda(x-a),\ x,a\in\R^{3}$. Then, $\lambda\cdot\lambda_{a}^{-1}$ and $\lambda_{a}^{-1}\cdot\lambda$ are $\textup{Gauss}_{\uP}$ transformations. This implies
\begin{align}
\label{eq:gaussimpliesgauge}
(\omega\circ\alpha_{\lambda})(f) & = (\omega\circ\alpha_{\lambda})(\alpha_{\lambda_{a_{f}}^{-1}}(f)) = (\omega\circ\alpha_{\lambda_{a_{f}}^{-1}\cdot\lambda})(f) \\ \nonumber
 & = \omega(f) \\[0.25cm]
(\omega\circ\alpha_{\lambda})(W_{S}(tn)) & = (\omega\circ\alpha_{\lambda})(\alpha_{\lambda_{a_{S,n}}^{-1}}(W_{S}(tn))) = (\omega\circ\alpha_{\lambda_{a_{S,n}}^{-1}\cdot\lambda})(W_{S}(tn)) \\ \nonumber
 & = \omega(W_{S}(tn)). \\[0.25cm]
(\omega\circ\alpha_{\lambda})(U(\lambda')) & = (\omega\circ\alpha_{\lambda})(\alpha_{\lambda_{a_{\lambda'}}^{-1}}(U(\lambda'))) = (\omega\circ\alpha_{\lambda_{a_{\lambda'}}^{-1}\cdot\lambda})(U(\lambda')) \\ \nonumber
 & = \omega(U(\lambda')),
\end{align}
where $f,W_{S}(tn),U(\lambda')$ are generators of $\textup{Gauss}_{\uP}\rtimes\mathfrak{A}_{\textup{LQG}}$, and we chose $a_{f}, a_{S,n}, a_{\lambda'}\in\R^{3}$ in accordance with the respective localization regions. The unitary implementability follows from a standard argument. The other statement follows from the same argument.
\end{Proof}
\end{Lemma}
The implementers of the (large) gauge transformations are unique up to phases in irreducible (or factorial) representations, i.e. w.r.t. to pure (or primary), $\textup{Gauss}_{\uP}$-invariant states $\omega$, of $\textup{Gauss}_{\uP}\rtimes\mathfrak{A}_{\textup{LQG}}$.\\[0.1cm]
In view of this result, and corollary \ref{cor:extendedrep}, we will use the algebra $\cG^{\textup{sa},0}_{\uP}\rtimes\mathfrak{A}_{\textup{LQG}}$ to discuss the spontaneous breakdown of the chiral symmetry and its relation to the topology of $\cG^{\textup{sa},0}_{\uP}$. To this end, we need a further result concerning the implementers of the (large) gauge transformations.
\begin{Proposition}[cp. \cite{MorchioChiralSymmetryBreaking}]
\label{prop:localimplementers}
In a GNS representation of a $\textup{Gauss}_{\uP}$-invariant state with $\textup{Gauss}_{\uP}$-invariant GNS-vacuum
\begin{align}
\label{eq:gaugeinvariantvacuum}
\pi_{\omega}(U(\lambda))\Omega_{\omega} & =\Omega_{\omega},\ \lambda\in\textup{Gauss}_{\uP},
\end{align}
the implementers $\pi_{\omega}(U(\lambda))$ of the gauge transformations $\lambda\in\cG^{\textup{sa},0}_{\uP}$ are of the form:
\begin{align}
\label{eq:localimplementer}
\pi_{\omega}(U(\lambda))\Omega_{\omega} & = C^{\omega}_{n_{[\lambda]}}\Omega_{\omega}.
\end{align}
The non-trivial elements $C^{\omega}_{n_{[\lambda]}}$ are central, and belong to the strong closure of $\pi_{\omega}(\cG^{\textup{sa},0}_{\uP}\rtimes\mathfrak{A}_{\textup{LQG}})$. Furthermore, we have
\begin{align}
\label{eq:localimplementeralgebra}
C^{\omega}_{n_{[\lambda]}}C^{\omega}_{n_{[\lambda']}} & = C^{\omega}_{n_{[\lambda]}+n_{[\lambda']}}.
\end{align}
\begin{Proof}
For $\lambda\in\cG^{\textup{sa},0}_{\uP}$, we (densely) define
\begin{align}
\label{eq:gaussimplementer}
S^{\omega}(\lambda)\pi_{\omega}(O)\Omega_{\omega} & = \pi_{\omega}(\alpha_{\lambda}(O))\Omega_{\omega},\ O\in\cG^{\textup{sa},0}_{\uP}\rtimes\mathfrak{A}_{\textup{LQG}}.
\end{align}
Clearly, $S^{\omega}(\lambda)$ is isometric on dense subspace of $\fH_{\omega}$, and extends to an unitary element of $\mathfrak{B}(\fH_{\omega})$, which we denote by $S^{\omega}(\lambda)$, as well. Then, by the same argument as in lemma \ref{lem:gaussgaugeinv}, we have
\begin{align}
\label{eq:stronglimit}
S^{\omega}(\lambda) & = \slim_{|a|\rightarrow\infty}\pi_{\omega}(U(\lambda^{-1}_{a}\cdot\lambda)).
\end{align}
This operator has the properties
\begin{align}
\label{eq:gaussproperties}
S^{\omega}(\lambda)\pi_{\omega}(O)S^{\omega}(\lambda)^{*} & = \pi_{\omega}(\alpha_{\lambda}(O)),\ O\in\cG^{\textup{sa},0}_{\uP}\rtimes\mathfrak{A}_{\textup{LQG}}, \\ \nonumber
S^{\omega}(\lambda)\Omega_{\omega} & = \Omega_{\omega}.
\end{align}
These allow us to define $C^{\omega}_{n_{[\lambda]}}=\pi_{\omega}(U(\lambda))S^{\omega}(\lambda)^{*}$, which are central and belong to the strong closure of \\ $\pi_{\omega}(\cG^{\textup{sa},0}_{\uP}\rtimes\mathfrak{A}_{\textup{LQG}})$. Clearly, the $C^{\omega}_{n_{[\lambda]}}$'s depend only on the topological quantities $n_{[\lambda]}$ and satisfy \eqref{eq:localimplementeralgebra}, since for any $\lambda,\lambda'\in\cG^{\textup{sa},0}_{\uP}$ with $n_{[\lambda]}=n_{[\lambda']}$ the operator $\pi_{\omega}(U(\lambda')^{*})\pi_{\omega}(U(\lambda))$ represents a $\textup{Gauss}_{\uP}$ transformation, which leaves $\Omega_{\omega}$ invariant. The (general) non-triviality of the elements $C^{\omega}_{n_{[\lambda]}}$ follow from \eqref{eq:largegaugeanomaly}.
\end{Proof}
\end{Proposition}
The proposition implies that the central elements $\{C^{\omega}_{n}\}_{n\in\mathds{Z}}$ represent the quotient $\cG^{\textup{sa},0}_{\uP}/\textup{Gauss}_{\uP}$. Similar to the preceding sections, we find non-trivial, central elements associated with the algebra $\cG^{\textup{sa},0}_{\uP}\rtimes\mathfrak{A}_{\textup{LQG}}$, reflecting the topology of the group of gauge transformations $\cG^{\textup{sa},0}_{\uP}$. The property \eqref{eq:largegaugeanomaly} of the chiral automorphisms leads to their spontaneous breakdown w.r.t. pure (or primary), $\textup{Gauss}_{\uP}$-invariant states, and the appearance of the $\theta$-sectors.
\begin{Corollary}[cp. \cite{MorchioChiralSymmetryBreaking}]
\label{cor:chiralsymmetrybreaking}
Given a pure (or primary), $\textup{Gauss}_{\uP}$-invariant state $\omega$ on $\cG^{\textup{sa},0}_{\uP}\rtimes\mathfrak{A}_{\textup{LQG}}$, the chiral automorphisms $\{\rho_{\zeta}\}_{\zeta\in\R}$ are necessarily spontaneouly broken. Moreover, every such state is labeled by an angle $\theta\in[0,\pi)$, $C^{\omega}_{n}=e^{i2n\theta}\cdot\mathds{1}_{\fH_{\omega}}$. The GNS representation of a chirally invariant, $\textup{Gauss}_{\uP}$-invariant state $\omega'$ admits a central decomposition, w.r.t. $C^{\omega'}_{1}$
\begin{align}
\label{eq:chiralcentraldecomposition}
\fH_{\omega'} & = \int_{[0,\theta)}\fH_{\theta}d\mu(\theta),\ C^{\omega'}_{n}\fH_{\theta}=e^{i2n\theta}\fH_{\theta},
\end{align}
with translation invariant measure $\mu$.
\begin{Proof}
Assume that the chiral symmetry is unbroken. Then, we find a $1$-parameter group of unitaries $\{U^{\omega}_{5}(\zeta)\}_{\zeta\in\R}$ that implements the symmetry by conjugation
\begin{align}
\label{eq:chiralimplementers}
\pi_{\omega}(\rho_{\zeta}(O)) & = U^{\omega}_{5}(\zeta)\pi_{\omega}(O)U^{\omega}_{5}(\zeta)^{*},\ O\in\cG^{\textup{sa},0}_{\uP}\rtimes\mathfrak{A}_{\textup{LQG}},\ \zeta\in\R.
\end{align}
This leads to a unique extension of the $\rho_{\zeta}$'s to the strong closure of $\pi_{\omega}(\cG^{\textup{sa},0}_{\uP}\rtimes\mathfrak{A}_{\textup{LQG}})$, and we find by \eqref{eq:largegaugeanomaly} and \eqref{eq:stronglimit} (since $n_{[\lambda]}=n_{[\lambda_{a}]}$):
\begin{align}
\label{eq:chiralgauss}
\rho_{\zeta}(S^{\omega}(\lambda)) & = S^{\omega}(\lambda),\ \zeta\in\R.
\end{align}
This implies, again by \eqref{eq:largegaugeanomaly} and the definition of $C^{\omega}_{n}$:
\begin{align}
\label{eq:centralchiraltransformation}
\rho_{\zeta}(C^{\omega}_{n_{[\lambda]}}) & = e^{-i2\zeta n_{[\lambda]}}C^{\omega}_{n_{[\lambda]}},\ \zeta\in\R,
\end{align}
which is incompatible with the purity (or primarity) of $\omega$, as this implies irreducibility (or factoriality) of $(\pi_{\omega},\fH_{\omega},\Omega_{\omega})$, and thus $C^{\omega}_{n_{[\lambda]}}=e^{i2n_{[\lambda]}\theta}\cdot\mathds{1}_{\fH_{\omega}},\ \theta\in[0,\pi)$.\\[0.1cm]
The central decomposition \eqref{eq:chiralcentraldecomposition} follows from the observation that \eqref{eq:chiralgauss} implies $\sigma(C^{\omega'}_{1})=\{e^{i2\theta}\ |\ \theta\in[0,\pi)\}$. The unitaries $U^{\omega'}_{5}(\zeta)$ act as intertwiners between the $\theta$-sectors:
\begin{align}
\label{eq:thetaintertwiners}
U^{\omega'}_{5}(\zeta)\fH_{\theta} & = \fH_{\theta-\zeta\ \textup{mod}\ \pi}.
\end{align}
\end{Proof}
\end{Corollary}
\section{Conclusions and perspectives}
\label{sec:conclusions}
To conclude the article, we comment on our findings in the various sections, and offer some future perspectives.\\[0.1cm]
Section \ref{sec:pre} mainly provided a review of the mathematical structures behind the (canonical) formulation of loop quantum gravity with two exceptions: Equation \eqref{eq:higherfluxcom}, which states an algebraic relation among the flux vector fields that affects the representation theory of the holonomy-flux algebra $\mathfrak{P}_{\textup{LQG}}$ in a non-trivial way (see section \ref{sec:KSreps}), and lemma \ref{lem:gaugeinvariantstates}, which shows that Hilbert space representations of $\mathfrak{P}_{\textup{LQG}}$ and its Weyl form $\mathfrak{A}_{\textup{LQG}}$ induced by gauge invariant states $\omega$ are necessarily discontinuous w.r.t. to the spin network functions, i.e. the two-point function ``$\omega(A(x)A(y))$'' of the ``quantum connection'' $A$ cannot exist in such representations. The latter result is in accordance with results in of quantum field theory in the temporal gauge \cite{LoeffelholzMathematicalStructureOf}, where the only alternative appears to be the use of (non-positive) Krein space representations, e.g. the Feynman-Gupta-Bleuler quantization of QED. Thus, it would be interesting, whether such an alternative is possible in loop quantum gravity, as well, and how it connects to the standard approach.\\[0.1cm]
In section \ref{sec:centers}, we focused on aspects of the representation theory of $\mathfrak{A}_{\textup{LQG}}$ with an emphasis on the presence of non-trivial central operators, and their relation to topological and geometrical structures of the structure group $\uG$. We found, that a non-trivial first homotopy group $\pi_{1}(\uG)$, supplemented by an associated bundle covering, can be related to the existence of a field algebra extension $\mathfrak{F}_{\textup{LQG}}$, that can be used to generate new, inequivalent representations from existing ones with the help of charge automorphisms defined by the adjoint action of unitary, charged fields. While this construction works well for Abelian structure groups, where it offers a new perspective on the Koslowski-Sahlmann representations and the $\varepsilon$-sectors of loop quantum cosmology, it is accompanied by further difficulties in the non-Abelian case, which are due to restrictive topological and geometrical properties of $\uG$ (see proposition \ref{prop:compactweylperfectness} \& \ref{prop:compactfluxperfectness}). Especially, there might exist no suitable unitary, charged fields in the extension $\mathfrak{F}_{\textup{LQG}}$ to define charge automorphisms. In the future, it could be gratifying to investigate the algebraic structure of $\mathfrak{A}_{\textup{LQG}}$ resp. $\mathfrak{P}_{\textup{LQG}}$ on a deeper level, e.g. its structure of ideals, its universal enveloping von Neumann algebra etc., to improve control on the representation theory and the possible dynamics supported by the algebra. Especially, in view of the deparametrizing models (see \cite{GieselScalarMaterialReference} for an overview), where $\mathfrak{A}_{\textup{LQG}}$ and $\mathfrak{P}_{\textup{LQG}}$ become algebras of elementary observables, instead of purely kinematical objects, such an analysis will offer immediate insight into physical questions.\\[0.1cm]
We continued our analysis of the Koslowski-Sahlmann representations, started in section \ref{sec:centers} for Abelian structure groups, in section \ref{sec:KSreps}, where we concentrated on the non-Abelian case. We showed, that the general line of thought, which places these representations into the framework of section \ref{sec:centers}, bifurcates for non-Abelian structure groups, and one is left with two possible interpretations:
\begin{itemize}
	\item[1.] The Koslowski-Sahlmann representations are defined for an (centrally) extended algebra, and the elementary operators $Y_{n}(S)$ are identified with the fluxes $E_{n}(S)$.
	\item[2.] The Koslowski-Sahlmann representations are defined for the holonomy-flux algebra, but the elementary operators $Y_{n}(S)$ are identified with shifted fluxes $E_{n}(S)+i\int_{S}*E^{(0)}(n)$.
\end{itemize}
This bifurcation is explained by the fact that the shift transformation $\rho_{E^{(0)}}:E_{n}(S)\mapsto E_{n}(S) + i\int_{S}*E^{(0)}(n)$ is not a *-automorphism of $\mathfrak{P}_{\textup{LQG}}$, but only an affine transformation, in the non-Abelian setting. Thus, the first point of view represents the idea to define a modified holonomy-flux algebra $\mathfrak{P}^{E^{(0)}}_{\textup{LQG}}$, s.t. \mbox{$\rho_{E^{(0)}}:\mathfrak{P}^{E^{(0)}}_{\textup{LQG}}\rightarrow\mathfrak{P}_{\textup{LQG}}$} becomes a *-isomorphism, while the second point of view changes the interpretation of the elementary operators $Y_{n}(S)$ of the Koslowski-Sahlmann framework. Clearly, the second option avoids the obstruction posed by corollary \ref{cor:nocentralterm}, and shows that the Koslowski-Sahlmann representations reduce to the Ashtekar-Isham-Lewandowski representation for the shifted fluxes, but it forbids the treatment of gauge and automorphism invariance along the lines of \cite{KoslowskiLoopQuantumGravity}, as well (see the summary at the end of section \ref{sec:KSreps}). The first option, which offers a richer mathematical structure, suffers from the fact that the ``central extension'' of the holonomy-flux algebra is only a necessary ingredient, but probably not sufficient due to further higher order commutation relations imposed by the basic commutation rule
\begin{align}
\label{eq:basicrule}
[Y_{n}(S),f] & = E_{n}(S)\cdot f.
\end{align}
Thus, this approach is weakened, because control on all higher order relation appears to be out of reach at the present stage.\\
Nevertheless, it is interesting to analyse the recent work on the Koslowski-Sahlmann representation, which is focused on the implementation of diffeomorphisms and possible applications to asymptotically flat scenarios \cite{VaradarajanTheGeneratorOf, CampigliaTheKoslowskiSahlmann1, CampigliaTheKoslowskiSahlmann2, CampigliaAQuantumKinematics, SenguptaQuantumGeometryWith, SenguptaAsymptoticFlatnessAnd}, in view of our findings. Especially, in \cite{CampigliaTheKoslowskiSahlmann2} it has been pointed out that it is possible to introduce a slightly modified algebra, $\fA^{\uB}_{\LQG}$, called holonomy-background exponential-flux algebra\index{algebra!holonomy-background exponential-flux|textbf}, which admits the Koslowski-Sahlmann representation as a true representation. The latter is possible because of a modification of the generators of $\fA^{\uB}_{\LQG}$ in comparison to $\fA_{\LQG}$, not only involving the fluxes, $E_{n}(S)$, but also the cylindrical functions, which are made background dependent, $E^{(0)}$, by means of so-called background exponentials
\begin{align}
\label{eq:backgroundexp}
\beta_{E^{(0)}}(A) & = e^{i\int_{\Sigma}E(A).} 
\end{align}
These background exponentials lead to an additional $U(1)^{N}$-factor accompanying the structure group $G$, on which the cylindrical functions are based ($N$ is the number of background fields). Thus, the modification of the fluxes can be realised by additional derivations that act on the $U(1)^{N}$-factors, avoiding our corollary \ref{cor:nocentralterm} on central extensions of $\fA_{\LQG}$. Put differently, the additional $U(1)^{N}$-structure resolves the obstruction posed by \eqref{eq:higherJacobiidentities}, rendering it invalid in $\fA^{\uB}_{\LQG}$.\\[0.1cm]
Finally, in section \ref{sec:theta}, we applied the general formalism of section \ref{sec:centers} to adapt the discussion of chiral symmetry breaking and $\theta$-vacua by Morchio and Strocchi \cite{MorchioChiralSymmetryBreaking} to the framework of loop quantum gravity. We showed that under the assumption of an anomalous, chiral symmetry \eqref{eq:largegaugeanomaly} this adaption is possible, and has some of the expected properties (a discussion of the Goldstone spectrum of the generator of the chiral symmetry is missing). Our analysis is intended to stimulate the discussion of gauge anomalies in loop quantum gravity, especially in the matter sector, because anomalies have important physical consequences for the matter content of the standard model. Thinking of the semi-classical limit of loop quantum gravity, it is necessary to make contact with the predictions of quantum field theory, and to offer an explanation of the consequences of anomalies in the latter, e.g. the solution of the $U(1)$-problem and the restriction of matter to so-called safe representations. Thus, in spite of the fact that an anomaly like \eqref{eq:largegaugeanomaly} appears to be a rather strong requirement, we would expect that a structure of this type arises in loop quantum gravity, at least in a limiting sense connected to the aforesaid semi-classical limit. A natural starting point for an investigation, of how anomalies could occur in loop quantum gravity, is suggested by symmetry generating currents of the form \eqref{eq:pointsplithol}, which, on the one hand, are natural objects in the framework of loop quantum gravity and, on the other hand, are the central objects in the study of anomalies in quantum field theory. More precisely, an understanding of the coincidence limit of these point-split currents, possibly in combination with a semi-classical limit, could offer first insights. At a preliminary stage, it might be easiest to consider these objects in the context of deparametrizing models (see \cite{GieselScalarMaterialReference} for an overview), which avoid complications due to the diffeomorphism and Hamiltonian constraints. A further simplification might be achieved, if the discussion was restricted to cosmological or other symmetry reduced models.
\section{Acknowledgements}
\label{sec:ack}
We thank Norbert Bodendorfer, Detlev Buchholz and Hanno Sahlmann for helpful comments and suggestions. Furthermore, we thank Stefan Hollands for pointing out the importance of anomalies in the relation to quantum field theory to one of us. AS gratefully acknowledges financial support by the Ev. Studienwerk e.V.. This work was supported in parts by funds from the Friedrich-Alexander-University, in the context of its Emerging Field Initiative, to the Emerging Field Project ``Quantum Geometry’’.
\bibliography{ncweylv2.bbl}
\bibliographystyle{elsarticle-num}

\end{document}